\documentclass{aa} 

\usepackage{natbib,twoopt}
\usepackage[varg]{txfonts}
\usepackage{supertabular}
\usepackage{multicol}
\usepackage[breaklinks=true]{hyperref} 
\bibpunct{(}{)}{;}{a}{}{,} 
\makeatletter
\newcommandtwoopt{\citeads}[3][][]{\href{http://adsabs.harvard.edu/abs/#3}%
{\def\hyper@linkstart##1##2{}%
\let\hyper@linkend\@empty\citealp[#1][#2]{#3}}}
\newcommandtwoopt{\citepads}[3][][]{\href{http://adsabs.harvard.edu/abs/#3}%
{\def\hyper@linkstart##1##2{}%
\let\hyper@linkend\@empty\citep[#1][#2]{#3}}}
\newcommandtwoopt{\citetads}[3][][]{\href{http://adsabs.harvard.edu/abs/#3}%
{\def\hyper@linkstart##1##2{}%
\let\hyper@linkend\@empty\citet[#1][#2]{#3}}}
\newcommandtwoopt{\citeyearads}[3][][]%
{\href{http://adsabs.harvard.edu/abs/#3}
{\def\hyper@linkstart##1##2{}%
\let\hyper@linkend\@empty\citeyear[#1][#2]{#3}}}
\makeatother
\usepackage{graphicx}
\usepackage{txfonts}
\usepackage[space]{grffile}
\usepackage{latexsym}
\usepackage{textcomp}
\usepackage{multirow,booktabs}
\usepackage{amsfonts,amsmath,amssymb}
\usepackage{url}
\hypersetup{draft}
\hypersetup{colorlinks=false,pdfborder={0 0 0}}
\usepackage[utf8]{inputenc}
\usepackage[english]{babel}
\usepackage[]{amsmath}
\usepackage[]{txfonts}
\usepackage[]{graphicx}
\usepackage{pdflscape}
\usepackage[]{subfigure}
\usepackage[]{color}
\usepackage[]{bm}
\usepackage[]{natbib}

\newcommand{\simgt}%
        {\,\hbox{\lower0.6ex\hbox{$\sim$}\llap{\raise0.6ex\hbox{$>$}}}\,}
\def\simlt{\lower.5ex\hbox{\ltsima}}
\def\ltsima{$\; \buildrel < \over \sim \;$}
\def\simlt{\lower.5ex\hbox{\ltsima}}
\def\gtsima{$\; \buildrel > \over \sim \;$}
\def\simgt{\lower.5ex\hbox{\gtsima}}
\def\kms{\mbox{km s$^{-1}$}}

\def\lsol{\mbox{L$_\odot$}}
\newcommand{\lint}{\mbox{$L_{int}$}} 
\def\msol{\mbox{M$_\odot$}}
\def\lbol{\mbox{$L_{\mbox{\tiny bol}}$}}

\def\arcsec{\hbox{$^{\prime\prime}$}}

\def\lul{\textsuperscript\textdagger}
\def\fp{\textsuperscript{$\diamondsuit$}}
\def\mic{$\mu$m}

\newcommand{\inside}{_{\mbox{\tiny i}}}

\begin{document}

   \title{Characterizing young protostellar disks with the CALYPSO IRAM-PdBI survey: large Class 0 disks are rare \thanks{Based on observations carried out with the IRAM Plateau de Bure Interferometer. IRAM is supported by INSU/CNRS (France), MPG (Germany), and IGN (Spain).}}

   \author{A. J. Maury
          \inst{1,2}
          \and Ph. Andr\'e\inst{1}
          \and L. Testi\inst{3,6}
          \and S. Maret\inst{4}
          \and A. Belloche\inst{5}
          \and P. Hennebelle\inst{1}
                  \and S. Cabrit\inst{8,4}
          \and C. Codella\inst{6}
          \and F. Gueth\inst{7}
          \and L. Podio\inst{6}
          \and S. Anderl\inst{4}
          \and A. Bacmann\inst{4}
          \and S. Bontemps\inst{9}
          \and M. Gaudel\inst{1}
              \and B. Ladjelate\inst{10}
          \and C. Lef\`evre\inst{7}
          \and B. Tabone\inst{8}
          \and B. Lefloch\inst{4}
          }

   \institute{AIM, CEA, CNRS, Universit\'e Paris-Saclay, Universit\'e Paris Diderot, Sorbonne Paris Cit\'e, F-91191 Gif-sur-Yvette, France
\email{anaelle.maury@cea.fr}
\and Harvard-Smithsonian Center for Astrophysics, Cambridge, MA 02138, USA
\and ESO, Karl Schwarzschild Strasse 2, 85748 Garching bei M\"unchen, Germany
\and Universit\'e Grenoble Alpes, CNRS, IPAG, 38000 Grenoble, France
\and Max-Planck-Institut f\"ur Radioastronomie, Auf dem H\"ugel 69, 53121 Bonn, Germany
\and INAF – Osservatorio Astrofisico di Arcetri, Largo E. Fermi 5, 50125 Firenze, Italy
\and Institut de Radioastronomie Millim\'etrique (IRAM), 38406 Saint-Martin d’H\`eres, France
\and LERMA, Observatoire de Paris, PSL Research University, CNRS, Sorbonne Universit\'e, UPMC Univ. Paris 06, 75014 Paris, France
\and OASU/LAB-UMR5804, CNRS, Universit\'e Bordeaux, 33615 Pessac, France
\and Institut de RadioAstronomie Millim\'etrique (IRAM), Granada, Spain
}

   \date{Received May 30 2018; accepted October 22 2018}
 
  \abstract
   {Understanding the formation mechanisms of protoplanetary disks and multiple systems and also their pristine properties are key questions for modern astrophysics. The properties of the youngest disks, embedded in rotating infalling protostellar envelopes, have largely remained unconstrained up to now.}
   %
   {We aim to observe the youngest protostars with a spatial resolution that is high enough to resolve and characterize the progenitors of protoplanetary disks. This can only be achieved using submillimeter and millimeter interferometric facilities. In the framework of the IRAM Plateau de Bure Interferometer survey CALYPSO, we have obtained subarcsecond observations of the dust continuum emission at 231 GHz and 94 GHz for a sample of 16 solar-type Class\,0 protostars.
   }
   {In an attempt to identify disk-like structures embedded at small scales in the protostellar envelopes, we modeled the dust continuum emission visibility profiles using Plummer-like envelope models and envelope models that include additional Gaussian disk-like components.}
   {Our analysis shows that in the CALYPSO sample, 11 of the 16 Class\,0  protostars are better reproduced by models including a disk-like dust continuum component contributing to the flux at small scales, but less than 25\% of these candidate protostellar disks are resolved at radii > 60 au. Including all available literature constraints on Class\,0 disks at subarcsecond scales, we show that our results are representative: most ($>$ 72\% in a sample of 26 protostars) Class\,0 protostellar disks are small and emerge only at radii < 60 au. We find a multiplicity fraction of the CALYPSO protostars $\simlt 57\% \pm 10\%$ at the scales 100--5000 au, which generally agrees with the multiplicity properties of Class I protostars at similar scales.}
   {We compare our observational constraints on the disk size distribution in Class\,0 protostars to the typical disk properties from protostellar formation models. If Class 0 protostars contain similar rotational energy as is currently estimated for prestellar cores, then hydrodynamical models of protostellar collapse systematically predict a high occurrence of large disks. Our observations suggest that these are rarely observed, however. Because they reduce the centrifugal radius and produce a disk size distribution that peaks at radii $<100$ au during the main accretion phase, magnetized models of rotating protostellar collapse are favored by our observations.}

\titlerunning{ CALYPSO view of Class\,0 protostellar disks}
\authorrunning{A.J. Maury et al.} 

   \maketitle
%
\section{Introduction}

\noindent{\it{Class 0 protostars and the formation of protostellar disks}}\\
Understanding the first steps in the formation of protostars and protoplanetary disks is a great unsolved problem of modern astrophysics. 
Observationally, the key to constraining protostar formation models lies in high-resolution studies of the youngest protostars. 
Class~0 objects, which were originally discovered at millimeter wavelengths, 
are believed to be the youngest known accreting protostars \citep{Andre93}. 
Because they are observed only $t \simlt 4-9 \times 10^{4}$ yr after the formation of a central hydrostatic protostellar object 
\citep{Evans09,Maury11} while most of their mass is still in the form of a dense core/envelope ($M_{env} > M_\star $),  
Class~0 protostars are believed to be representative of the main accretion phase of protostellar evolution 
and are likely to retain detailed information on the initial conditions of protostellar collapse (see review by \citealt{Andre00,Dunham14}).

High-resolution studies of Class~0 protostars are also key to constraining theoretical models for the formation of protostellar disks.
At the simplest level, 
the formation of circumstellar disks is a natural consequence of the conservation of angular momentum during the collapse 
of rotating protostellar envelopes in the course of the main accretion phase \citep{Cassen81,Terebey84}. 
Hydrodynamic simulations show that in the absence of magnetic fields, rotationally supported disks form and quickly grow 
to reach large radii $> 100$ au after a few thousand years \citep{Yorke99}. 
These hydrodynamical disks are often massive enough to be gravitationally unstable, and their fragmentation has been suggested 
to contribute to the formation of brown dwarfs and multiple stellar systems \citep{Stamatellos07c,Vorobyov10}.
On the other hand, early ideal magnetohydrodynamics (MHD) numerical simulations describing the protostellar collapse of magnetized envelopes 
had difficulties to form rotationally supported disks at scales $r>10$ au because of strong magnetic braking: the increase in magnetic energy as 
field lines are dragged inward during protostellar collapse \citep{Galli06,Hennebelle09}  reduces the envelope rotation and delays the formation of large disks.
This so-called ``magnetic braking catastrophe'' was  quickly mitigated, however, by including non-uniform initial conditions such as a
magnetic field that is misaligned with the core rotation axis, transonic turbulent cores, or the treatment of radiative transfer \citep{Joos12,Seifried12,Bate14,Machida14}. 
More realistic numerical simulations including non-ideal MHD physics that allows dissipating and/or decoupling magnetic fields from the inner protostellar environment 
have recently been developed by several groups \citep{Machida11,Dapp12,Masson12,LiPPVI}.
Most numerical MHD studies now agree that including one or several  resistive or dissipative effects (e.g., ambipolar diffusion, Ohmic dissipation, or the Hall effect) allows small ($10<r<100$ au) rotationally supported disks to be formed during magnetized protostellar collapse \citep{Machida14,Tsukamoto15b,Masson16}.
The exact ingredients responsible for early disk properties during the main accretion phase remain widely debated, however.

From the observational point of view, young stellar objects (YSOs) have been studied in great detail in recent years, showing that large disks 
with radii $\simgt 100$ au are common in Class~II \citep[e.g.][]{Andrews09,Isella09,Ricci10,Spezzi13} and 
Class~I objects \citep{Wolf08,Jorgensen09,Takakuwa12,Eisner12,Lee-F16,Sakai16}. 
However, we still lack good constraints on the properties 
of the progenitors of these disks during earlier phases of evolution: it has been difficult to observationally characterize the first stages of disk formation around Class 0 protostars because emission from 
the protostellar envelope dominates at most scales that are probed by single-dish telescopes \citep{Motte01a} or early interferometric observations \citep{Looney00}. 
 Long-wavelength observations are required to image deeply embedded disks and peer through dense protostellar envelopes. Subarcsecond resolution 
is needed to match the disk sizes at the typical distances of nearby star-forming clouds ($d \sim \,$100--400 pc), 
as is high sensitivity to detect the weak fluxes of the youngest disks.
Until recently, the small number of Class 0 protostars in nearby clouds and their relatively weak emission on small scales has restricted the millimeter interferometric studies that are required to reach subarcsecond (or $\la 100\,$au) resolution to the most extreme objects. 
For example, a survey of bright 
Class 0 and Class I protostars with the Sub-Millimeter Array (SMA) by \citet{Jorgensen09} attributed the detection of compact dust continuum 
emission components, all unresolved at the $\sim 2\arcsec$ scales that these data probe, to the potential presence of disks with masses 
between 0.002 and 0.5 $M_{\sun}$ during the Class 0 phase. 
This simple interpretation was questionable since modeling of the millimeter continuum emission from Class 0 protostars sometimes 
indicates that an irregular density structure (e.g., complex envelope substructure or radial density enhancements at small scales) 
can lead to additional compact continuum emission in Class 0 protostars without the need of a disk component \citep{Chiang08,Maury14}.
Only the recent advent of powerful interferometric facilities with kilometer baselines at (sub)millimeter (submm) wavelengths has allowed
the inner envelopes of Class 0 protostars to be explored
at resolutions and sensitivities that are sufficient to distinguish envelope emission from resolved disk emission at the relevant scales (20--200 au).

A pilot high-resolution study of five Class 0 protostars in Taurus and Perseus 
was carried out by \citet{Maury10} with the IRAM Plateau de Bure Interferometer (PdBI). 
No large $r>100$ au disks or protobinaries with separations $50<a<500$ au were detected. 
\citet{Maury10}  concluded that the formation of protostellar disks and the fragmentation of dense cores into multiple systems at scales 50-500 au 
might be largely modified by magnetic fields during the main accretion phase.The apparent lack of large $r>100$ au Class 0 disks could only be reproduced when magnetic-braking effects were included \citep{Hennebelle08b}, while pure hydrodynamical models produced too many large $r>100$ au disks. 
Subsequently, \citet{Maury14} showed that any disk component in the NGC1333-IRAS2A protostar would need to be $\simlt40$ au 
to reproduce the radial profile of the millimeter dust continuum emission observed with the PdBI. 
\citet{Maret14} modeled the kinematics observed in methanol emission lines toward the same source
and found that no significant rotation pattern was detected on similar scales. This 
confirmed the absence of a large disk in this Class~0 protostar.

The fast improvement of interferometric facilities such as ALMA has recently allowed a few other high-resolution studies to be made. 
A handful of Class 0 protostars have been proposed to have resolved disk-like rotation in their inner envelopes: while VLA 1623, HH212, and L1448-NB have 
been suggested to harbor Keplerian-like kinematics at scales $40<r<100$ au \citep{Murillo13,Codella14b,Tobin16b}, the most convincing case for a resolved protostellar disk was found in the L1527 Class 0/I protostar, where \citet{Ohashi14} found a transition from a rotating infalling envelope 
to Keplerian motions in a disk at radii $50-60$ au. 
On the other hand, \citet{Yen15b} concluded that the disk in the Class 0 protostar B335 must have a radius  $r\simlt10$ au 
to reproduce the absence of a Keplerian pattern from the velocity field observed with ALMA at subarcsecond scale in this edge-on source. 

To summarize, disks in young Class 0 protostars have largely remained elusive up to now.
The intrinsic difficulty in distinguishing individual components in envelope-dominated objects 
has precluded any statistical observational constraints on the distribution of disks sizes and masses. 
Only a survey providing high angular resolution observations for a large sample of Class 0 protostars can shed light 
on the controversy about the pristine characteristics of protostellar disks and ultimately on the importance of magnetic fields in regulating disk formation during protostellar formation \citep{LiPPVI}.\\

\begin{table*}[!t]
\centering
\caption{CALYPSO sample of target protostars}
\let\center\empty
\let\endcenter\relax
\begin{tabular}{lcccccc}
\hline
 \hfill & \hfill & \hfill & \hfill & \hfill & \hfill \\ 
 {\bf{Protostar}}                       & {Distance}    & {L$_{\rm{int}}$} & {M$_{\rm{env}}$}      & {Outflow P.A.}        & References \\
 {}                                     & {(pc)}        & (L$_{\odot}$) & (M$_{\odot}$)           & ($^{\circ}$)  & \\
 $^{[1]}$ & $^{[2]}$ & $^{[3]}$ & $^{[4]}$ & $^{[5]}$  & $^{[6]}$\\ 
\hfill & \hfill & \hfill & \hfill & \hfill & \hfill \\ 
\hline
\hfill & \hfill                 & \hfill & \hfill & \hfill & \hfill \\ 
{L1448-2A}                              & 235   & 3.0   & 1.2           & -63     & \citet{Olinger99} / \citet{Enoch09}  \\
{L1448-NB}                              & 235   & 2.5   & 3.1           & -80     & \citet{Curiel90} / \citet{Sadavoy14}  \\
{L1448-C}                               & 235   & 7.0   & 1.3           & -17     & \citet{Anglada89} / \citet{Sadavoy14}  \\
{IRAS2A}                                & 235   & 30    & 5.1           & +205    & \citet{Jennings87} / \citet{Karska13} \\
{SVS13B}                                & 235   & 2.0   & 1.8           & +167    & \citet{Grossman87} / \citet{Chini97} \\
{IRAS4A}                                & 235   & 3.0   & 7.9           & +180    & \citet{Jennings87} / \citet{Sadavoy14}\\
{IRAS4B}                                & 235   & 1.5   & 3.0           & +167    & \citet{Jennings87} / \citet{Sadavoy14} \\
{IRAM04191}                     & 140   & 0.05  & 0.5           & +200  & \citet{Andre99} / \citet{Andre00} \\
{L1521F}                                & 140   & 0.035 & 0.7-2         & +240    & \citet{Mizuno94} / \citet{Tokuda16} \\
{L1527}                                 & 140   & 0.9   & 1.2           & +60     &  \citet{Ladd91} / \citet{Motte01a} \\
{SerpM-S68N}                    & 415   & 10    & 10            & -45   &  \citet{Casali93} / \citet{Kaas04} \\
{SerpM-SMM4}                    & 415   & 2             & 7             & +30     & \citet{Casali93} / \citet{Kaas04} \\
{SerpS-MM18}                    & 260   & 16    & 3             & +188  &  \citet{Maury11} / \citet{Maury11} \\
{SerpS-MM22}                    & 260   & 0.2   & 0.5           & +230           & \citet{Maury11} / \citet{Maury11}  \\
{L1157}                                 & 250   & 2.0   & 1.5           & 163     & \citet{Umemoto92} / \citet{Motte01a} \\
{GF9-2}                                 & 200   & 0.3   & 0.5           & 0       & \citet{Schneider79} / \citet{Wiesemeyer97} \\
\hfill & \hfill & \hfill & \hfill & \hfill & \hfill \\
\hline
\end{tabular}
\tablefoot{The sources are ordered by increasing right ascension of the targeted core (see phase centers of our observations in Table \ref{table:observations}).
Column 1: Name of the protostar.
Column 2: Distance of the cloud where the protostar lies. The Taurus distance is taken from a VLBA measurement estimating distances from 130 pc to 160 pc \citep{Torres09} depending on the location in the cloud: we adopt a mean value of 140 pc.
The distance of Perseus is taken following recent VLBI parallax measurements that have determined a distance to the NGC 1333 region $235\pm18$ pc \citep{Hirota08} and a distance to the L1448 cloud of $232\pm18$ pc \citep{Hirota11}.
The distance of the Serpens Main cloud (SerpM sources) follows VLBI measurements in \citet{Dzib10}, who have determined a distance of 415 pc for the Serpens Main core, while a distance 230-260 pc was widely used before. Therefore, both L$_{int}$ and M$_{env}$ reported here are larger by a factor of 2.8 than those listed in the Bolocam and {\it{Spitzer}} literature before 2011.
The distance of Serpens South (SerpS sources) is still subject of debate, since no VLBI measurements toward the Serpens South filament exist: for consistency, we use here the distance of 260 pc adopted in \citet{Maury11} and in {\it{Herschel}} studies of  \citet{Konyves10, Konyves15}.
Column 3: Internal luminosity of the protostar. The internal luminosities come from the analysis of {\it{Herschel}} maps obtained in the framework of the Gould Belt survey (HGBS, see, e.g., \citealt{Andre10} and Ladjelate et al. in prep.).
Column 4: Protostellar envelope mass from the literature (associated references in Col. 6).
Column 5: {{Position angle (P.A. of the blue lobe counted east of north) of the high-velocity emission from $^{12}$CO or SiO tracing the protostellar jet component}}, from our CALYPSO molecular line emission maps when detected. For some protostars, the blue and red lobes are not well aligned, and/or several jets are detected in our CALYPSO maps: in these cases, we report the P.A. of the blue lobe, originating from the primary protostar position (see Table \ref{table:continuum-sources}). Information on individual sources is reported in Appendix \ref{section:appindivsour}), and the global jet properties in the CALYPSO sample will be provided in Podio \& CALYPSO (in prep).
Column 6: References for values reported here: protostar discovery paper, then the reference for the envelope mass.
}
\label{table:sample}
\end{table*}

\noindent{\it{The CALYPSO survey}}\\
The Continum And Lines in Young ProtoStellar Objects (CALYPSO, see {\url{http://irfu.cea.fr/Projets/Calypso/}}) IRAM Large Program is a survey of 16 Class 0 protostars, carried out with the IRAM Plateau de Bure (PdBI) interferometer in three spectral setups (around 94\,GHz, 219 GHz, and 231\,GHz). 
CALYPSO was crafted as an effort to make progress in our understanding of the angular momentum problem for star formation through high angular resolution observations ($0.3\arcsec$, i.e., $\simlt 100$ au) of a significant sample of the youngest protostars. 
The main goals of the PdBI observations are to improve our understanding of (1) the formation of accretion disks and multiple systems during protostellar collapse, (2) the role of Class 0 jets and outflows in angular momentum extraction, and (3) the kinematics and structure of the inner protostellar environment.
The 16 Class~0 objects of the CALYPSO sample are among the youngest known solar-type protostars \citep{Andre00}, with M$_{env} \sim $~0.5--10~M$_{\odot}$, 
and internal luminosities L$_{int} \sim $~0.03--30 L$_{\odot}$ (see Table \ref{table:sample}). 
This sample comprises most of the pre-{\it Herschel} confirmed Class 0 protostars located in nearby ($d < 420$~pc) clouds  that could be observed in shared tracks from the IRAM-PdBI location.
Several papers have been published based on the CALYPSO survey and discuss remarkable properties of individual protostars: IRAS2A  (\citealt[][]{Maret14,Codella14a,Maury14}), IRAS4A (\citealt[][]{Santangelo15}), L1157  (\citealt[][]{Podio16}), or the Class~I protostar SVS13A (\citealt[][]{Lefevre17}).
Two analyses of the molecular line emission in CALYPSO subsamples have also been carried out \citep{Anderl16,DeSimone17}.\\

\noindent{\it{Goals and plan of the present paper}}\\
This paper is the first of a series of statistical analyses (Belloche et al. in prep., Maret et al. in prep., Podio et al. in prep, Gaudel et al. in prep.) that are carried out for the whole CALYPSO sample. They address the three cornerstone questions described above that lie at the heart of the scientific motivations of CALYPSO.

This paper focuses on the inner density structure(s) of the CALYPSO Class 0 protostars, analyzing the dual-frequency dust continuum emission visibilities to probe the structure of protostellar envelopes down to radii $\sim 30$ au (for the Taurus sources) to $\sim 90$ au (for the Serpens Main sources), with
special emphasis on characterizing  candidate protostellar disks and multiple systems.
We show the dust continuum emission datasets, obtained at 1.37~mm and 3.18~mm, for the whole CALYPSO sample (see Table \ref{table:sample}) in Sect. 2 and the dust continuum sources detected in our maps in Sect. 3. 
Section 4 is dedicated to describing the analysis of the dust continuum visibility dataset we performed to test whether candidate disk components are detected in each of the primary targets of our sample.
In Sect. 5.1 we discuss the occurrence and properties of the candidate embedded disks in Class 0 protostellar envelopes from our sample and the literature. In Sect. 5.2 we explore current predictions from theoretical models of protostellar collapse, and the constraints our results bring to  these models. Finally, we briefly discuss in Sect. 5.3 the multiplicity of Class 0 protostars in our sample and in the literature and compare them to the multiplicity properties of YSOs at more advanced evolutionary stages.

\section{Observations and data reduction}

\begin{table*}[!h]
\begin{center}
\caption{Properties of the PdBI continuum emission maps}
\label{table:continuumobs}
\begin{small}
\begin{tabular}{|c|cc|cc|}
\hline
\hfill & \hfill & \hfill & \hfill & \hfill \\ 
{\bf{Field}} & \multicolumn{2}{|c|}{Synthesized beam FWHM ($\arcsec$), P.A. ($^{\circ}$)} & \multicolumn{2}{|c|}{$\sigma$ (mJy/beam)} \\
\hfill          & 231 GHz              & 94 GHz                 & 231 GHz                 & 94 GHz  \\
\hfill & \hfill & \hfill & \hfill & \hfill \\ 
\hline
\hfill & \hfill & \hfill & \hfill & \hfill \\ 
{L1448-2A}              & 0.56 $\times$ 0.37, 36        & 1.41 $\times$ 0.93, 40      & 0.31  &  0.06 \\
\hfill & \hfill & \hfill & \hfill & \hfill \\ 
{L1448-NB}              & 0.56 $\times$ 0.40, 38        & 1.42 $\times$ 0.97, 36      & 1.1   &  0.23  \\
\hfill & \hfill & \hfill & \hfill & \hfill \\ 
{L1448-C}                       & 0.58 $\times$ 0.36, 28        & 1.45 $\times$ 0.96, 43        & 0.6   &  0.09 \\
\hfill & \hfill & \hfill & \hfill & \hfill \\ 
{IRAS2A}                        & 0.62 $\times$ 0.45, 45        & 1.40 $\times$ 0.98, 38        & 0.97  &  0.14 \\
\hfill & \hfill & \hfill & \hfill & \hfill \\ 
{SVS13}                         & 0.54 $\times$ 0.33, 29        & 1.52 $\times$ 0.96, 24        & 1.0   &  0.17  \\
\hfill & \hfill & \hfill & \hfill & \hfill \\ 
{IRAS4A}                & 0.57 $\times$ 0.35, 29        &  1.53 $\times$ 0.98, 25        & 3.1   & 0.8   \\
\hfill & \hfill & \hfill & \hfill & \hfill \\ 
{IRAS4B}                        & 0.58 $\times$ 0.36, 30        & 1.52 $\times$ 0.99, 25        & 2.3   &  0.83  \\
\hfill & \hfill & \hfill & \hfill & \hfill \\ 
{IRAM04191$^{\star}$}           & 1.00 $\times$ 0.85, -160    & 1.90 $\times$ 1.57, 48        & 0.10  & 0.023    \\
\hfill & \hfill & \hfill & \hfill & \hfill \\ 
{L1521F$^{\star}$}                      & 1.15 $\times$ 0.87, 25    &  2.19 $\times$ 1.90, 93       & 0.06  & 0.03   \\
\hfill & \hfill & \hfill & \hfill & \hfill \\ 
{L1527}                 & 0.53 $\times$ 0.35, 37     &  1.49 $\times$ 0.97, 33      &  0.45         &   0.07     \\
\hfill & \hfill & \hfill & \hfill & \hfill \\ 
{SerpM-S68N}            & 0.82 $\times$ 0.35, 22     &  1.84 $\times$ 1.11, 32      &  0.6          & 0.09         \\
  \hfill & \hfill & \hfill & \hfill & \hfill \\ 
{SerpM-SMM4}            & 0.74 $\times$ 0.33, 25     &  1.73 $\times$ 0.85, 28      &  1.5          &   0.25    \\
\hfill & \hfill & \hfill & \hfill & \hfill \\ 
{SerpS-MM18}            & 0.74 $\times$ 0.33, 24      & 1.73 $\times$ 0.87, 27      & 1.3   & 0.2   \\
\hfill & \hfill & \hfill & \hfill & \hfill \\ 
{SerpS-MM22}            & 0.88 $\times$ 0.35, 20      &  1.80 $\times$ 0.90, 27      & 0.39          & 0.05      \\
\hfill & \hfill & \hfill & \hfill & \hfill \\ 
{L1157}                         & 0.51 $\times$ 0.40, 4         & 1.36 $\times$ 1.04, 64        & 0.6   &   0.09   \\
\hfill & \hfill & \hfill & \hfill & \hfill \\ 
{GF9-2$^{\star}$}                       & 0.85 $\times$ 0.68, 18        & 1.43 $\times$ 1.02, 66          &    0.14  &   0.025   \\
\hfill & \hfill & \hfill & \hfill & \hfill \\ 
\hline 
\end{tabular}\\
\tablefoottext{${\star}$}{Maps of the weakest continuum emission sources in the sample, IRAM04191, L1521F, and GF9-2, were produced using a natural weighting to maximize the sensitivity to point sources, see text in Sect. 2.2.
}
\end{small}
\end{center}
\end{table*}

\subsection{PdBI observations and calibration}

Observations of the CALYPSO sources were carried out with the IRAM PdBI between September 2010 and March 2013 (see details in Table~\ref{table:observations}).
We adopted a multiple configuration strategy, using the six-antenna array in the most extended configuration (A array) and in an intermediate antenna configuration (C array). This provided a fairly dense coverage of the uv-plane with 30 baselines ranging from 16\,m to 760\,m.
We used the WideX\footnote{see \url{www.iram.fr/widex}} correlator to cover a 3.8 GHz broadband spectral window for each of the three spectral setups, which were observed separately: at 1.29~mm for observations with a WideX central frequency of 231 GHz, at 1.37~mm for observations with a WideX central frequency of 219 GHz, and at 3.18~mm for observations with a central WideX frequency of 94~GHz.
A higher spectral resolution correlator was placed onto a handful of molecular emission lines, but we present here only the analysis of the continuum emission extracted from the WideX dataset.
The proximity of some of the sources on the sky allowed us to use common gain calibrators for several groups of sources and therefore to time-share a total of 37 tracks of $\sim$8~hr on the 16 sources. 
Each track was divided unequally among the sources, roughly in inverse proportion to the source luminosities so as to obtain the most homogeneous signal-to-noise ratios (S/N) in the sample.
For each track, nearby amplitude and phase calibrators were observed to determine the time-dependent complex antenna gains. 
The calibration was performed in the GILDAS/CLIC\footnote{\url{www.iram.fr/IRAMFR/GILDAS/}} environment.
The correlator bandpass was calibrated on strong quasars (e.g., 3C273 and 3C454.3), while the absolute flux density scale was usually derived from observations of MWC349 and 3C84. The absolute flux calibration uncertainty is estimated to be $\sim$10$\%$ at 94\,GHz and $\sim$15$\%$ at 231\,GHz. 

For sources where the peak of the dust continuum emission is detected at >40\,$\sigma$ (roughly equivalent to all sources with peak flux $>80$ mJy/beam at 231~GHz, see Table~\ref{table:continuumobs}), we also carried out self-calibration of the continuum emission visibility dataset to improve the S/N at the longest baselines. 

\subsection{PdBI continuum emission maps}
After a cross-check of the absolute flux calibration consistency of the observing dates, the visibility datasets obtained with the A and C configuration were combined, for each of the three frequency setups independently.
The continuum visibilities were generated from the WideX units, avoiding channels in which line emission was detected at a level $>5\, \sigma$ in the spectrum integrated over 2$\arcsec$ around the peaks of the continuum emission reported in Table \ref{table:continuum-sources}. Examples of wide-band spectra are shown in \citet{Maury14}, while the analysis of the molecular content of the whole sample will be presented in Belloche et al. (in prep). 
In order to produce images and visibility curves with optimum sensitivity and best uv-coverage at 1.3 mm, we merged the continuum visibility data that were independently obtained at 219~GHz and 231~GHz: we used the spectral index computed from the shortest common baseline of our PdBI observations at 94\,GHz and 231\,GHz ($\alpha_{20\rm{k}\lambda}$, see Col. 6 of Table \ref{table:continuum-pdbi-sd}) for each individual source to scale the visibility amplitudes obtained at 219~GHz to 231~GHz. In the following, all 231\,GHz fluxes, maps, and visibility profiles stem from the combined data.

\begin{figure*}[!h]
\centering
\includegraphics[trim={0 0 0 0},clip,width=0.68\textwidth]
{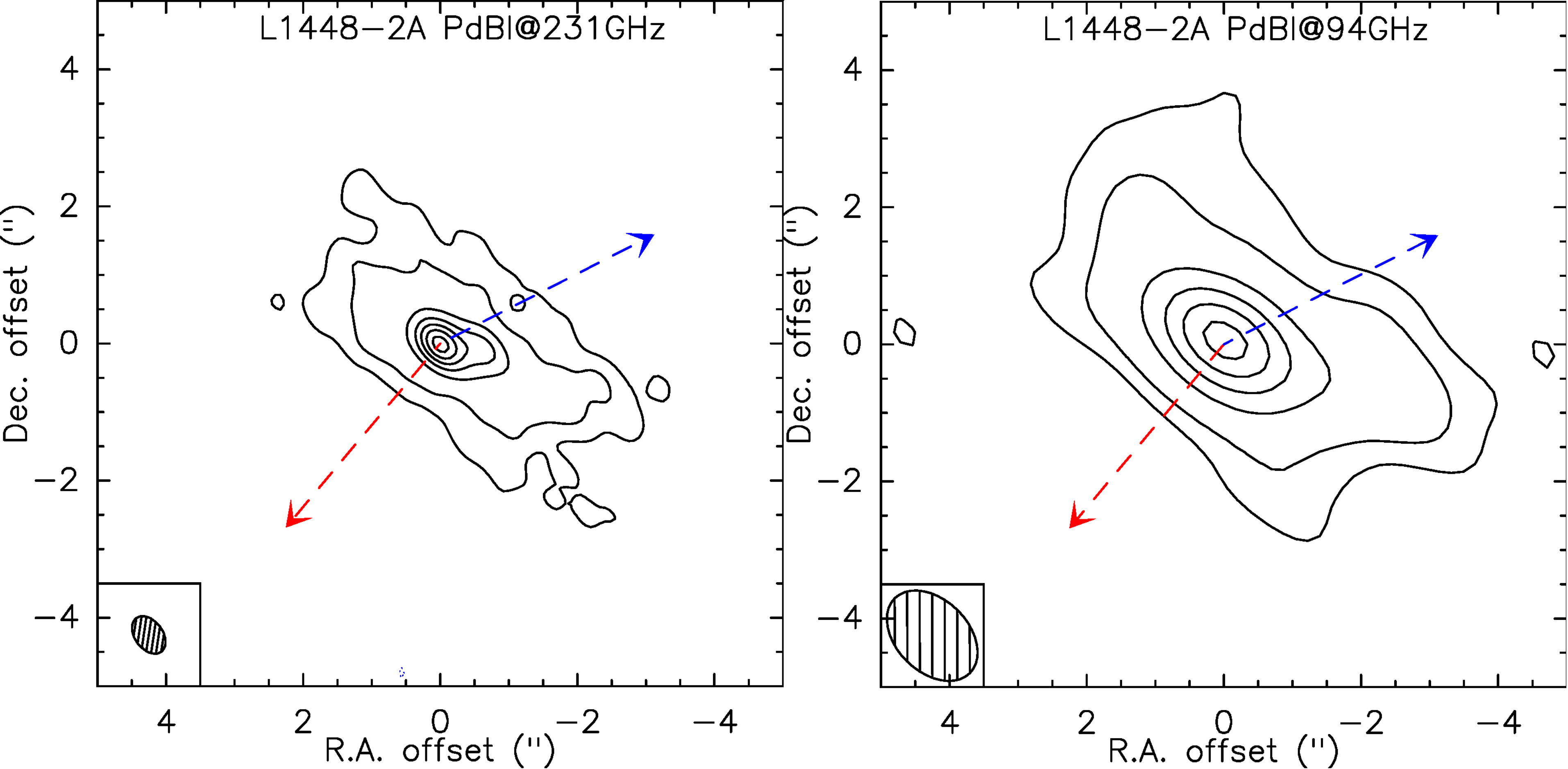}
\caption{\label{fig:l14482a_maps} 
1.3~mm (231 GHz, left) and 3.3~mm (94 GHz, right) PdBI dust 
continuum emission maps of L1448-2A, corrected for primary beam attenuation. The ellipses in the bottom left corner show the respective synthesized beam sizes. The contours are levels of -3$\sigma$ 
(dashed), 5$\sigma$, and 10$\sigma,$ then from 20$\sigma$ in steps of 20$\sigma$ (rms noise computed in the map before primary beam correction, reported in Table \ref{table:continuumobs}). The blue and red arrows show the direction of the protostellar jet(s) either as found in our CALYPSO dataset or from the literature (see Table \ref{table:sample}).
}
\end{figure*}

\begin{figure*}[!h]
\centering
\includegraphics[trim={0 0 0 0},clip,width=0.68\textwidth]
{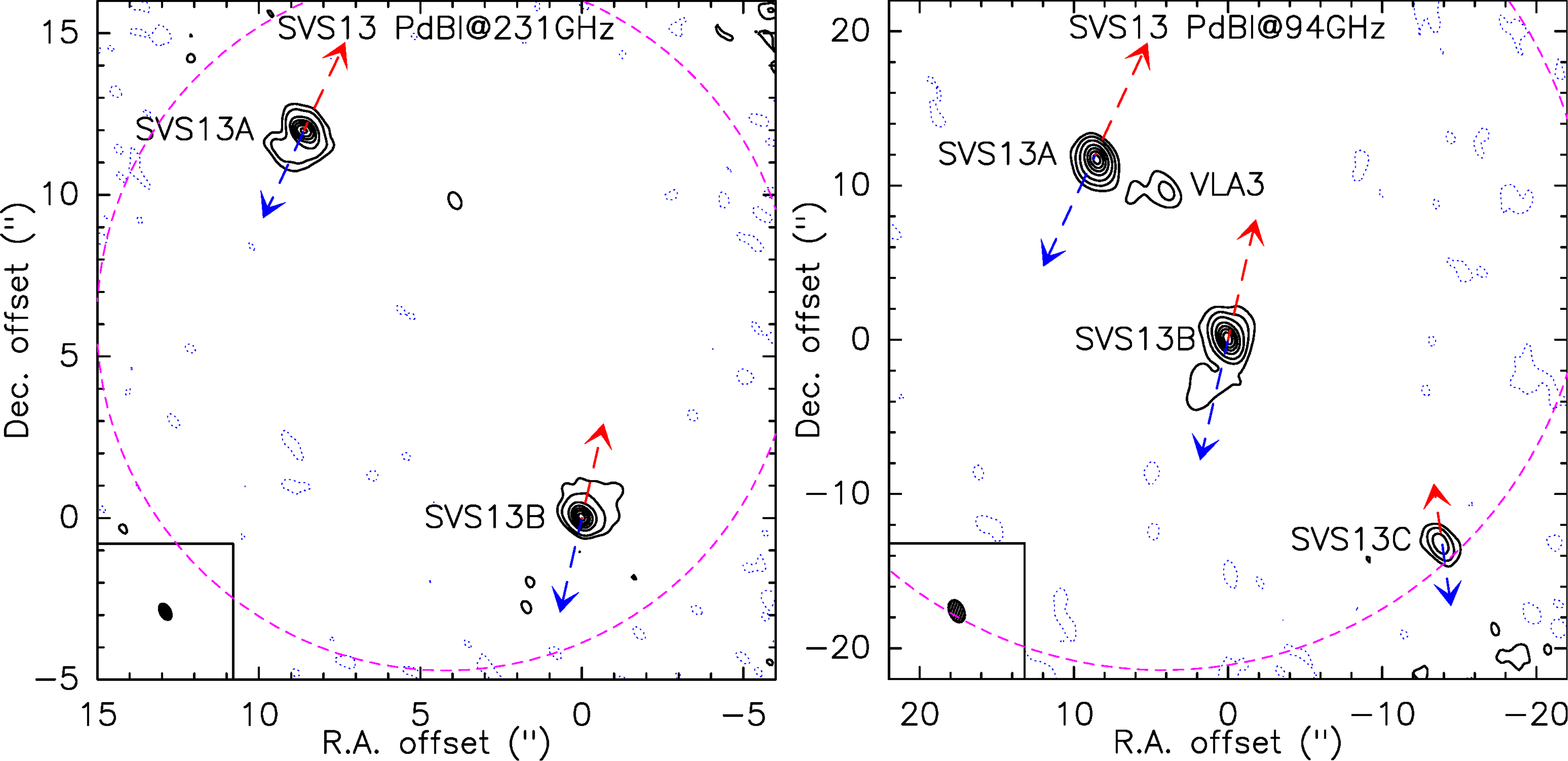}
\caption{\label{fig:svs13b_maps} 
Same as Fig. \ref{fig:l14482a_maps} for the SVS13 region. The dashed pink contour 
shows the PdBI primary beam at each frequency. The contours show levels 
of -3$\sigma$ (dashed), 5$\sigma$, and 10$\sigma$, then from 20$\sigma$ in steps of 
20$\sigma$ (see Table \ref{table:continuumobs}). The blue and red arrows show the direction of the protostellar jets from our CALYPSO data for SVS13A/B (Podio et al. in prep.), while the SVS13C outflow P.A. stems from the CARMA map \citep{Plunkett13}.
}
\end{figure*}

\begin{figure*}[!h]
\centering
\includegraphics[trim={0 0 0 0},clip,width=0.68\textwidth]
{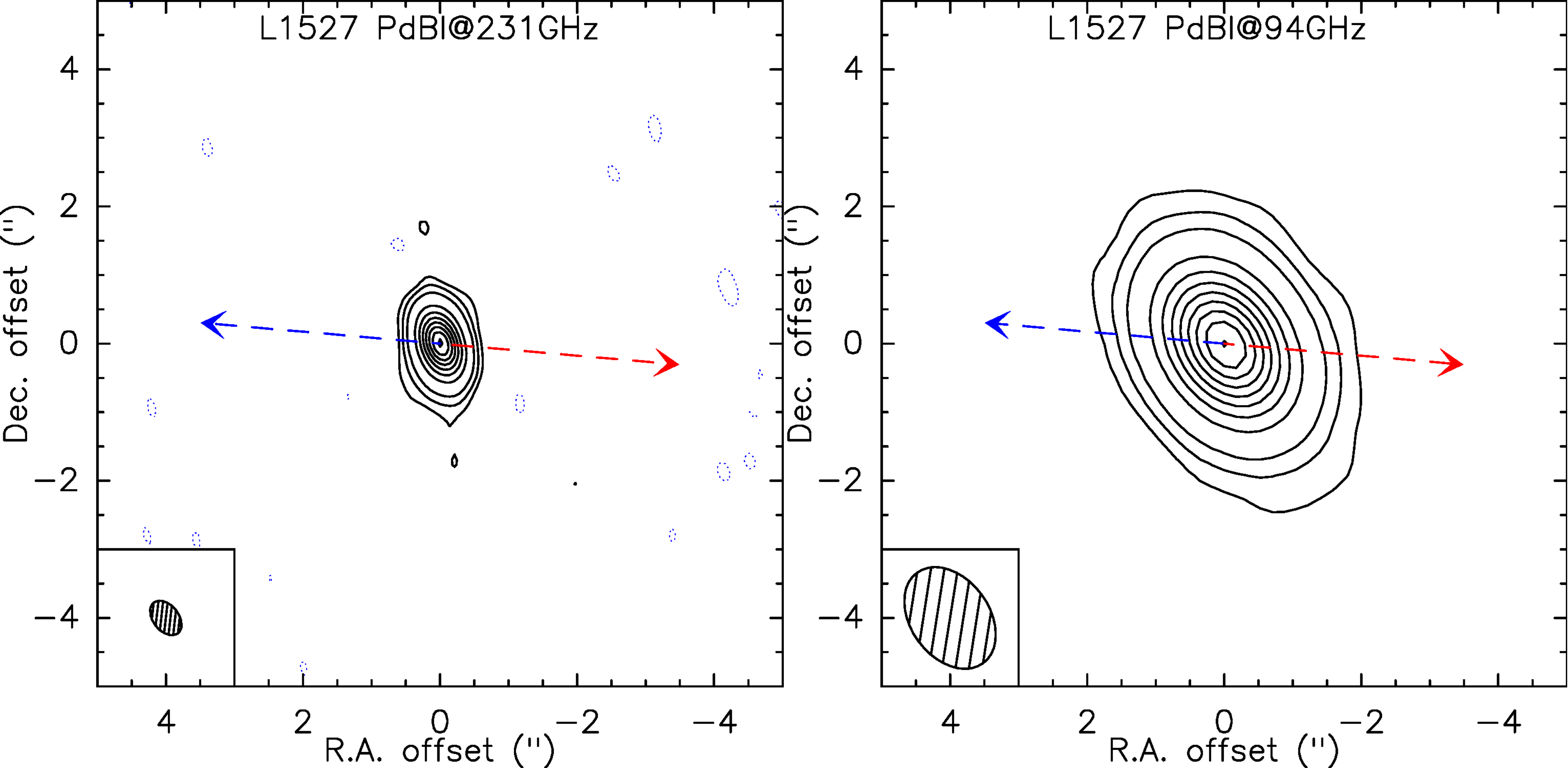}
\caption{\label{fig:l1527_maps} 
Same as Fig. \ref{fig:l14482a_maps} for L1527. The contours show 
levels of -3$\sigma$, 5$\sigma$, and 10$\sigma,$ then from 20$\sigma$ in steps of 
40$\sigma$ (see Table \ref{table:continuumobs}) .
}
\end{figure*}

Imaging of the continuum visibility tables was carried out using a robust scheme for weighting that allowed us to improve the resolution and lower the side-lobes without exceedingly degrading the overall sensitivity. The robust \citep{Briggs95} threshold we adopted was moderate (r=0.3\footnote{The robust parameter is set so that if the sum of natural weights in a given uv cell is lower than this threshold, natural weights are kept; if it is higher, the weight is set to this value. For more information on weighting schemes performed by the GILDAS/MAPPING software, see \url{http://www.iram.fr/IRAMFR/IS/IS2002/html_1/node156.html}}) and allowed us to reach a good compromise between imaging quality and sensitivity while enhancing the contribution of the high spatial frequencies. This resulted in typical full widths at half-maximum (FWHMs) of the synthesized beams $\simlt 0.5\arcsec$ at 231 GHz. Exceptions were made for the low-luminosity sources IRAM04191, L1521F, and GF9-2, where natural weighting was used to maximize the sensitivity for these faint objects (producing synthesized beams $\sim 1.0\arcsec$ at 231 GHz). The maps were subsequently cleaned using the Hogbom CLEAN algorithm provided in the GILDAS/MAPPING software, with a cleaning threshold set to twice the rms noise in the map.
The properties of the resulting CLEANed maps (synthesized HPBWs, rms noises) are reported in Table~\ref{table:continuumobs}. 
The dust continuum emission maps obtained at 231~GHz and 94~GHz are shown in Figs. \ref{fig:l14482a_maps}, \ref{fig:svs13b_maps}, and \ref{fig:l1527_maps} for two of the Perseus sources (L1448-2A and SVS13B) and a Taurus source (L1527), while the remaining dust continuum maps for the 13 other fields in the CALYPSO sample are shown in Appendix \ref{section:appmaps} (Figures \ref{fig:l1448n_maps} to \ref{fig:gf92_maps}).


\section{Dust continuum sources in the CALYPSO maps}

Dust continuum emission was successfully detected at 231~GHz and 94~GHz toward all the Class 0 protostars we targeted in our survey: 
the positions and flux densities of the continuum sources detected in the fields observed with the PdBI are reported in Table\,\ref{table:continuum-sources}. 
The target Class 0 protostars are labeled as "primary protostars" in the "source nature" column.
 
Multiple dust continuum emission components are sometimes detected in our PdBI maps.
For each field that was observed with CALYPSO, we built a first sample of millimeter continuum sources from our 231~GHz PdBI dataset and selected components whose dust continuum emission was detected above the $5\sigma$ level in the maps (within the $21\arcsec$ FWHM primary beam). 
This detection threshold was chosen so as to limit contamination from spurious sources created by deconvolution of data sampling only partially the uv-plane.
We note that throughout this paper, we always report and use rms noises at phase center, but the maps we show and the fluxes we report are corrected for the primary beam 
attenuation (the phase centers used in our observations are given in Table \ref{table:observations}).
At the positions of the 231 GHz continuum sources, we checked the PdBI 94~GHz maps for independent detection of a counterpart at lower frequency. All the sources detected in the CALYPSO 231~GHz maps have a counterpart detected at $>3\sigma$ in the 94~GHz maps when it is possible to resolve them (the synthesized beam of the 94~GHz data is 2.5 times larger than the beam of the 231~GHz data). 

Moreover, since the PdBI 94~GHz maps have a larger primary beam than the 231~GHz maps, four sources detected at $>11\arcsec$ from the main protostar in the 94 GHz map have no 231~GHz counterpart since they fall outside the primary beam of the higher frequency observations. 
In these cases (e.g., SVS13-C, see Fig.\ref{fig:svs13b_maps}), the 94 GHz continuum source is reported only if the source was detected at another wavelength in the literature.
Applying this method to all fields observed in the CALYPSO survey, we detected 
a total of 30 dust continuum emission sources. We report the coordinates and peak flux densities in Table~\ref{table:continuum-sources} and discuss their nature in Section \ref{section:naturecont}.

\begin{table*}
\caption{Sources detected in the CALYPSO millimeter continuum maps.}
\label{table:continuum-sources}
\begin{tabular}{|c|cc|cc|c|c|c|}
\hline
\hfill & \hfill & \hfill & \hfill & \hfill & \hfill & \hfill & \hfill   \\ 
{Source} & {$\alpha_{2000}$} & {$\delta_{2000}$} & {F$^{\rm{peak}}_{\rm{231GHz}}$} & {F$^{\rm{peak}}_{\rm{94GHz}}$} & {Source}             & {$a$} & {Other}\\
\hfill & \hfill & \hfill & \hfill & \hfill & {nature} & \hfill & {names}   \\ 
\hfill & {$\left[ \rm{h:m:s} \right]$} & {[$\degr$:$\arcmin$:$\arcsec$]} & \multicolumn{2}{c|}{$\left[ \rm{mJy/beam} \right]$}                                   & {}      & {[au]} & {} \\
$^{[1]}$ & $^{[2]}$ & $^{[3]}$ & $^{[4]}$ & $^{[5]}$  & $^{[6]}$ & $^{[7]}$ & $^{[8]}$ \\ 
\hline
\hfill & \hfill & \hfill & \hfill & \hfill & \hfill & \hfill & \hfill   \\ 
{L1448-2A }     & 03:25:22.405 & 30:45:13.26    & 23$\pm$2              & 5.8$\pm$1                       & primary protostar             & {}    & {Per-emb-22A}\\
{L1448-2Ab}             & 03:25:22.360 & 30:45:13.20    & 11$\pm$1              & 4.5$\pm$2\tablefootmark{$\star$} & CE protostar ?       & 130   & {Per-emb-22B}  \\
\hfill & \hfill & \hfill & \hfill & \hfill & \hfill & \hfill & \hfill   \\ 
{ L1448-NA}     & 03:25:36.498 & 30:45:21.85    & 46$\pm$4              & 6.7$\pm$0.3                     & CE Class I            & 1700  & {L1448-IRS3A} \\ 
{ L1448-NB1}    & 03:25:36.378 & 30:45:14.77    & 146$\pm$6             & 69$\pm$2                        & primary protostar             &               & {L1448-IRS3B}\\
{ L1448-NB2}    & 03:25:36.315 & 30:45:15.15    & 69$\pm$3              & $<$25$\pm$5\tablefootmark{$\star$}  & CE protostar ?    & 210   & {L1448-IRS3B-a}               \\
{ L1448-NW}     & 03:25:35.670 & 30:45:33.86    & -             & 6.4                   & SE protostar            & 4900  & {L1448-IRS3C}         \\
\hfill & \hfill & \hfill & \hfill & \hfill & \hfill & \hfill & \hfill   \\ 
{ L1448-C}      & 03:25:38.875 & 30:44:05.33    & 123$\pm$5             & 19$\pm$1                        & primary protostar             & {}    & {L1448-mm}\\
{ L1448-CS}     & 03:25:39.132 & 30:43:58.04    & 8$\pm$2               & 1.6$\pm$0.1                     & CE Class I            & 1900  & {} \\
\hfill & \hfill & \hfill & \hfill & \hfill & \hfill & \hfill & \hfill   \\ 
{ IRAS2A1}      & 03:28:55.570 & 31:14:37.07    & 132$\pm$5             & 20$\pm$1                        & primary protostar\tablefootmark{$\star\star$}                 & {}    & {Per-emb-27} \\
\hfill & \hfill & \hfill & \hfill & \hfill & \hfill & \hfill & \hfill   \\ 
{ SVS13B}       & 03:29:03.078 & 31:15:51.74    & 127$\pm$7             &         22$\pm$1                        & primary protostar             & {}      & {} \\
{ SVS13A}       & 03:29:03.756 & 31:16:03.80    & 120$\pm$7             &         21$\pm$1                        & SE Class I                    & 2330    & {} \\
{ SVS13C}       & 03:29:01.980 & 31:15:38.14    & -             &       5.8$\pm$2                       & SE protostar            & 3550  & {} \\
{ VLA3}       & 03:29:03.378 & 31:16:03.33      & 9$\pm$1               &         2.6$\pm$0.5                     & unknown                       & 1390    & {A2} \\
\hfill & \hfill & \hfill & \hfill & \hfill & \hfill & \hfill & \hfill   \\ 
{ IRAS4A1 }     & 03:29:10.537 & 31:13:30.98    & 481$\pm$10            & 148$\pm$6                       & primary protostar\tablefootmark{$\star\star$}                 &  {}   & {Per-emb-12}  \\
{ IRAS4A2}      & 03:29:10.432 & 31:13:32.12    & 186$\pm$8             & $<$34$\pm$10\tablefootmark{$\star$}     & CE protostar          & 420   & {} \\ 
\hfill & \hfill & \hfill & \hfill & \hfill & \hfill & \hfill & \hfill   \\ 
{ IRAS4B}       & 03:29:12.016 & 31:13:08.02    & 278$\pm$6             & 75$\pm$3                        & primary protostar             & {}    & {Per-emb-13}  \\
{ IRAS4B2}      & 03:29:12.841 & 31:13:06.84    & 114$\pm$4             & 31$\pm$1                        & SE protostar ?                & 2500  & {IRAS 4BII} \\
\hfill & \hfill & \hfill & \hfill & \hfill & \hfill & \hfill & \hfill   \\ 
{ IRAM04191}  & 04:21:56.899 & 15:29:46.11              & 4.7$\pm$0.8           & 0.31$\pm$0.09                   & primary protostar                                     & {}      & {} \\
\hfill & \hfill & \hfill & \hfill & \hfill & \hfill & \hfill & \hfill   \\ 
{ L1521F}      & 04:28:38.941 & 26:51:35.14     & 1.6$\pm$0.2           & 0.27$\pm$0.05                   & primary protostar                                     & {}      & {MC27}  \\
\hfill & \hfill & \hfill & \hfill & \hfill & \hfill & \hfill & \hfill   \\ 
{ L1527}        & 04:39:53.875 & 26:03:09.66    & 129$\pm$8             & 23$\pm$1                        & primary protostar                                     & {}      & {}  \\
\hfill & \hfill & \hfill & \hfill & \hfill & \hfill & \hfill & \hfill   \\ 
{SerpM-S68N}    & 18:29:48.091 & 01:16:43.41    & 35$\pm$3              & 5.3$\pm$0.5                     & primary protostar             & {}    & {Ser-emb8}  \\
{SerpM-S68Nb}   & 18:29:48.707 & 01:16:55.53    & -             & 2.9$\pm$0.4                   & SE protostar            & 6400  & {Ser-emb8(N)} \\
{SerpM-S68Nc}   & 18:29:48.811 & 01:17:04.24    & -             & 2.7$\pm$0.5                   & SE Class I  ?           & 9700  & {} \\
\hfill & \hfill & \hfill & \hfill & \hfill & \hfill & \hfill & \hfill   \\ 
{SerpM-SMM4a}    & 18:29:56.716 & 01:13:15.65   & 184$\pm$11    & 48$\pm$2                                 & primary protostar                     & {}      & {} \\
{SerpM-SMM4b}    & 18:29:56.525 & 01:13:11.58   & 27$\pm$4      & 9$\pm$1                                 & CE protostar ?                & 2000  & {} \\
\hfill & \hfill & \hfill & \hfill & \hfill & \hfill & \hfill & \hfill   \\ 
{SerpS-MM18a}   & 18:30:04.118 & -02:03:02.55   & 148$\pm$9             & 20$\pm$1                        & primary protostar                     & {}      & {CARMA-7} \\
{SerpS-MM18b}   & 18:30:03.541 & -02:03:08.33   & 62$\pm$4              & 7.8$\pm$0.8                     & CE protostar ?                                & 2600    & {CARMA-6} \\
\hfill & \hfill & \hfill & \hfill & \hfill & \hfill & \hfill & \hfill   \\ 
{SerpS-MM22}    & 18:30:12.310 & -02:06:53.56   & 20$\pm$2              & 2.8$\pm$0.7                     & primary protostar             & {}    & {} \\
\hfill & \hfill & \hfill & \hfill & \hfill & \hfill & \hfill & \hfill   \\ 
{L1157}                 & 20:39:06.269 & 68:02:15.70    & 117$\pm$9             & 18$\pm$1                        & primary protostar             & {}    & {} \\
\hfill & \hfill & \hfill & \hfill & \hfill & \hfill & \hfill & \hfill   \\ 
{GF9-2}         & 20:51:29.823 & 60:18:38.44    & 9.9$\pm$1             & 1.6$\pm$0.4                     & primary protostar             & {}    & {L1082 C} \\
\hfill & \hfill & \hfill & \hfill & \hfill & \hfill & \hfill & {}\\ 
\hline
\end{tabular}\tablefoot{[1] Name of the primary protostar and its individual mm components. Columns 2 and 3 are the equatorial coordinates of the dust continuum peak position in the CALYPSO 231~GHz maps. Columns 4 and 5 list the 231~GHz and 94~GHz dust continuum peak flux densities in the PdBI synthesized beams reported in Table~\ref{table:continuumobs}. Column 6 shows the putative nature of the source (see Sect. \ref{section:naturecont} and comments on individual sources in Appendix \ref{section:appindivsour}). SE is used for a candidate protostellar companion with a separate envelope, and CE  is used when the (candidate) protostellar companion is within a common envelope with the primary protostar. We use the sizes computed from single-dish observations (see "Single-dish constraints" in Appendix \ref{section:appindivsour}) as envelope radii. Column 7 shows the projected separation (in the plane of the sky) between the secondary and the primary protostar, translated into physical units using the distances reported in Table \ref{table:sample}. Column 8 lists other names used in the literature for the associated object (when previously reported). 
\tablefoottext{$\star$}{Unresolved from the primary source because of the larger 94 GHz synthesized beam: the measured peak flux is largely contaminated by the flux belonging to the primary source.}
\tablefoottext{$\star\star$}{IRAS4A3 \citep{Santangelo15}, as well as IRAS2A2 and IRAS2A3 \citep{Maury14,Codella14a} are not reported here as individual sources (see Sect. \ref{section:naturecont} for further details). A secondary source $0.4\arcsec$ south of IRAS2A is detected with VLA \citep{Tobin15b} and ALMA (Maury et al. in prep) but is not detected with our CALYPSO data because the high resolution data for IRAS2A was obtained prior to the CALYPSO program (see Appendix \ref{section:appiras2a}).}
}
\end{table*}


\section{Analysis of dust continuum source structures}

\subsection{Building a proper visibility dataset for each primary target protostar}
\label{section:buildvisib}

Each secondary source reported in Table \ref{table:continuum-sources} was removed from the visibility data by subtracting a Gaussian source model with FWHM at most twice the synthesized beam size (to remove only compact components and avoid affecting the envelope emission), or a point source if separated from the primary continuum component by less than two synthesized beams. During this subtraction process, the coordinates and peak fluxes of secondary components were fixed to the values reported in Cols. 2 and 3 of Table \ref{table:continuum-sources}.  
Subtracting the visibilities that are due to secondary components allowed us to perform a focused analysis of the continuum emission that originates from each main protostar that was targeted by our observing program. 
We stress, however, that this operation does not preclude the possible presence of tighter multiple components at scales that cannot be resolved with PdBI.  
For example, the real parts  of the visibilities of L1448-NB, IRAS2A, and L1157 
show oscillations at long baselines in circularly averaged data, suggesting that additional structure or components (at $a<0.4\arcsec$) that are not centered on the phase center may remain in these three sources.
From now on, we discuss and analyze the millimeter continuum emission originating solely from the primary protostar in each field.

For all primary protostars targeted with CALYPSO, we report the integrated flux densities (above a 5$\sigma$ level) that we recovered with our PdBI observations in Table~\ref{table:continuum-pdbi-sd}.
Since the largest scale sampled by the PdBI observations depends on the frequency $\nu_{obs}$ and the shortest baseline in the array $B_{min}$ (maximum recoverable scale $MRS\sim0.6c/(\nu_{obs}\times B_{min})$, i.e., $14\arcsec$ at 94 GHz and $6\arcsec$ at 231 GHz),
we also report in Table~\ref{table:continuum-pdbi-sd} the matching visibility flux values, obtained at the shortest common $B=20\rm{k}\lambda$ baseline (value averaged in a 10 $\rm{k}\lambda$ bin). 
We note that the shortest baseline sampled at 94 GHz is $\sim 8$ k$\lambda$, so the integrated fluxes from the 94\,GHz maps can sometimes be significantly higher than the 20 k$\lambda$ visibility flux values at 94 GHz. 
For the continuum data at 231 GHz, and especially for strong sources, 
the flux recovered in the CLEANed maps is sometimes significantly lower than the flux measured at the 20 $\rm{k}\lambda$ baselines. This
illustrates the difficulty of reliably reconstructing interferometric maps from data with incomplete uv-coverage, and it justifies our approach to carry out modeling in the uv-plane rather than in the image plane.

\subsection{Large-scale constraints from single-dish observations of the dust continuum emission}

\begin{figure}
\begin{center}
\includegraphics[width=0.9\linewidth]{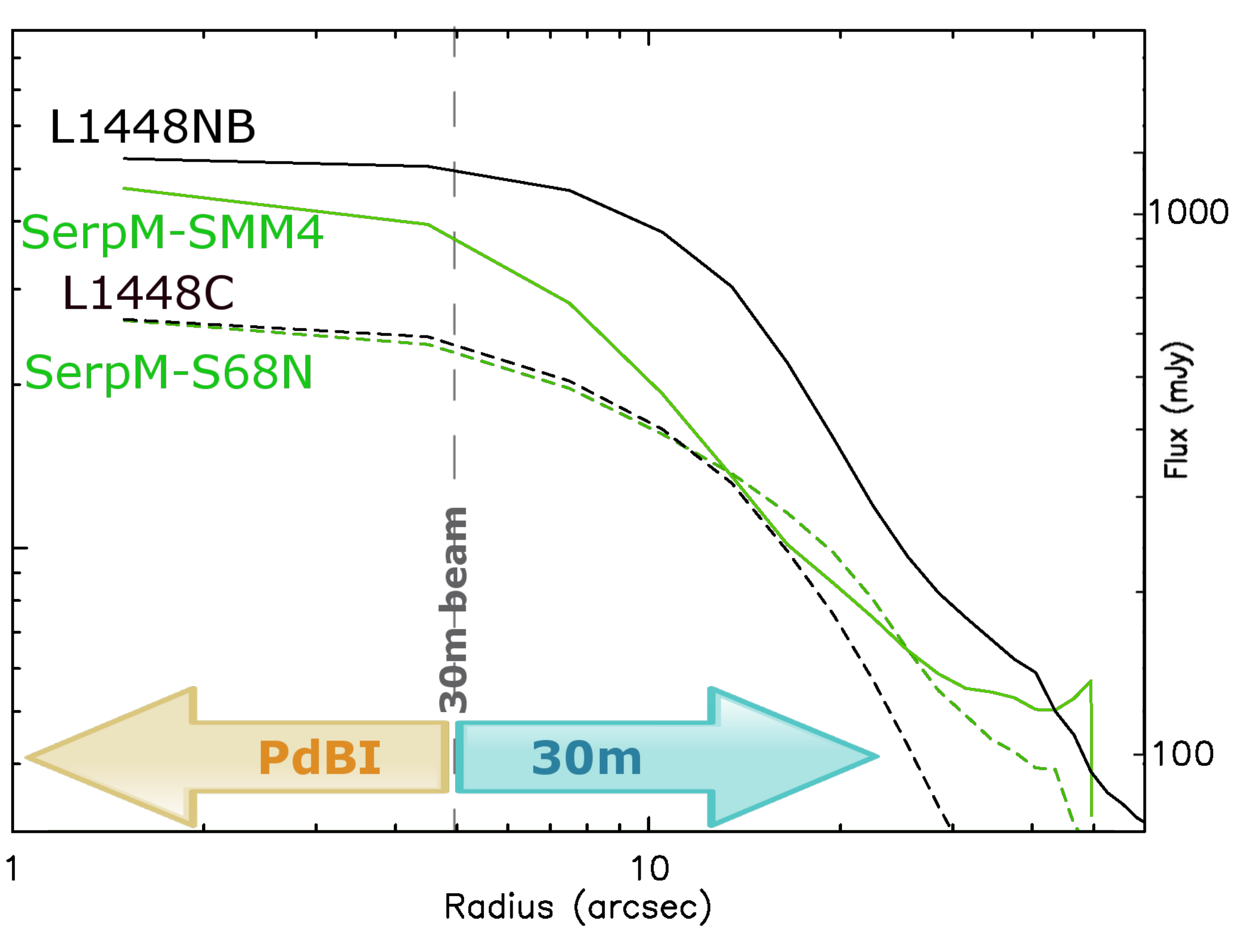}
\caption{Examples of radial intensity profiles from single-dish maps obtained at the IRAM-30m telescope that are used as large-scale constraint for the envelope modeling. These profiles were obtained from the dust continuum emission maps acquired using the IRAM-30m and are published in \citet{Motte01a} and \citet{Kaas04}.
\label{fig:SD-L1448-SerpM}
}
\end{center}
\end{figure}

We used a combination of PdBI configurations to observe the continuum emission down to baselines of  $17\,\rm{k\lambda}$ at 231 GHz ($8\, \rm{k\lambda}$ at 94 GHz), 
probing spatial scales up to $\sim 6\arcsec$ at 231 GHz ($14\arcsec$ at 94 GHz).
Protostellar envelopes typically extend to larger angular scales at the distance of our sources \citep[][]{Motte01a}, and therefore cannot always be completely sampled using our PdBI observations alone (see Figure \ref{fig:SD-L1448-SerpM} for an example of IRAM-30m intensity profiles for four of the protostars in our sample). 
We therefore collected information from the literature 
on single-dish continuum observations of the target protostars that probe envelope scales $\simgt 10\arcsec$, so as to better constrain the outer envelope density profiles through (i) dust continuum fluxes that are
integrated at envelope scales reported in Col.~8 of Table~\ref{table:continuum-pdbi-sd} and (ii) peak flux densities obtained with 
single-dish telescopes that trace the material at the scales of the single-dish beam size, reported in Col.~10 of Table~\ref{table:continuum-pdbi-sd}. 

We can extrapolate the dust continuum fluxes found in the literature, which are often obtained at slightly different frequencies than our PdBI setups, when we assume a simple power-law dependence of the flux density on frequency by scaling the flux by an average spectral index at envelope scales. 
We used the spectral index computed from the shortest common baseline of our dual frequency PdBI observations at $20\, \rm{k}\lambda$ (see Col. 6 of Table \ref{table:continuum-pdbi-sd}) for each individual source to extrapolate the single-dish fluxes to the frequency of our PdBI observations. 
When secondary components were present in the field (see Table \ref{table:continuum-sources}), we subtracted their flux densities as estimated in the PdBI u--v plane from the extrapolated single-dish flux of the primary sources. 
The resulting extrapolated total envelope fluxes and peak flux densities at 231 GHz are reported in Cols.~13 and 14 of Table~\ref{table:continuum-pdbi-sd}, respectively,  for the sources that are resolved by single-dish observations. 
Values for the total fluxes at 94 GHz were obtained by scaling the 231 GHz fluxes assuming the spectral index indicated in Col.~6.

Because single-dish continuum observations are broadband and the spectral index of the dust continuum emission in the envelope at larger scales is not well constrained 
by our PdBI dual-frequency observations (since the minimum common baseline at 94 and 231 GHz only probes scales up to  $\la 6\arcsec$), 
the fluxes at baselines $< 10 k\lambda$ extrapolated from single-dish data are uncertain.
When we used them in our modeling of the envelope emission, we therefore allowed the total envelope flux to vary within a range $\pm 30\%$.

None of the four sources in NGC 1333 (IRAS2A, IRAS4A, IRAS4B, and SVS13B) is resolved by single-dish observations: the MAMBO bolometer-array studies by \citet{Motte01a} and \citet{Chini97} measured peak flux densities that are roughly equal to the integrated fluxes, integrated in areas twice the beam size. We therefore constrained the envelope fluxes of these sources to match the single-dish peak fluxes ($\pm 40\%$) in an outer envelope radius smaller than or equal to the single-dish beam ($\pm 40\%$). 
For L1521F, IRAM-30m/MAMBO observations suggest a total flux 500-1000 mJy in a radius $30\arcsec$ at 243~GHz (see \citealt{Motte01a, Tokuda16}) 
and a peak flux 120 mJy in a $14\arcsec$ beam (similar to the value reported by \citealt{Crapsi04} in the nominal IRAM-30m beam of $11\arcsec$). 
Although the L1521F envelope was resolved with MAMBO, the total flux is somewhat uncertain as the observations were affected by relatively strong sky noise. 
We therefore only fixed the peak flux in our model fit, allowing the total integrated flux to vary by up to $\pm 50 \%$ around the single-dish value.
For GF9-2 we used the IRAM-30m fluxes from \citet{Wiesemeyer97}: 60 mJy/beam peak flux and 315 mJy integrated up to a radius $35\arcsec$ at 240 GHz, extrapolated to our frequencies. Using a spectral index $\alpha_{20k\lambda}=2.1$ between 94\,GHz and 231\,GHz (see Table \ref{table:continuum-pdbi-sd}), we expect an integrated flux of 291 mJy at 231 GHz (55 mJy peak flux) and 44 mJy at 94 GHz (8 mJy peak flux). Since these extrapolated envelope-scale fluxes are quite uncertain, we let them quite loose during the fitting process with an error bar of $50\%$ at both frequencies.

\begin{landscape}
\begin{table}
\begin{center}
\begin{small}
\caption{Properties of the dust continuum emission from the primary sources in our sample: PdBI data and single-dish observations}
\label{table:continuum-pdbi-sd}
\let\center\empty
\let\endcenter\relax
\begin{tabular}{|c||cccc|cc|c||c|cc|cc|c||cc|}
\hline
\hfill &  \multicolumn{7}{|c||}{PdBI} &  \multicolumn{6}{|c||}{Single dish} &  \multicolumn{2}{|c|}{Single-dish fluxes} \\
 \hfill & \multicolumn{4}{|c|}{231\,GHz} & \multicolumn{2}{|c|}{94\,GHz} & & \multicolumn{6}{|c||}{~} &  \multicolumn{2}{|c|}{extrapolated to 231 GHz} \\  
{\bf{Source}} & {F$_{\rm{int}}$} & {F$_{20k\lambda}$} & {$\rm{T}_{B}^{peak}$} & {$\rm{T}_{B}^{peak}$/$\rm{T}_{dust}^{peak}$} & {F$_{\rm{int}}$} & {F$_{20k\lambda}$} & {$\alpha_{20k\lambda}$} & {Frequency} & \multicolumn{2}{c|}{Integrated flux} &  \multicolumn{2}{c|}{Peak flux}  & {Ref} & {F$_{\rm{int}}^{\rm{SD}}$} & {F$_{\rm{peak}}^{\rm{SD}}$} \\
 {} & {(mJy)} & {(mJy)} & {(K)} & & {(mJy)} & {(mJy)} & & {(GHz)}       & {(mJy)}         & {R$_{\rm{int}}$ ($\arcsec$)}  & {(mJy/beam)} & {$\theta$ ($\arcsec$)}    & & {(mJy)}     & {(mJy/beam)}   \\
 \hfill & \hfill & \hfill & \hfill & \hfill & \hfill & \hfill & \hfill & \hfill & \hfill & \hfill & \hfill & \hfill & \hfill & \hfill & \hfill \\
 $^{[1]}$ & $^{[2]}$ & $^{[3]}$ & $^{[4]}$ & $^{[5]}$  & $^{[6]}$ & $^{[7]}$ & $^{[8]}$ & $^{[9]}$ & $^{[10]}$ & $^{[11]}$ & $^{[12]}$ & $^{[13]}$ & $^{[14]}$ & $^{[15]}$ & $^{[16]}$ \\ 
\hline
 \hfill &  \hfill& \hfill & \hfill & \hfill & \hfill & \hfill & \hfill & \hfill & \hfill & \hfill & \hfill & \hfill & \hfill & \hfill & \hfill \\ 
{ L1448-2A }            & 129$\pm$18            & 141$\pm$14    & 2 & 0.03      & 13$\pm$2                & 8$\pm$1               & 3.1$\pm$0.3   & 268   & 930     & 20            &  730  & 31 & $(1)$    & 580   & 450   \\
 \hfill & \hfill& \hfill & \hfill & \hfill & \hfill & \hfill & \hfill & \hfill & \hfill & \hfill & \hfill & \hfill & \hfill & \hfill & \hfill \\ 
{ L1448-NB1}            & 763$\pm$50    & 713$\pm$40    & 13 & 0.20     & 89$\pm$8                & 72$\pm$6              & 2.6$\pm$0.2                   & 243     & 3700  & 20            & 1400  & 12  & $(2)$   & 3240  & 1160   \\
 \hfill & \hfill& \hfill & \hfill & \hfill & \hfill & \hfill & \hfill & \hfill & \hfill & \hfill & \hfill & \hfill & \hfill & \hfill & \hfill \\ 
{ L1448-C}              & 178$\pm$13    & 215$\pm$20    & 11 & 0.16 & 20$\pm$4                 & 21$\pm$3              & 2.6$\pm$0.3   & 243   & 910   & 14              & 620   & 12  & $(2)$   & 827   & 560    \\
 \hfill & \hfill& \hfill & \hfill & \hfill & \hfill & \hfill & \hfill & \hfill & \hfill & \hfill & \hfill & \hfill & \hfill & \hfill & \hfill \\ 
{IRAS2A}                        & 386$\pm$36    & 420$\pm$32    & 12 & 0.13      & 43$\pm$6              & 31$\pm$5              & 2.9$\pm$0.3           & 243     & -             & -             & 875   & 12  &  $(2)$  & 780   & -                \\
 \hfill & \hfill& \hfill & \hfill & \hfill & \hfill & \hfill & \hfill & \hfill & \hfill & \hfill & \hfill & \hfill & \hfill & \hfill & \hfill \\ 
 {SVS13B}               &  318$\pm$20   & 269$\pm$21    & 12 & 0.23      &  43$\pm$4             & 29$\pm$4              & 2.5$\pm$0.3           & 243     & -             & -             & 900   & 11  & $(3)$   & 817   & -                 \\
 \hfill & \hfill& \hfill & \hfill & \hfill & \hfill & \hfill & \hfill & \hfill & \hfill & \hfill & \hfill & \hfill & \hfill & \hfill & \hfill \\
 {IRAS4A1 }             & 2710$\pm$110  & 2978$\pm$90   & 44 & 0.7      & 314$\pm$19      & 253$\pm$16    & 2.7$\pm$0.2           & 243   & -             & -               & 4100  & 12  & $(2)$   & 3260  & -               \\
 \hfill & \hfill& \hfill & \hfill & \hfill & \hfill & \hfill & \hfill & \hfill & \hfill & \hfill & \hfill & \hfill & \hfill & \hfill & \hfill \\ 
{IRAS4B1}                       & 937$\pm$90    & 1310$\pm$80   & 25 & 0.5       & 127$\pm$16    & 128$\pm$10    & 2.6$\pm$0.2           & 243   & -             & -       & 1470  & 12  & $(2)$   & 1290  & -              \\
\hfill & \hfill& \hfill & \hfill & \hfill & \hfill & \hfill  & \hfill & \hfill & \hfill & \hfill & \hfill & \hfill & \hfill & \hfill & \hfill  \\ 
{ IRAM04191}            &  4.5$\pm$1.3          & 5.2$\pm$0.9           & 0.4 & 0.013 & 0.3$\pm$0.1               & 0.3$\pm$0.1           & 3.0$\pm$0.4           &  243    & 650   & 30            & 110   & 11 & $(2)$    & 585   & 99             \\
 \hfill & \hfill& \hfill & \hfill & \hfill & \hfill & \hfill & \hfill & \hfill & \hfill & \hfill & \hfill & \hfill & \hfill & \hfill & \hfill \\ 
{ L1521F}               & 3.7$\pm$1             & 6.4$\pm$1             & 0.14 & 0.05 & 0.2$\pm$0.08              & 0.45$\pm$0.09         & 2.9$\pm$0.4                 & 243   & -             & -             & 120           & 14  & $(4)$     & -     & 95             \\
 \hfill & \hfill & \hfill & \hfill & \hfill & \hfill & \hfill & \hfill & \hfill & \hfill & \hfill & \hfill & \hfill & \hfill & \hfill & \hfill \\ 
{ L1527}                        & 220$\pm$18    & 262$\pm$14    & 12 & 0.20 & 27$\pm$3              & 26$\pm$3              & 2.6$\pm$0.2                   & 243     & 1500  & 30            & 375   & 12 & $(2)$    & 1350  & 340   \\
\hfill & \hfill & \hfill & \hfill & \hfill & \hfill  & \hfill & \hfill & \hfill & \hfill & \hfill & \hfill & \hfill & \hfill & \hfill & \hfill  \\ 
{ SerpM-S68N}           & 111$\pm$8     & 208$\pm$9     & 3 & 0.05 &  10$\pm$2          & 10$\pm$3                & 3.3$\pm$0.3                           & 243   & 1030    & 15            & 550   & 11  & $(8)$   & 870   & 465   \\
 \hfill & \hfill & \hfill & \hfill & \hfill & \hfill & \hfill & \hfill & \hfill & \hfill & \hfill & \hfill & \hfill & \hfill & \hfill & \hfill \\
{ SerpM-SMM4}           & 596$\pm$50    & 620$\pm$40    & 17 & 0.35     & 60$\pm$8                & 60$\pm$6              & 2.6$\pm$0.2                   & 243     & 2550  & 20            & 1000  & 11  & $(8)$   & 2192  & 840   \\
 \hfill & \hfill & \hfill & \hfill & \hfill & \hfill & \hfill & \hfill & \hfill & \hfill & \hfill & \hfill & \hfill & \hfill & \hfill & \hfill \\
{ SerpS-MM18}   & 453$\pm$55     & 483$\pm$45   & 13 & 0.17     &  46$\pm$6             & 29$\pm$3                & 3.1$\pm$0.2                   & 243   & 2505  & 9               & 1376  & 12  & $(5)$   & 2180  & 1230   \\
\hfill & \hfill & \hfill & \hfill & \hfill & \hfill  & \hfill & \hfill & \hfill & \hfill & \hfill & \hfill & \hfill & \hfill & \hfill & \hfill  \\
{ SerpS-MM22}   & 30$\pm$4              & 50$\pm$7              & 1.8 & 0.05 & 3.3$\pm$0.9   & 3.5$\pm$0.8           & 3.0$\pm$0.3                   & 243     & 261   & 10            & 129   & 12 & $(5)$    & 225   & 111   \\
 \hfill & \hfill & \hfill & \hfill & \hfill & \hfill & \hfill & \hfill & \hfill & \hfill & \hfill & \hfill & \hfill & \hfill & \hfill & \hfill \\
{L1157}                         & 262$\pm$15    & 260$\pm$18    & 11 & 0.20  & 34$\pm$3             & 24$\pm$3              & 2.6$\pm$0.3                   & 243     & -     & -             & 630   & 12 & $(2)$    & 569   & -     \\
 \hfill & \hfill & \hfill & \hfill & \hfill & \hfill & \hfill & \hfill & \hfill & \hfill & \hfill & \hfill & \hfill & \hfill & \hfill & \hfill \\ 
{GF9-2}                         & 13$\pm$3              & 13$\pm$3              & 1  & 0.02 & 1.7$\pm$0.8         & 2.0$\pm$0.8           & 2.1$\pm$0.3                   & 243     & 315   & 35            & 60            & 12 & $(7)$    & 291   & 55      \\
 \hfill & \hfill & \hfill & \hfill & \hfill & \hfill & \hfill & \hfill & \hfill & \hfill & \hfill & \hfill & \hfill & \hfill & \hfill & \hfill \\ 
\hline
\end{tabular}
\tablefoot{Column 1: Primary protostar name. Columns 2 and 6 indicate the PdBI integrated fluxes (integrated above the 5$\sigma$ level in the clean maps) at 231 GHz and 94\,GHz, respectively. Columns 3 and 7 indicate the visibility fluxes at 20 k$\lambda$ (shortest common baseline for the 94\,GHz and 231\,GHz data), averaged in a $20 k\lambda$ baseline bin centered at $20 k\lambda$ that we used to compute the  spectral index $\alpha_{20k\lambda}=log(F^{231GHz}_{20k\lambda}/F^{94GHz}_{20k\lambda})/log(231/94)$, given in Col. 8. 
This reflects the properties of dust emission at the largest envelope 
scales probed by the PdBI observations at 231\,GHz ($\sim 10\arcsec$).
We note that the 231 GHz visibility fluxes shown here stem from the 231\,GHz data alone, before they were merged with the rescaled 219\,GHz data (making use of the $\alpha_{20k\lambda}$).
Cloumn 4: Brightness temperature from the peak flux at 231 GHz. [5] Ratio T$_B$/T$_{dust}$ at $0.5\arcsec$ scales (using Equ.\ref{eq:temp} for T$_{dust}$ with r the physical radius probed by the synthesized beam) indicative of the optical thickness of the continuum emission at the smallest scales probed in our study. Column
9: Frequency of the single-dish observations used to extrapolate the envelope-scale fluxes (zero spacing).
Column 10: Single-dish integrated flux at the frequency indicated inCol. 9. Column 11: Angular radius of the area used to compute the integrated envelope flux in single-dish maps.
Column 12: Singl-dish peak flux at the frequency indicated in Col. 9.
Column 13: FWHM of the single-dish beam used to compute the peak flux in single-dish maps.
Column 14: Reference reporting the single-dish observations.
(1)~\citet{Enoch07} CSO Bolocam 1.1 mm; (2) \citet{Motte01a} IRAM-30m-MPIfR 
arrays 1.3 mm; (3) \citet{Chini97} IRAM-30m MPIfR array 1.3 mm; (4) 
\citet{Crapsi04} IRAM-30m MAMBO II 1.3 mm; 
(5) \citet{Maury11} IRAM-30m MAMBO II 1.3 mm; (6) \citet{Enoch11} CSO 
Bolocam 1.1 mm; (7) \citet{Wiesemeyer97} IRAM-30m MPIfR array 1.3 mm; 
(8) \citet{Kaas04} IRAM-30m MPIfR array 1.3 mm.
Column 15: Envelope-integrated dust continuum flux from the single-dish maps, extrapolated to 231\,GHz in order to be used as zero-spacing constraint in our modeling.
Column 16: Envelope dust continuum flux at the scale that the single-dish beam probes, extrapolated to 231\,GHz in order to be used as either zero-spacing constraint in our modeling (if the source is unresolved by single-dish observations) or as an additional point at the equivalent baseline.\\
}
\end{small}
\end{center}
\end{table}
\end{landscape}

\subsection{Comparison to protostellar envelope models}

From the continuum visibility dataset constructed for each primary protostar (see Sect.~\ref{section:buildvisib}), 
we extracted the flux density of the protostellar dust continuum emission as a function of spatial frequency by averaging the real parts of the observed visibilities in bins of baseline lengths
(bin widths of $20 \rm{k}\lambda$ at 231 GHz, and bin widths of $10 \rm{k}\lambda$ at 94 GHz).
We obtained visibility curves sampling baselines [20--590] $\rm{k}\lambda$ at 231~GHz and [10--240] $\rm{k}\lambda$ at 94~GHz.
Error bars on visibility amplitudes were estimated from the individual weights ($w_i$) of the interferometric visibility amplitudes ($y_i$), which were then averaged in bins of uv-distance: the error bar is the error on the mean value in an individual bin $(y_{err}=\Sigma(w_i^2\times(y_i - w_{mean})^2 ) / (\Sigma w_i )^2$ and $w_{mean}=\Sigma(y_i\times w_i)/\Sigma w_i$).
The continuum visibility curves for two of the Perseus sources (L1448-2A and SVS13B) and a Taurus source (L1527) are shown in Figures \ref{fig:l14482a_popgmodels} to \ref{fig:l1527_popgmodels}, while the visibility curves of the other sources of the sample can be found in Appendix C  (Figs \ref{fig:l1448nb1_bothmodels} to \ref{fig:gf92_bothmodels}). 
Our PdBI visibility profiles probe the radial distribution of the dust continuum emission from spatial scales $\theta=0.35\arcsec$ to $\theta=12\arcsec$. The envelope emission at larger scales is constrained by the single-dish dust continuum profiles, which are mostly obtained with the IRAM-30m telescope (beam FWHM $11\arcsec$ at 1.3mm).

We compared these dust continuum emission visibility curves to simple models of protostellar envelopes 
in a way similar to what was carried out in \citet{Maury14} to analyze the preliminary CALYPSO data for the NGC1333 IRAS2A protostar. 
The envelope models considered here and the results of their comparison with  
the dust continuum observations are described below.

\subsubsection{Parameterized envelope models}

For protostars in which a hydrostatic core has formed at the center, self-similar collapse solutions \citep{Shu77,Whitworth85} suggest that the radial density profile of the envelope at large radii $r>R\inside$  
is expected to range between $\rho(r)\propto r^{-3/2}$ in the inner dynamical free-fall region (in {{spherical}} similarity solutions) and $\rho(r)\propto r^{-2}$ in the outer region where the initial conditions found in prestellar cores would have been conserved, where $R\inside$ can be the centrifugal radius if a disk develops or the radius of an inner cavity in the envelope. Since the radius of the expansion wavefront grows at the sound speed ($\sim 0.2\, \kms$ at 10 K), reaching $\sim 2000$ au in about $5 \times 10^4$ years (Class 0 lifetime), it seems reasonable to adopt a single power-law envelope model to describe the inner envelope probed by the PdBI observations (the largest scales probed are 500 to 2000 au).
Since our sample consists of objects at different evolutionary stages from very early Class 0 to borderline Class 0/I sources (e.g., L1527) that are
embedded in different environments, we chose to explore a range of density profiles from $\rho(r) \propto r^{-2.5}$ to the more shallow $\rho(r) \propto r^{-1}$, 
in agreement with published millimeter continuum studies of resolved protostellar envelopes \citep[][Shirley et al. 2002]{Motte01a,Looney03}.

We adopted a simple spherically symmetric parameterized model with a Plummer-like density profile \citep{Plummer11, Whitworth01} to describe the protostellar density structure.
In this model, the inner envelope has a nearly constant central density inside $R\inside$, while the outer 
envelope has a density profile approaching a power law $r^{-p}$, where $p$ is a constant index:
\begin{equation}
\label{eq:rho}
\rho(r) \; = \; \rho\inside \left[ \frac{R\inside}
{\left( R\inside^2 + r^2 \right)^{1/2}} \right]^p \equiv \; \frac{\rho\inside}
{\left(1 + \left(r/R\inside\right) ^2 \right)^{p/2}}
.\end{equation}
The model has three free parameters: 
a truncation radius at $R_{\rm out}$ to account for the finite size of the envelope,
the index $p,$ which fixes the logarithmic slope of the density profile at radii $R_{\rm out} > r > R\inside$, and the inner radius $R\inside$ 
 , which marks the end of the approximately uniform density of the inner region ($0 < r \leq R\inside$). 
In practice, $R\inside$ may be interpreted as the radius at which rotational motions start to dominate envelope infall motions. 

The dust temperature in the inner region of a protostellar envelope is governed by infrared radiation from the release of gravitational energy of the material that accretes onto the central object.
We ignore here the contribution of the interstellar radiation field, which mostly contributes to heating the outer layers of the envelope, 
although this contribution may be significant in the very low luminosity objects IRAM04191 and L1521F. 
We note that the dust continuum emission tracing the inner envelopes in our sample is optically thin at the two wavelengths we probed, as checked from the peak and integrated brightness temperatures from our PdBI maps (see Cols.~4 and 5 of Table~\ref{table:continuum-pdbi-sd}), except for IRAS4A where the emission is partially optically thick at $\sim 0.5\arcsec$ scales.
Following \citet{Butner90} and \citet{Terebey93}, the dust temperature distribution $T(r)$ in the power-law envelope at radii $r > R\inside$ that is due to the central heating of the protostellar envelope
may be approximated as follows in the case of optically thin dust continuum emission:
\begin{equation}
T(r) = 60 \left(\frac{r}{13400~{\rm au}}\right)^{-q} \left(\frac{\lint}{10^5~{\rm L}_\odot}\right)^{q/2}~~~{\rm K},
\label{eq:temp}
\end{equation}
where $r$ represents the radius from the central source of luminosity \lint. 
We varied the index  $q$ of the power-law  temperature dependence ($T(r) \propto r^{-q} $)
 between $q = 0.3$ to $q = 0.5$ in our model fits.

The emerging intensity distribution of the envelope results from the combination of the dust density and temperature distributions 
of Eqs.~\ref{eq:rho} and~\ref{eq:temp}. 
For instance, in the Rayleigh-Jeans regime, the radial intensity distribution of a power-law envelope is expected to scale as $I(r) \propto r^{-(p+q-1)}$ (cf. Adams 1991). 
In practice, we therefore considered the following model intensity distribution:
\begin{equation}
\label{eq:int}
I(\bar{r}) \; = \;  \frac{I_0}{\left(1 + \left(\bar{r}/R\inside\right) ^2 \right)^{(p+q-1)/2}},
\end{equation}
where $I_0$ is the intensity from radii $r < R\inside$ and $\bar{r}$ represents the projected radius.
Considering the combination of likely density and temperature distributions presented above, 
we let our model span a range from $p+q=1.3$ to $p+q=2.9$ between $R\inside$ and $R_{\rm out}$.

\subsubsection{Fitting the dust continuum emission visibilities with envelope models}

To model our interferometric observations, we converted the intensity distribution of the Plummer envelope model into a 1D visibility curve as a function of baseline $b=\sqrt{u^2+v^2}$
using a Hankel transform (corresponding to the 2D Fourier transform of a circularly symmetric function, see \citealt{Bracewell65, Berger07}) :
\begin{equation}
 V(b) = 2\pi \int^\infty_0 I_\nu(r_{\rm b})J_0\left(2\pi r_{\rm b}b\right)
 r_{\rm b}d r_{\rm b},
\label{eq:TF}
\end{equation}
where $J_0$ is the zeroth-order Bessel function,
\begin{equation}
 J_0(z) = \frac{1}{2\pi}\int^{2\pi}_0 \exp\left(-iz\cos\theta\right) d\theta.
\label{eq:Bessel}
\end{equation}
For example, 
this interferometric transform turns a spherically symmetric power-law intensity distribution $I(r) \propto r^{-(p+q-1)}$ 
into a power-law visibility distribution as a function of u-v distance, 
$V(b) \propto b^{p+q-3}$,
solely determined by the power-law indices of the temperature and density profiles $p$ and $q$
in the envelope at $r>R\inside$.

We performed a least-squares fit of the observed visibility data
with the interferometric transform of the parameterized envelope model described by Eqs.~\ref{eq:rho} and~\ref{eq:temp}. 
The envelope power-law index $(p+q)$ and the inner cavity radius $R\inside$ were left as free parameters within the previously mentioned ranges, when we adjusted the observed profiles. The estimated parameter uncertainties were computed from the covariance matrix (see the description of the mpcurvefit in \citealt{Markwardt09}).
The best-fit set of envelope parameters that reproduces the PdBI continuum visibility distribution for each of the 16 CALYPSO Class 0 sources is reported in Table \ref{table:popg-fits}. The modeled envelope visibility profiles are shown for three sources of our sample as a black curve overlaid on the observed visibilities in Figures \ref{fig:l14482a_popgmodels} to \ref{fig:l1527_popgmodels}, while the models for other sources of the sample are shown in Appendix \ref{section:appindivsour} (Figs \ref{fig:l1448nb1_bothmodels}  to \ref{fig:gf92_bothmodels}).

The radial distributions of the millimeter dust continuum emission of 10 of the 16 Class 0 protostars in our sample (L1448-2A, IRAS2A, SVS13B, IRAS4A1, IRAM04191, L1521F, SerpM-S68N, SerpS-MM18, SerpS-MM22, and L1157) can be satisfactorily reproduced with models of Plummer-like protostellar envelopes 
with reduced $\chi^2$ values $\leq 3$ (see the model parameters and reduced $\chi^2$ values reported in the first two lines of Table \ref{table:popg-fits} for each source) in the two continuum bands that are probed in CALYPSO observations. 
Moreover, the model envelope sizes and power-law indices $p+q$ for these 10 sources that we 
report in Table \ref{table:popg-fits} show a good agreement in the two continuum bands (within 15\%). This validates our choice of not fitting the two visibility profiles at the two frequencies jointly, and it shows that when a Plummer envelope model is satisfactory, there is no need to fix the envelope parameters to reproduce the dust continuum emission distribution at the longest baselines that are only probed by the 231\,GHz data.

\subsection{Comparison with envelope models including a disk-like component}

For three protostars in our sample (L1448-C, L1448-NB1, and L1527; see Table \ref{table:popg-fits}), no single Plummer envelope model was able to satisfactorily reproduce the continuum emission data at both 231~GHz and 94~GHz (reduced $\chi^2 > 3$). At these two frequencies, the curvature of the radial intensity profile for these sources does not follow a simple power-law trend, or at uv-distances longer than 100 k$\lambda,$ their PdBI visibility curve does not decrease quickly enough to be reproduced by one envelope model alone. 
In addition, while the 94\,GHz visibility profiles of IRAS4B and GF92 are satisfactorily described by single-envelope models, their 231~GHz visibility profiles are inconsistent with single Plummer-like envelope models. This suggests that emission at baselines probed only by the 231~GHz data (250-550 k$\lambda$) differs from the best-fit envelope model we found to reproduce the 94~GHz continuum emission. 
Finally, while the 231 GHz profile can be modeled quite satisfactorily by a Pl model (reduced $\chi^2 \leq3$) for SerpM-SMM4, this does not hold for the 94 GHz profile.

The failure to reproduce one or both of the visibility profiles using circularly symmetric Plummer-like envelope models may reflect the presence in the source of either (i) asymmetric features in the envelope structure or (ii) an additional density or temperature component at small scales that is not properly modeled, such as a protostellar disk that is embedded in the envelope\footnote{Note that our approximation to model the envelope contribution with a single power-law Plummer profile is not expected to affect our detection of disk-like structures from the visibility curves at long baselines since a flatter inner density power law index produces a steeper visibility curve.}. 
To check whether the visibility profiles of these six objects could include an additional continuum emission component originating from a disk that is embedded in the envelope, we also fit the visibility profile of each CALYPSO protostar 
with a two-component model consisting of a Plummer-like envelope (cf. Eq.~\ref{eq:rho}) and an additional compact circular Gaussian component located at phase center, whose flux $F_{\rm{Gauss}}$ and FWHM $\Theta_{\rm{Gauss}}$ were left as free parameters (hereafter PG model for Plummer$+$Gaussian model). For these sources where the Gaussian component is spatially resolved, the value of $R_{\rm{i}}$ is enforced to be higher than the value of $\Theta_{\rm{Gauss}}$ during the minimization process.
The visibility function of the Gaussian disk component was calculated as
\begin{equation}
\label{eq:Gauss}
V(b) \propto \exp(-\pi\times\Theta_{\rm{Gauss}}\times b^2/4ln2).
\end{equation}

\begin{figure*}[!h]
\centering
\includegraphics[trim={0 0 0 0},clip,width=0.72\textwidth]{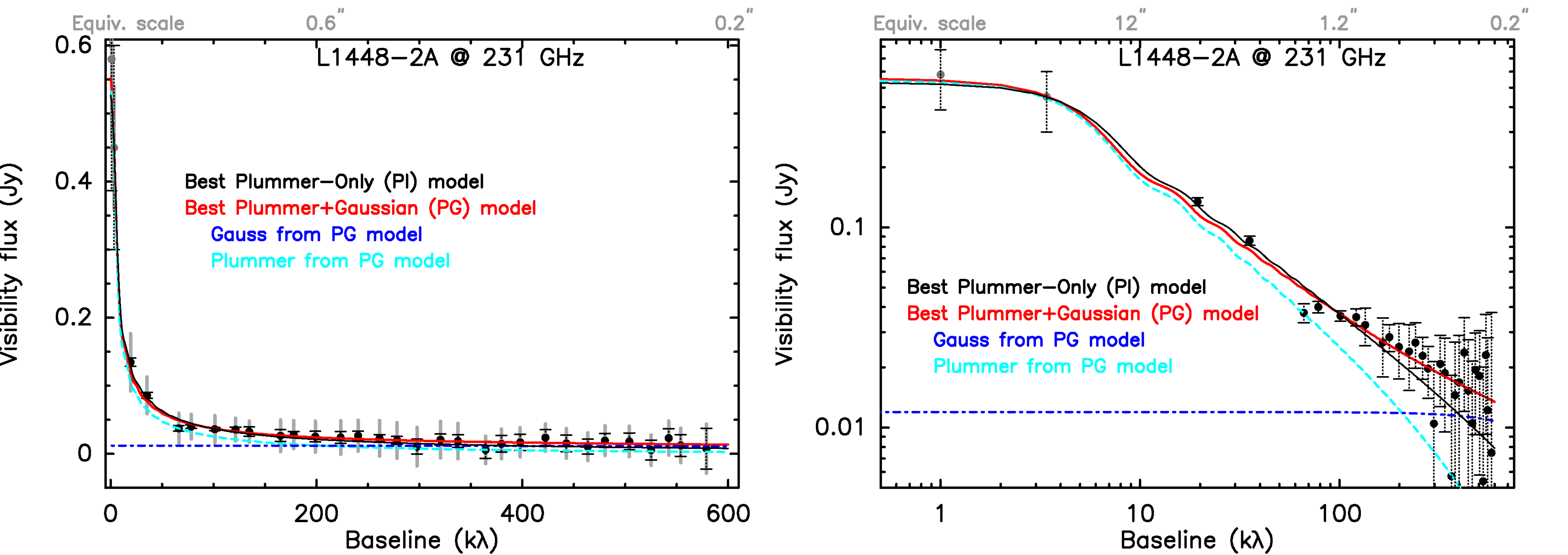}
\includegraphics[trim={0 0 0 0},clip,width=0.72\textwidth]{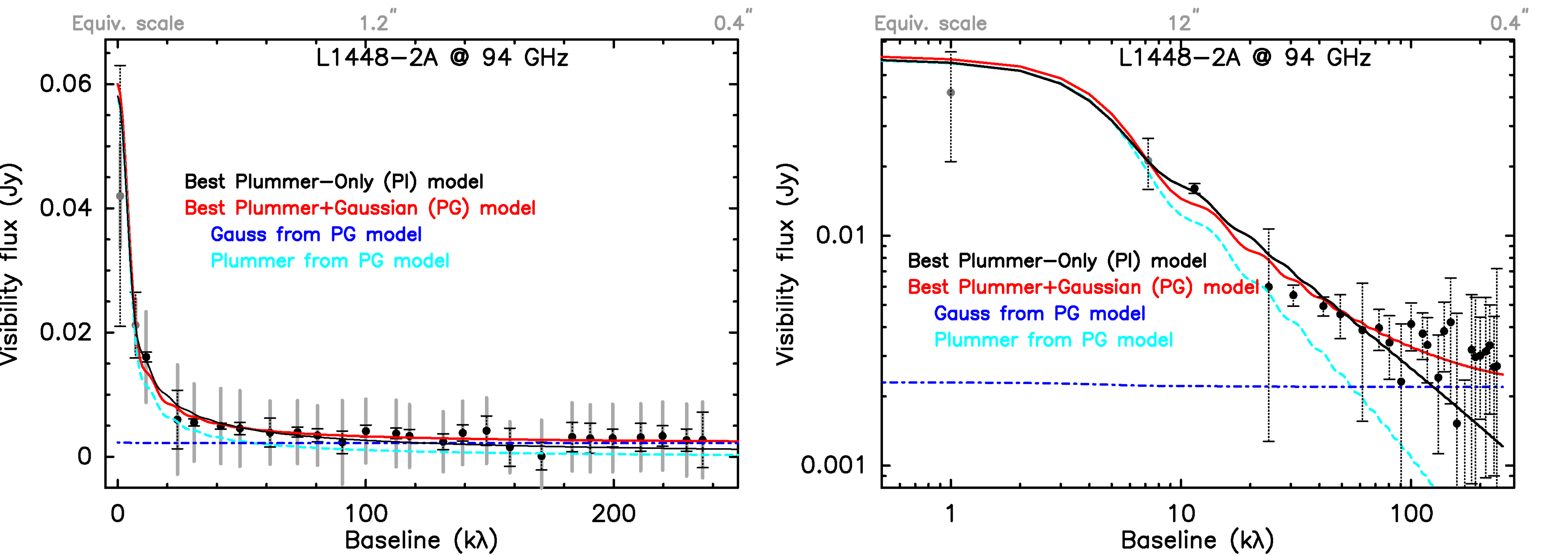}
\caption{Top: 1.3~mm (231 GHz) PdBI dust continuum emission visibility real parts as a function of baseline length (circularly averaged in 20k$\lambda$ bins) for L1448-2A. The left panel uses a linear scale, while the right panel shows the same data and models in logarithmic scale to enhance the visibility of the long baseline points. The equivalent scales probed by the baselines in the image space are indicated at the top of each panel (computed as $0.6\times\lambda/b$). In both panels, the plain lines show the best-fit Plummer-only envelope (Pl, black) and Plummer+Gaussian (PG, red) models we found to reproduce the visibility profile of the dust continuum emission in this source. The dashed lines show the two components included in the best-fit PG model: the Plummer-envelope component in dashed light blue and the additional Gaussian component in dot-dashed dark blue (see Table \ref{table:popg-fits} for more information on the two models). The Plummer+Gaussian (PG)  model is not statistically better than the Plummer-only (Pl) for L1448-2A.
Bottom: Same as the top panels for the 94\,GHz visibility profiles.}
\label{fig:l14482a_popgmodels}
\end{figure*}
\begin{figure*}[!h]
\centering
\includegraphics[trim={0 0 0 0},clip,width=0.72\textwidth]{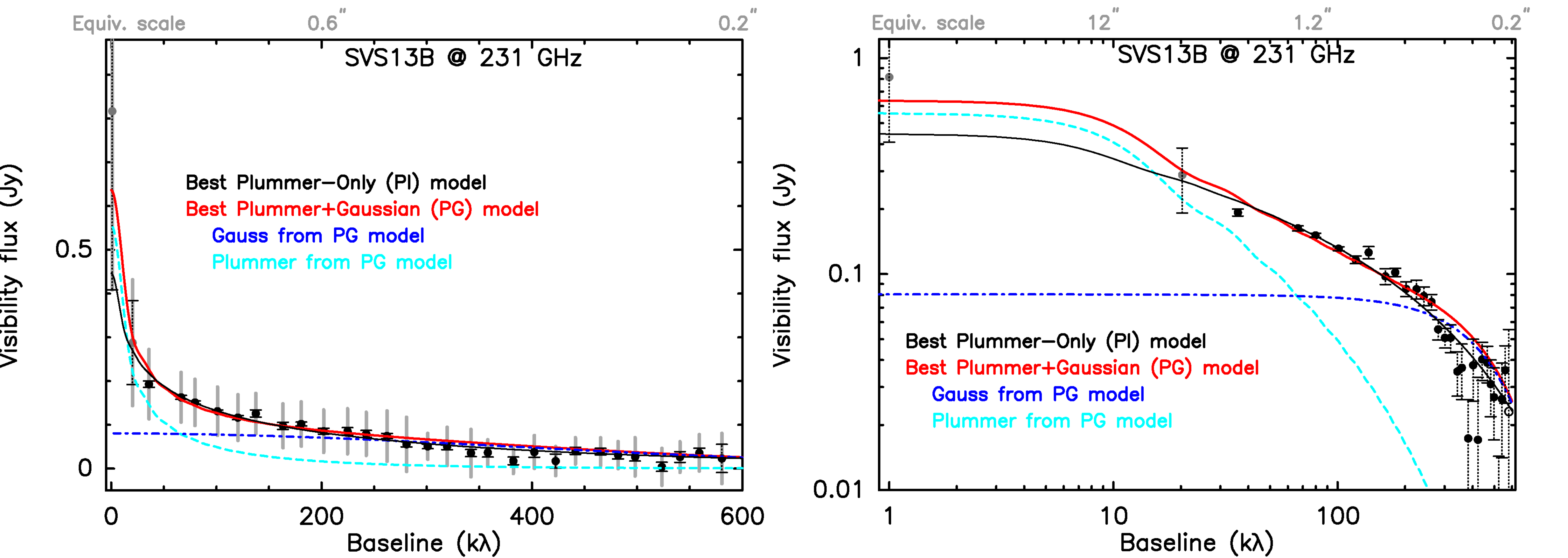}
\includegraphics[trim={0 0 0 0},clip,width=0.72\textwidth]{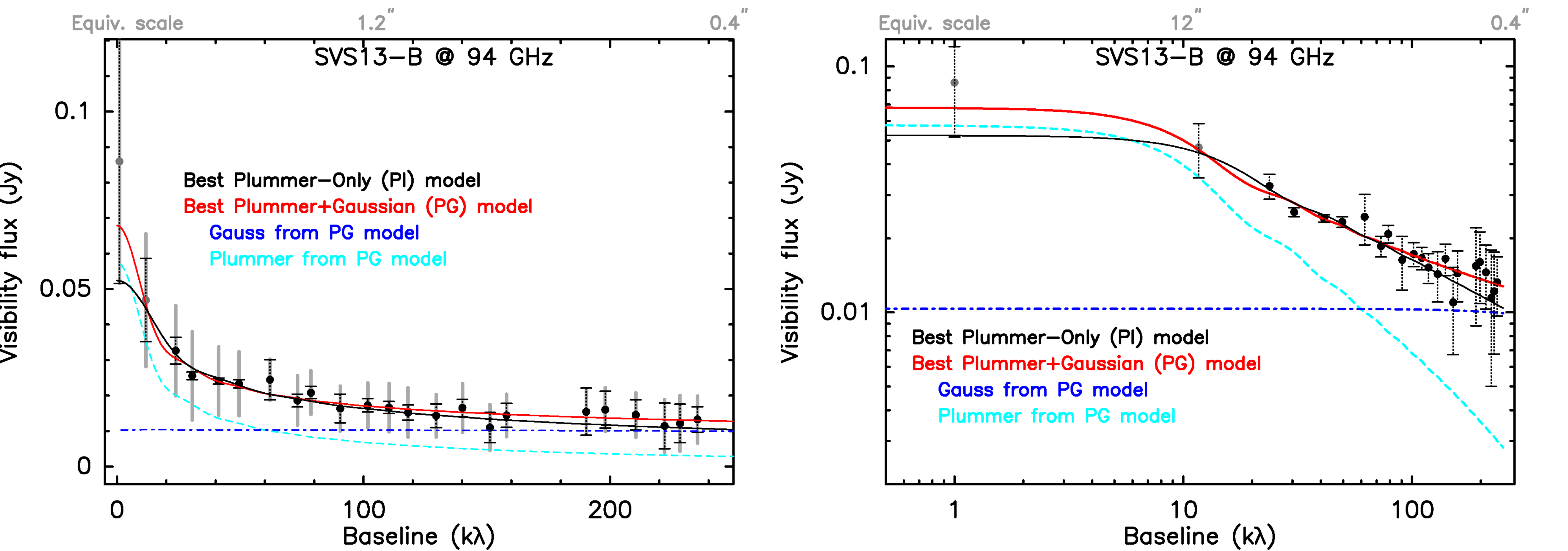}
\caption{Same as Fig. \ref{fig:l14482a_popgmodels} for SVS13B. The red Plummer+Gaussian (PG) model that includes a marginally resolved $0.2\arcsec$ FWHM Gaussian component is as good as the black Plummer-only (Pl) model in reproducing the 231 GHz visibility profile, but it provides a more reasonable value of the $p+q$ index. 
Bottom: Same as the top panels for the 94\,GHz visibility profiles. The PG model does not perform any better than the Pl model for the 94 GHz visibility profile (see Table \ref{table:popg-fits}).
}
\label{fig:svs13b_popgmodels}
\end{figure*}
\begin{figure*}[!h]
\centering
\includegraphics[trim={0 0 0 0},clip,width=0.78\textwidth]{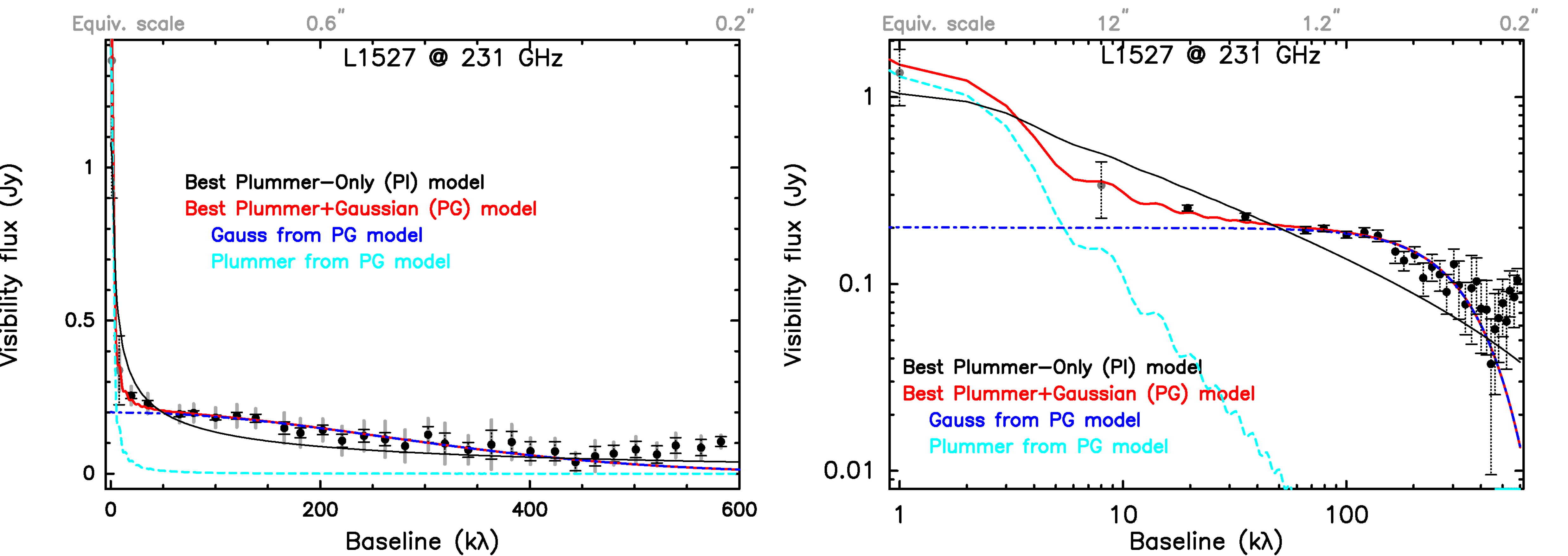}
\includegraphics[trim={0 0 0 0},clip,width=0.78\textwidth]{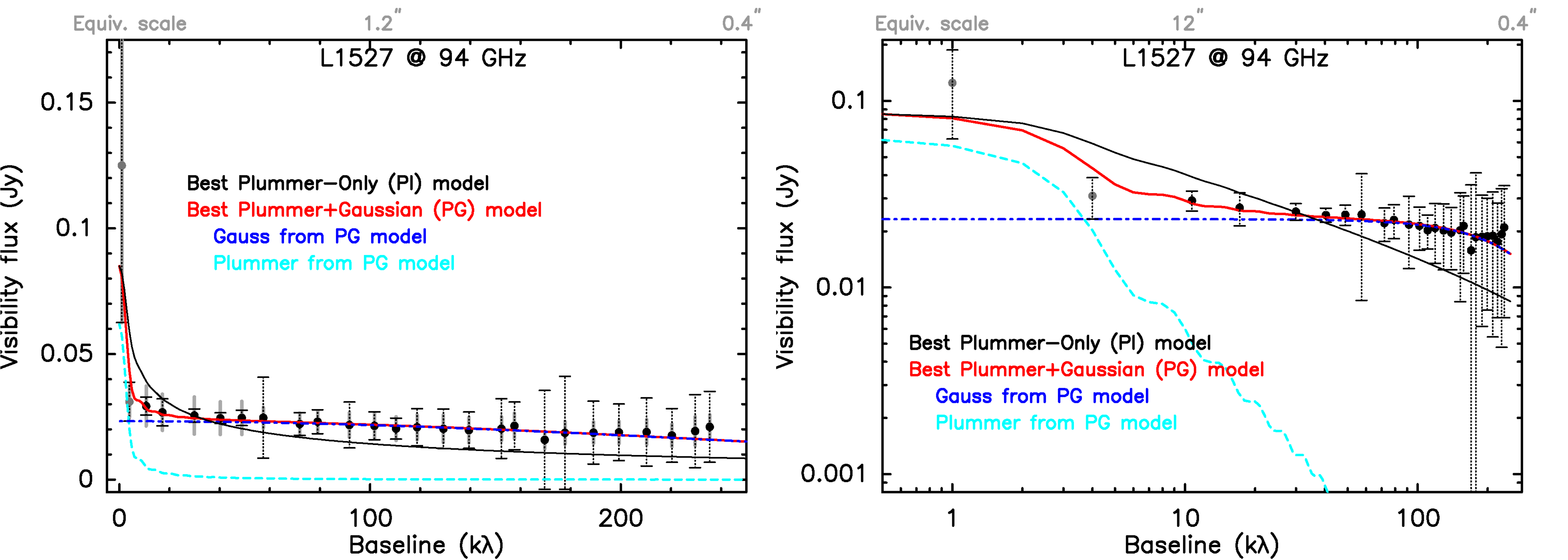}
\caption{Same as Fig. \ref{fig:l14482a_popgmodels} for L1527. The red Plummer+Gaussian (PG) model that includes a $0.4\arcsec$-FWHM Gaussian component (see Table \ref{table:popg-fits} for more information on the two models) is statistically better than the black Plummer-only (Pl) model.
Bottom: Same as the top panels for the 94\,GHz visibility profiles.
}
\label{fig:l1527_popgmodels}
\end{figure*}
\begin{figure*}[!h]
\centering
\includegraphics[trim={0 0 0 0},clip,width=0.78\textwidth]{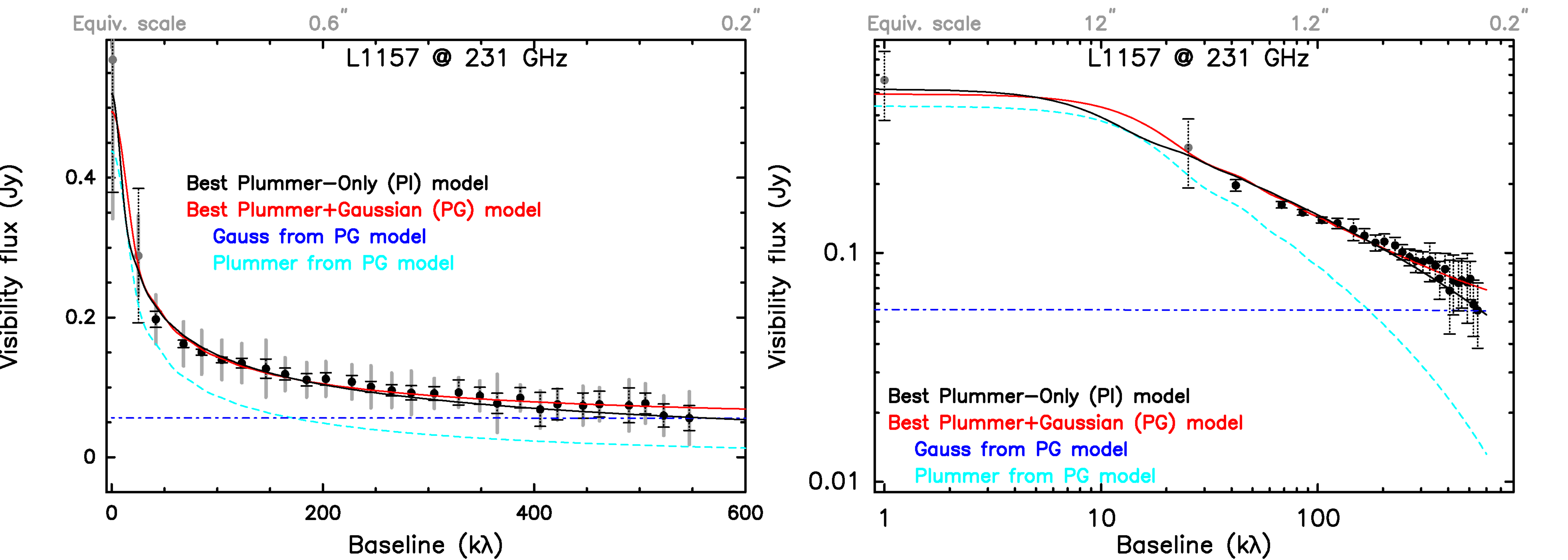}
\includegraphics[trim={0 0 0 0},clip,width=0.78\textwidth]{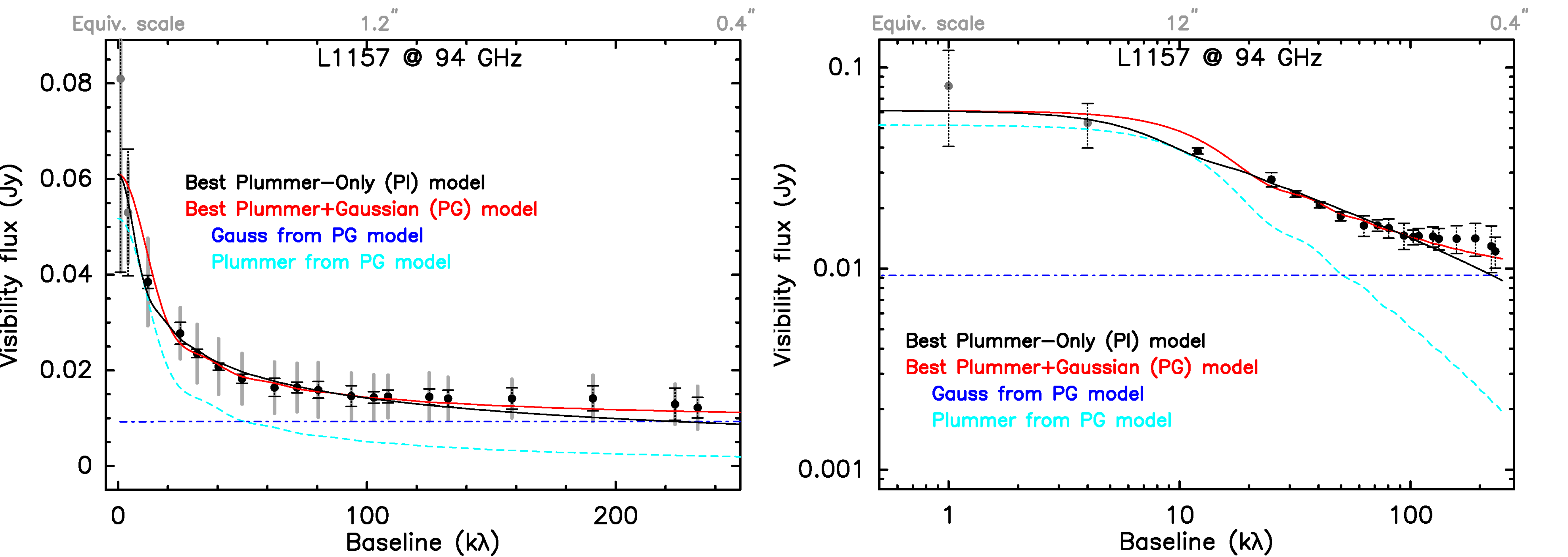}
\caption{Same as Fig. \ref{fig:l14482a_popgmodels} for L1157. The black Plummer-only (Pl) model is statistically better than the red Plummer+Gaussian (PG) model that includes an unresolved Gaussian component (see Table \ref{table:popg-fits} for more information on the two models).
Bottom: Same as the top panels for the 94\,GHz visibility profiles.
}
\label{fig:l1157_popgmodels}
\end{figure*}

The parameters of the best-fit PG models are reported in Table \ref{table:popg-fits}. The figures \ref{fig:l1448nb1_bothmodels} to \ref{fig:gf92_bothmodels} in Appendix \ref{section:appindivsour} show the Pl and PG models over the 231\,GHz and 94\,GHz visibility curves of all the sources in our sample.

We performed a Fisher \footnote{For more information on the Fisher test, see for example \citet{Donaldson} and references therein.} statistical test (later referred to as an F-test) to decide which of the best-fit Plummer-only model (Pl) and the best Plummer+Gauss (PG) models we obtained for each source in our sample was better. The goal was to avoid overmodeling the visibility profiles of sources with complex structure 
using a model that reproduced the observations better only coincidentally, because it includes two additional free parameters, such as the PG model. 
The results of the F-test (computing P values from the F distribution to test if the PG models are statistically better than the Pl models despite their two additional free parameters), which compared the best-fit Pl and PG models we obtained for each protostar are reported in Column 10 of Table \ref{table:popg-fits}: we indicate a "yes" if the F-test suggests that the PG model is statistically better than the Pl model, and a "no" otherwise.

For all ten sources that have satisfactory (reduced $\chi^2 < 3$) Pl models at the two frequencies (231 GHz and 94 GHz), none of the PG models provided a better fit for the two frequencies, except for SerpS-MM22.
However, for five out of the six sources with unsatisfactory Pl models, the F-test suggests at either 231 GHz (L1448-C, L1527, IRAS4B, and GF92) or 94\,GHz (L1448-C, L1527 and SerpM-SMM4) that the PG models reproduce the millimeter dust continuum visibility profiles  significantly better that are not well described by the Pl models.

For three protostars in our sample that are better reproduced with PG models at the two frequencies (L1527, SerpM-SMM4 and SerpS-MM22), our modeling suggests an additional Gaussian component is detected, that is well resolved in our 231~GHz PdBI data. 
While the inclusion of a Gaussian component significantly improved the minimization
at both frequencies for two other protostars (L1448-C and GF9-2), 
the FWHMs of these components are very similar to the scale probed by our
longest baseline (570 k$\lambda,$ i.e., probing radii down to 0.26$\arcsec$). This indicates that the additional Gaussian components suggested by our modeling are at best marginally resolved considering the increase in phase noise at the longest baselines, and that their sizes are therefore highly uncertain (see Table \ref{table:popg-fits}).
For IRAS4B, the inclusion of a large and bright Gaussian component significantly improves the minimization for the 231 GHz visibility profile, which was unsatisfactorily modeled with a single Plummer envelope, but does not improve the modeling of our 94 GHz data.
For five protostars that are well modeled at both frequencies using Pl models (SVS13B, IRAM04191, L1521F, SerpS-MM18, and L1157), the inclusion of unresolved Gaussian components in the PG models allows obtaining a significantly better minimization at 231\,GHz, where we reach the best angular resolution, but does not improve the modeling of the 94\,GHz visibility profiles. When we consider the null flux level dependence at the longest baselines in our PdBI observations, the decreasing slope of such low fluxes with increasing baseline at the longest baselines is questionable, and we therefore conservatively assume that these Gaussian components are detected but unresolved.

Finally, none of the models we attempted performed very well for L1448-NB1 (with reduced $\chi^2$ values always higher than or equal to 3), 
probably because the emission at long baselines ($>200$ k$\lambda$) shows wiggles that cannot be reproduced 
with a simple model such as we adopted here. In the specific case of this protostar, other models centered either on L1448-NB2 or in between 
the two millimeter sources L1448-NB1 and L1448-NB2 were attempted. They tentatively suggested the presence of 
an additional circumbinary structure (see Appendix \ref{section:appl1448n}).

Figures \ref{fig:l14482a_popgmodels} to \ref{fig:l1157_popgmodels} show examples of Plummer-only (Pl, black curve) and Plummer+Gauss (PG, red curve) model fits for four sources of our sample: L1448-2A, L1157, and SVS13B, which are satisfactorily described by an envelope-only model, and L1527, for which the inclusion of an additional central Gaussian source allows us to reproduce the visibility profile better than an envelope-only model.

\begin{table*}
\begin {center}
\caption{Properties of candidate disk-like structures detected in the PdBI visibility profiles of  the CALYPSO Class~0 sources.}
\label{table:disks-properties}
\begin{tabular}{|c|cc||c|c|c|}
\hline
  \hfill & \hfill & \hfill & \hfill & \hfill & \hfill   \\ 
 {\bf{Source}} & \multicolumn{2}{|c||}{Candidate disk} & {Disk } & {Disk flux} & {Disk/envelope}  \\
  \hfill & \multicolumn{2}{|c||}{detected ~~/~~ resolved} & {radius (au)} & {at 231 GHz (mJy)} & {flux ($\%$)}  \\ 
        {\tiny{[1]}} & {\tiny{[2]}} & {\tiny{[3]}} & {\tiny{[4]}} & {\tiny{[5]}} & {\tiny{[6]}}  \\ 
\hline
\hfill & \hfill & \hfill & \hfill & \hfill & \hfill  \\ 
{ L1448-2A }            & No & -                & $<40$                 & $<12$           & $<2$  \\
\hfill & \hfill & \hfill & \hfill & \hfill & \hfill   \\ 
{ L1448-NB1 $^{\star}$}                 & No & -                & $<40$                 & $<38$   & $<2$  \\
\hfill & \hfill & \hfill & \hfill & \hfill & \hfill  \\ 
{ L1448-C}              &  Yes & Marginally     & $ 37\pm12$    & $130\pm5$                 & $15\pm5$              \\
\hfill & \hfill & \hfill & \hfill & \hfill & \hfill  \\ 
{ IRAS2A1}              &  Yes (only at $$94GHz) & No             & $\simlt50$                 & $52\pm5$              & $ 6.7 \pm 2$  \\
\hfill & \hfill & \hfill & \hfill & \hfill & \hfill  \\ 
{ SVS13B}               &  Yes (only at $$231GHz) & No            & $\simlt60$                 & $80\pm7$              & $ 10\pm20^{\lul}$      \\
\hfill & \hfill & \hfill & \hfill & \hfill & \hfill  \\ 
{ IRAS4A1}              & No & -                & $<75$                 & $<350$                  & $<12$ \\
\hfill & \hfill & \hfill & \hfill & \hfill & \hfill  \\ 
{ IRAS4B}               &  Yes (only at $$231GHz) & Yes   & $125\pm25$    & $645\pm35$              & $50\pm10$             \\
\hfill & \hfill & \hfill & \hfill & \hfill & \hfill  \\ 
{ IRAM04191$^{\star\star\star}$}        &  Yes (only at $$231GHz) & No            & $\simlt60$              & $3.6\pm1$             & $0.6\pm0.3$   \\
\hfill & \hfill & \hfill & \hfill & \hfill & \hfill \\ 
{ L1521F$^{\star\star\star}$}           &  Yes (only at $$231GHz) & No            & $\simlt60$              & $1.3\pm0.4$                   & $0.14\pm0.1$  \\
\hfill & \hfill & \hfill & \hfill & \hfill & \hfill  \\ 
{ L1527}                &  Yes & Yes    & $54\pm10$     & $215\pm14$    & $16\pm2$                \\
\hfill & \hfill & \hfill & \hfill & \hfill & \hfill   \\ 
{ SerpM-S68N}   & No & -        & $<50$                 & $<28$                 & $<4$ \\
\hfill & \hfill & \hfill & \hfill & \hfill & \hfill   \\ 
{ SerpM-SMM4}   & Yes & Yes     & $290\pm40$    & $595\pm35$    & $27\pm4$ \\
\hfill & \hfill & \hfill & \hfill & \hfill & \hfill   \\ 
{ SerpS-MM18}   &  Yes (only at $$231GHz) & No    & $\simlt45$            & $76\pm4$                & $3.5\pm1$ \\
\hfill & \hfill & \hfill & \hfill & \hfill & \hfill   \\ 
{ SerpS-MM22}   & Yes & Yes     & $ 65\pm10$            & $31\pm4$              & $14\pm35$ \\
\hfill & \hfill & \hfill & \hfill & \hfill & \hfill   \\ 
{L1157}                 & Yes (only at $$231GHz) & No     & $\simlt 50$           & $56\pm6$                & $10$ \\
\hfill & \hfill & \hfill & \hfill & \hfill & \hfill   \\ 
{GF9-2}                 & Yes & Marginally      & $36\pm9$              & $12\pm1$                & $4\pm2$ \\
\hfill & \hfill & \hfill & \hfill & \hfill & \hfill  \\ 
\hline
\end{tabular}
\end {center} 
\tablefoot{
Column 1 reports the name of the primary protostar. 
Column 2 reports ''Yes'' or ''No'' : whether adding a disk-like Gaussian component improves the modeling at the respective frequency. 
Column 3 reports ''Yes'' if a disk-like component is detected and resolved by at least one of the two frequencies, ''Marginally'' if a disk-like component is detected at both frequencies but is only marginally resolved by our 231 GHz data, and ''No'' if a disk-like component is detected by at least one of the two frequencies but is not resolved by any. 
Column 4 reports the disk radius. If a disk-like component is detected and resolved (at least one "Yes" in Cols. 2 and 3), it is the FWHM of the additional Gaussian component. 
When no disk component is detected at either frequency ( two "No" in Cols. 2 and 3), we report the maximum disk-like component that can be added to the Plummer model.
Column 5 reports either the flux of the disk-like component added to the envelope model at 231 GHz (if "Yes" in Col. 2 and "Yes" in Col. 4) or the maximum flux of the disk-like component that can be added to the best-fit envelope model (if "No" in Col. 2 and "No" in Col. 4), see text.
Column 6 reports the disk-to-envelope flux at 231 GHz (total envelope flux obtained from single-dish observations) for each source.
\\$^{\lul}$ This value uses the single-dish envelope flux from \citet{Chini97}, but could be up to three times higher if the total envelope flux is closer to the lower value reported by \citet{Lefloch98}. See Appendix \ref{section:appsvs13b} for further details.
\\$^{\star}$ L1448-NB1 might be embedded within a $\sim 200$ au circumbinary structure together with L1448-NB2, see comments on this individual source in Appendix \ref{section:appl1448n}, but the individual protostellar disk of NB1 cannot be stronger and larger than the remaining structure once this circumbinary structure and the surrounding Plummer envelope are subtracted. We here report these upper-limit values.
\\$^{\star\star\star}$ The low S/N of binned visibilities for IRAM04191 and L1521F does not allow us to reliably determine the size from the disk-like component detected at 231 GHz in these two sources. Since all the error bars overlap at baselines $>300$ k$\lambda,$ we report this equivalent size as the maximum disk radius.
}
\end{table*}

\subsection{Characterization of disk-like components in the sample}

For a 2D Gaussian distribution, most of the radiation (90\%) is emitted from within a radius $r < FWHM$, thus we conservatively defined the candidate disk radius to be the estimated FWHM size of the additional Gaussian component when the PG model provided a better model than the Pl model (as suggested by the F-test, see previous section and Col. 10 of Table \ref{table:popg-fits}). 

For sources where the Pl model reproduces the visibility profile better at both frequencies (L1448-2A, IRAS4A, SerpM-S68N, and SerpS-MM18), we performed a new minimization of the PG model with fixed envelope parameters from the best-fit Pl model (PGf models in Appendix \ref{section:appindivsour}). The Gaussian parameters of this PGf model were compared to the Gaussian parameters of the best-fit PG model (with free envelope parameters): the highest values provide upper limits to the disk size and flux in the source. 
For sources for which the PG model reproduces the two the visibility profiles better (L1448-C, L1527, SerpM-SMM4, SerpS-MM22, and GF9-2), the parameters of the Gaussian component at 231 GHz  (best angular resolution) are taken as the candidate disk size and disk flux in the source (or upper limits if the Gaussian component in the PG model is unresolved).
For sources for which only the 231\,GHz visibility profile (probing the smallest spatial scales) is better reproduced by a PG model (SVS13B, IRAS4B, IRAM04191, L1521F, and L1157), we used the properties of the Gaussian component of the PG231 model (see Table \ref{table:popg-fits}) as candidate disk size and flux, or upper limits if the Gaussian component is unresolved.
IRAS2A is the only source in the sample that is better reproduced by the PG model at 94 GHz, but not at 231 GHz. However, the Pl model at 94 GHz is also satisfactory (reduced chi square of 0.9) and the Gaussian component of the PG model is unresolved: hence we consider it likely that an unresolved candidate disk is detected in our 94 GHz data and report the candidate disk as unresolved in Table \ref{table:disks-properties}, with the upper limits on its size and flux stemming from the best-fit PG model at 231 GHz (consistent with the parameters of the best-fit PG model at 94 GHz).
Following this method, we report in Table \ref{table:disks-properties} the detection of candidate protostellar disk-like structures in our CALYPSO data together with their radii and dust continuum fluxes at 231 GHz.

We are limited in the disk sizes we can probe (set by the spatial resolution of our data) and the surface brightness of disks we can detect (set by the sensitivity of our data at the longest baselines). In our PG modeling the Gaussian size and flux are both free parameters: hence if a large but faint disk component were to emerge above our sensitivity threshold, we would be able to model it with this description. Table 5 shows that when the visibility profile can be well modeled by a Pl model, the maximum flux that can be added in a disk-like component is generally low. Although it is not impossible that disks fainter than these upper limits would extend to larger radii than the upper-limit radii we report, it seems rather unlikely because (i) we would then detect their emission at our shorter baselines where our sensitivity is significantly better, and (ii) from a physical point of view, it seems unrealistic that $<1\%$ of the envelope mass could reside in an extremely low-mass thin disk that is rotationally supported up to large radii at the center of an infalling envelope 100-200 times its mass. Thus the maximum parameters of dust disk-like components provided by our simple models can be used robustly as upper limits on the sizes and fluxes of candidate disks.

We note that the two Gaussian parameters (flux and FWHM) are somewhat degenerate when no Gaussian component is clearly detected (sensitivity and resolution wise): for the sources where the Pl model is best, we also performed additional PG models in which we tied the disk flux $F_{\rm{disk}}$ to $10\%$ the envelope flux $F_{\rm{env}}$. This ratio is similar to those observed in the resolved disk candidates in our sample. Minimizing these PGt models to the visibility profiles (the disk size and envelope parameters were let free to vary) allowed us to set an upper limit to the radius of the disk-like component that can be added in each protostar that is well described by an envelope-only model. Because these upper-limit radii stem from a strong hypothesis (obtained when we fixed $F_{\rm{disk}}/F_{\rm{env}}=0.1$), we do not report them in the Table \ref{table:disks-properties}. We do report their parameters (see PGt models) for each source individually in Appendix \ref{section:appindivsour}, however. 
We stress that these PGt models are always worse than either the Pl or PG models that are performed with free parameters only, and the size of the Gaussian component is always found to be smaller than the best-fit PG model (where we did not constrain the ratio $F_{\rm{disk}}/F_{\rm{env}}$). This exercise only points out that the protostellar disks in the sources that are well reproduced by the Pl model are in any case much less massive and much smaller than the resolved disk candidates found in the sources that are better reproduced with the PG model.

Moreover, we also carried out an analysis of the visibility profiles obtained in sectors of the uv-coverage to model only the dust continuum emission in the equatorial plane (at $90^{\circ}$ from the outflow axis reported in Table 1) for all the strong sources in our sample. The parameters we found for this specific modeling (models Pleq and PGeq) are reported in Appendix \ref{section:appindivsour} for each source with $F_{\rm{peak}}>80$ mJy/beam at 231\,GHz: the equatorial plane modeling produces results similar to the modeling using the full uv-coverage.

The choice of a Gaussian model to describe a possible disk contribution is also the simplest model that can be used to describe additional emission of unknown nature at small scales. We carried out tests using disks models with truncated power-law surface density profiles added to our Plummer envelope model for the two sources where extended or strong candidate-disk emission is detected (Serp-SMM4 and L1527), and we found very similar candidate disk properties as had been derived using Gaussian components (size and flux compatible within the error bars).  Since such power-law disk models introduce two more free parameters in an already quite degenerate modeling, they are not well adapted to our PdBI data but may be used to perform 2D modeling in the uv-plane with future observations that will provide better resolution and sensitivity.

%
%
%
%
%

\section{Discussion}

Here, we discuss the occurrence of large Class 0 disks from our analysis of the CALYPSO sample. We show that less than $25\%$ of the CALYPSO Class 0 protostars include a resolved disk-like component with radii $>60$ au. Taking into account all available literature on resolved Class 0 disks (or upper limits obtained with interferometers), we also show that a similar fraction ($<28\%$) of Class 0 protostars studied so far may harbor large disks with radii $>60$ au. 
We also discuss the multiplicity of the CALYPSO protostars and compare to the literature on Class~0, Class I, and Class II multiplicity.
Finally, we argue that only magnetized protostellar collapse models can reproduce the disk size distribution that is found in the CALYPSO sample. 

%
\subsection{Occurrence of large Class~0 disks}

\subsubsection{In the CALYPSO sample}

Our sample of Class 0 protostars was observed with sufficient resolution and sensitivity, at wavelengths that are sensitive to the bulk of dense circumstellar material, so that we probe the pristine properties of the progenitors of protoplanetary disks and multiple systems at the typical scales where they are observed at later stages of evolution.
Based on the visibility analysis, we can detect continuum disk-like structures down to radii $0.15\arcsec-0.2\arcsec$ at 231 GHz depending on the S/N at our longest baselines. 
We find the following (see Table \ref{table:disks-properties}): three low-mass Class 0 protostars have a well-resolved disk-like continuum structure at radii $> 60$ au (L1527, Serp-SMM4, and SerpS-MM22) that is detected at both 94\,GHz and 231\,GHz. 
Two protostars (GF9-2 and L1448-C) are better described at these two frequencies by a model that includes a marginally resolved candidate disk component at scales 30-50 au. 
Six protostars have indications of a disk-like component that is only detected at 231 GHz (IRAS4B, SVS13B, IRAM04191, L1521F, SerpS-MM18, and L1157): of these six sources, only the disk candidate in IRAS4B is resolved by our 231 GHz observations.
IRAS2A is marginally better described with an unresolved disk-like component at 94 GHz, and models with and without a Gaussian component are equally good at 231 GHz. In this case, we consider that an unresolved disk component might be present, although it is not formally detected. We report the maximum size (unresolved) and flux of the disk-like component from the Plummer+Gaussian model at 231 GHz.
Finally, four  protostars (L1448-2A, L1448-NB1, IRAS4A1, and SerpM-S68N) show no indication of a disk-like component at our sensitivity and spatial resolution.

In the case of the L1448-NB1/NB2 system, the individual protostars are not found to harbor individual disk-like structures at radii larger than $\simgt$ 50 au. However, a structure of radius $\sim 200-250$ au is tentatively detected centered either on NB2 or in between the two millimeter sources (see Appendix \ref{section:appl1448n}): this emission could trace the circumbinary structure reported in ALMA observations by \citet{Tobin16b}. 

Furthermore, the case of IRAS4B is quite peculiar: if we do not constrain the total envelope flux to contain at least half the single-dish peak flux, then the PdBI visibilities of this source seem well described by a Gaussian component alone. Considering that this source is also unresolved by single-dish studies (and our modeling suggests a very compact envelope structure, if any), its Class 0 nature may be questioned and it may be more typical of the Class I stage (as, e.g., SVS13A) still embedded in its native cloud (resolved out by our interferometric observations, hence the Gaussian-like profile). It is also possible that the 231 GHz dust emission at PdBI scales is optically thick, but this is not suggested by either (i) the temperature brightness of the 231 GHz PdBI dust continuum emission peak or (ii) the similarly small envelope size found when modeling the optically thin 94 GHz dust continuum profile.
Finally, we stress that the remarkably large size of the disk-like component in Serp-SMM4 (290 au) is quite intriguing and calls for further analysis of the kinematics at these scales to characterize the nature of the dust continuum emission, which does not follow a standard envelope power-law radial distribution.

When we assume that all resolved continuum structures we detected in our sample trace protostellar disks, the fraction of candidate protostellar disks with radii $>$ 60 au is 4 out of 16 (IRAS4B, L1527, SerpS-MM22, and SerpM-SMM4), i.e $\sim 25\%$. 
Moreover, our modeling shows that the candidate disk radii are $< 100$ au in 14 out of 16 Class 0 protostars. 
Since these disk-like continuum structures need to be confirmed kinematically, the analysis of the dust continuum emission in the CALYPSO sample provides an upper limit to the occurrence of large (with radii $\simgt 60$ au) disks of  $\leq 25\%$ during the main accretion phase. Our results show that most Class\,0 protostars have embedded disks ($75\%$ of the sources in our sample are better described when a disk component is included), but  most of these young protostellar disks are small in size and flux.

\subsubsection{Comparison with other works}

Several groups have recently tried to characterize the properties of disks around Class 0 protostars. For example, \citet{SeguraCox16,SeguraCox18} used the VANDAM 8mm continuum VLA survey of 43 Class 0 and Class 0/I protostars in Perseus. They found that only 15 of these Class 0 and 0/I sources are resolved at 15 au, and the candidate disks of only 10 of the VANDAM Class 0 protostars are resolved at radii $>12$ au. All of these 10 disk Class 0 candidates have radii $r <$45 au, and the disk component is not resolved in 67\% of the VANDAM protostars at a 12 au scale. When the VANDAM and CALYPSO samples are combined, only 4 out of 52 Class 0 and 0/I protostars, that is, $\leq 8\%$, appear to have dust continuum circumstellar disks with radii $\simgt$ 60 au.
We did not include these disk sizes in our statistics for Class\,0 protostars : as stated by the authors themselves, they may only represent lower limits because the 8 mm dust continuum emission size could be biased by a population of large dust grains that drift inward. 
We note that our CALYPSO results, which are obtained at shorter wavelengths, are less strongly subject to dust property limitations \footnote{The 1.3mm dust continuum emission remains optically thin at most radii we probe here, and was never shown to be preferentially tracing large grains, although grain growth at scales $<500$ au cannot be excluded.}, but we nevertheless also find that most Class 0 disks have small radii $r <60$ au, with radius values consistent for the sources in common to both samples. 

\begin{figure}[!h]
\centering
\includegraphics[trim={0.8cm 2cm 4cm 2cm},clip,width=0.99\linewidth]{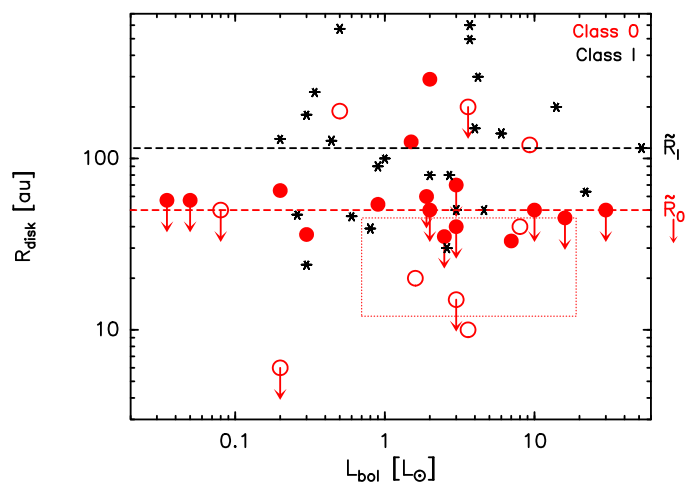}
\caption{Protostellar disk radii $R_{\rm{disk}}$ as a function of the protostellar bolometric luminosity $\lbol$ for all Class 0 (red circles) and Class I (black stars) protostars observed between 0.7\,mm and 2.7\,mm with a spatial resolution better than 50 au from the CALYPSO sample and the literature. The Class 0 disks shown here are from the CALYPSO sample (filled circles) and those characterized in the literature (contour circles, see text for names and references). When only upper limits on  disk sizes could be estimated, the upper-limit values are indicated with arrows pointing down. The dotted red box shows the location of the ten Class 0 disk radii from the 8mm continuum observations of the VANDAM sample \citep{SeguraCox18}. 
The individual Class I disks shown are those described in the text of Sect. \ref{sec:diskYSOs}.
The dashed lines show the median disk radii: in red from the sample of 25 Class 0 disks with  $\tilde{R_0}<$50 au (including upper limit radii), and in black the median radius from the sample of 25 Class I disks with $\tilde{R_I}=115$ au.
}
\label{fig:Rdisk-Lbol}
\end{figure}

Other recent results for the quest of rotationally supported circumstellar disks in Class 0 protostars include L1527 (Ohashi et al. 2014, ALMA), which has a centrifugal radius of $\sim60$ au, and VLA 1623 \citep{Murillo13}, which was suggested to show Keplerian disk motions up to $r \sim 189$ au. The latter requires confirmation, however, because the model that was used is complex. In the HH212 protostar, \citet{Lee-F18} found a candidate Keplerian disk at radius $\sim40$ au that is embedded in a dusty disk-like structure at radius $\sim60$ AU. 
Recent SMA observations of the Class 0 protostar BHR7 suggest a compact dust continuum component at $r \sim120$ au, although the kinematics obtained within the same dataset do not confirm Keplerian rotation at similar scales \citep{Tobin18}. 
\citet{Gerin17} analyzed ALMA 0.8mm dust continuum observations of two Class~0 protostars, B1b-N and B1b-S. They found that compact disk-like components could only contribute to the intensity profiles at scales $\simlt$ 50 au. The 0.87mm ALMA observations by \citet{Tokuda17} enabled the detection of a small 10 au radius disk in L1521F.
For the B335 protostar, observations of the kinematics in the inner envelope with ALMA provided an upper limit to the disk radius of B335 $<15$ au \citep{Yen15b}.
The two Class\,0 protostars IRAS16253 and IRAS15398 have been observed with ALMA by \citet{Yen17}, who found a candidate disk radius of 20 au in IRAS15398 and an upper-limit disk radius of 6 au in IRAS16253. 
The Class 0 L1455 IRS1 was observed with the SMA, and the size of its protostellar disk was constrained to $<200$ AU \citep{Chou16}. The HH211 Class\,0 protostar was observed with ALMA, and its disk radius was constrained to $\simlt 10$ au \citep{Lee-F18b}.
Finally, in an SMA survey of envelope kinematics toward a sample of 14 Class 0 protostars, \citet{Yen15a} found that three sources (L1448-NB, Per-emb 9, and IRAS 03292+3039) harbored kinematics that were consistent with rotational motions dominating infalling motions in their inner envelopes at scales $\sim100-200$ au. Two Class 0 protostars of their sample, L1527 and HH212, show rotationally dominated motions close to Keplerian at radii 50-80 au: rotationally supported disks with radii $\sim40-60$ au have indeed be confirmed by other groups in these two sources. The remaining 8 Class 0 protostars of the \citet{Yen15a} sample do not show any indication of rotationally dominated motions at scales sampled by the SMA observations ($1.5\arcsec$ to $5\arcsec$ synthesized beams, and heterogeneous source distances).
Our CALYPSO analysis triples the number of Class 0 protostars that are analyzed at high angular resolution to seek for disk-like structures at wavelengths that probe the bulk of circumstellar material ($<3$mm). When we combine the CALYPSO sample with the recent ALMA results that probe radii $\simlt 60$ au, only 5-7 out of 26 Class 0 protostars, that is, $< 28\%$, show either confirmed or candidate disks at radii $>60$ au. The disk radii as a function of the protostellar bolometric luminosity for all Class\,0 disks that are characterized with millimeter-wavelength observations that probe scales $\simgt 60$ au from CALYPSO and the literature are shown as red symbols in Figure \ref{fig:Rdisk-Lbol}.

The main caveat in characterizing candidate disk structures from continuum visibilities stems from the uncertain estimate of the large-scale envelope contribution. 
We stress that we left the envelope parameters free to vary in our modeling to account for the variety of envelopes that is observed in Class 0 protostars. We also emphasize that all of our best-fit models suggest envelope parameters that are physically reasonable. 
If the large-scale envelope deviates from the spherical symmetry assumed here, the estimated properties of the candidate disks may change slightly. However, the spherical envelope model has the advantage of simplicity and accurately reproduces the observed visibilities at short baselines. Our additional modeling, which used only the equatorial dust continuum profiles, suggests that the possible asymmetry of the envelope structure does not significantly affect our results.

We also acknowledge that the full extent of rotationally supported disks may be larger than the extent seen in millimeter dust continuum emission. We stress, however, that this is not the case in L1527, for example, where our estimate of the disk radius from dust continuum emission ($56\pm10$ au) agrees with the radius that was determined from the molecular gas (centrifugal radius $\sim$ 54-74 au, \citealt{Ohashi14, Aso17}) based on an analysis of the CO kinematics at an angular resolution similar to our CALYPSO data. We note, however, that Keplerian motions were tentatively detected up to $r_{c}\sim 100$ au in the SO emission of L1527 (CALYPSO data, Maret et al. in prep.).

Our study suggests that many Class 0 disks still remain uncharacterized because they lie at small scales that are not yet probed by current surveys in the (sub-)millimeter regime, where the structure of Class 0 objects is best characterized.
In the future, it will be of paramount importance to obtain accurate disk size and disk mass distributions at scales that are yet unexplored by most observational studies to make progress in our understanding of the formation and early evolution of stars and protoplanetary disks. Such studies will allow us to probe the disk kinematics and ultimately constrain the central protostellar mass and track its growth against the evolution of the envelope$+$disk system. 

\subsubsection{Comparison to disk properties at later YSO stages}
\label{sec:diskYSOs}

Early millimeter interferometric observations of T-Tauri stars suggested that they are surrounded by rather large Keplerian protoplanetary disks with radii between 20 and 450 au (mean size of 165 au; \citealt{Isella09,Andrews10,Guilloteau11}) and disk masses in the range 0.004--0.055 $M_{\odot}$ around 0.2--2.5 $M_{\odot}$ stars \citep{Andrews10}. 
Surveys of the dust continuum emission from protoplanetary disks also suggest a trend that more massive T-Tauri stars tend to host more massive disks \citep{Andrews13,Ribas17}.
While the distribution of disk radii is not well characterized for Class I protostars, large ($\ge 100$ au) disks are generally found toward more evolved sources where $M_{\star}/M_{\rm tot} > 0.65$ (where $M_{\rm tot} = M_{\star} + M_{\rm env} + M_{\rm disk}$) \citep{Barenfeld17}. Individual studies of Class\,I disks that are characterized by either dust or molecular lines observations between 0.7\,mm and 2.7\,mm with a spatial resolution better than 50 au include SVS13A, R-CrA, L1551NE, L1551-IRS1, L1455-IRS1, Lupus3-MMS, L1489, IRS43, IRS63, TMR1, TMC1, TMC1A, and L1536 \citep{Lindberg14, Yen17, Takakuwa14, Aso15, Harsono14}, Elias29, WL12 \citep{Miotello14}, and the ten Taurus Class\,I from \citet{Sheehan17}. They are shown in Figure \ref{fig:Rdisk-Lbol} as black stars. We stress that some radii are estimated based on the kinematics of molecular lines (centrifugal radius), while some are determined from continuum-size measurements. Although there is no trend of systematic radius differences between the samples characterized by the two methods, we refer to the individual papers quoted for each source for more information on individual disk properties. The radii of these 25 Class\,I disks suggest that the median radius of Class\,I disks (115 au) is slightly larger than the median radius of Class\,0 disks ($<$50 au).

Recent observations of the dust continuum emission from large samples of T-Tauri stars with new interferometric facilities (PdBI, ALMA) suggest that a population of faint and small (radii $<60$ au) disks might be the bulk of circumstellar disks \citep{Pascucci16,Tripathi17}. This population of small dusty disks may have remained hidden so far because of earlier detectability limits that skewed the disk size distribution of Class II YSOs toward the largest disks. 
This is the case for example in the study by \citet{Tripathi17}, who measured the sizes of 50 Class II disks (Taurus and Ophiuchus star-forming regions): they found a size distribution ranging from 19 to 182 au, with a median of 48 au. Moreover, \citet{Tazzari17} observed 22 Class II disks in Lupus, measuring effective radii ranging from 18 to 129 au, with a median of 55 au.
Finally, in Chamaeleon I, \citet{Pascucci16} showed that they were able to resolve only half of the disks in their sample (ALMA observations at 100 au resolution), with an indication that smaller dust disks are found around lower-mass stars. 

The disk size distribution around Class 0 protostars remains to be sampled properly, but our results for Class 0 protostars suggest that most disks might start very small ($<60$ au, see Figure \ref{fig:Rdisk-Lbol}). Although the viscous evolution of disks over a time span of a few Myr is highly unconstrained, if the pristine disk size distribution is indeed dominated by a large population of small disks, it would provide a natural explanation for the revised estimates of disk size distribution in more evolved YSOs.
However, the dusty extent of T-Tauri disks might also reflect the evolution of the dust during the pre-main-sequence phase (dust drift and coagulation), while their gas extent might have been affected by viscous evolution with time \citep{Ansdell18}.
We examine in the following possible scenarii to explain the lack of large Class~0 disks that is suggested by our CALYPSO observations.

\subsection{Implications for theoretical models of protostellar disk formation}

The collapse of rotating dense cores naturally produces rotationally supported disks as a result of angular momentum conservation \citep{Terebey84,Shu87}. For low-mass protostars, disks are expected to grow quickly in mass and size with time, reaching rather large size ($r \sim 100$ au, \citealt{Cassen81,Basu97,Matsumoto03,Bate18}) on timescales  of $<10^4$ years (the centrifugal radius grows with $t^3$ in the Shu model). 

Star-forming clouds are permeated by large-scale magnetic fields, as recently pointed out by the analysis of the polarized dust emission obtained with the {\it{Planck}} spatial observatory \citep{Planck16}. 
Moreover, multiple results in the literature \citep{Matthews09,Girart09,Hull14} reported the detection of magnetic fields in protostellar cores. A recent SMA survey of low-mass protostars \citep{Galametz18} shows that magnetic fields are detected in all low-mass protostellar cores: early non-detections of the magnetization of cores probably were due to sensitivity limitations and the intrinsic difficulty of properly observing polarized dust emission. 

Even modest values of the magnetic field strength significantly modify the collapse models because the radial and toroidal components of the magnetic field are amplified by the differential motions within the collapsing core. This is especially true in models with low turbulent energy, which are representative of low-mass cores: in the following, we discuss relevant magnetized models that include rotation (rotational over gravitational energy $\beta \sim 1\%-5\%$) and low levels of turbulence (subsonic).
In moderately magnetized models with a ratio of mass-to-flux over critical mass-to-flux ratio, $\mu= (M/\Phi) / (M/\Phi)_c < 10$ (with $(M / \Phi)_c = c_1 / 3 \pi \times (5/G)^{1/2}$, see \citealt{Mouschovias76}), the collapse primarily occurs along the field lines, thus reducing the angular momentum that is transported into the inner envelope, whereas at the same time, strong magnetic braking occurs \citep{Galli06,Krasnopolsky10}. 
As a consequence, no centrifugally supported disks forms in these ideal MHD numerical models \citep{Hennebelle08a,Mellon08}: only when the magnetic braking drops efficiency because the surrounding envelope dissipates do these models predict the rapid growth of disks at up to a few hundred au near the end of the main accretion phase \citep{Machida11}.
To mitigate this effect, magnetized numerical simulations of core collapse were carried out using different initial conditions in an attempt to weaken the magnetic braking effect, notably by introducing an angle between the core rotation axis and the magnetic field direction at core scale. These models showed that for a low mass-to-flux ratio ($\mu<3$), magnetic braking still completely inhibits the formation of rotationally supported disks, while for weakly magnetized cases $\mu>5,$ the formation of small Keplerian disks would become possible if the difference in angle <$\vec{j},\vec{B}$> is large enough (typically $>45^{\circ}$) (see \citet{Joos12,Krumholz13,Ciardi10}. 
Under some specific conditions, models that include turbulence also form large disks because turbulence can induce some misalignment that reduces the magnetic braking efficiency \citep{SantosLima13,Seifried15}.

Recently, non-ideal MHD models were carried out. They included the effects of different mechanisms that allowed the field to diffuse under some conditions (for a review, see \citealt{LiPPVI}). 
Generally, the inclusion of diffusive physics in numerical models weakens the magnetic braking in the inner envelope and might allow the formation of rotationally supported disks, although these are systematically much smaller than disks that formed in purely hydrodynamical models. The exact disk properties from non-ideal MHD simulations of protostellar collapse are still heavily debated, however, see for example the diverging results regarding ambipolar diffusion: \citet{Mellon09}, \citet{Li11}, and \citet{Dapp12} concluded that AD fails to enable the formation of rotationally supported disks, while \citet{Zhao16} proposed that the depletion of small grains could increase the ambipolar diffusivity and hence allow the formation of such disks. 
A recent analytical model by \citet{Hennebelle16}, which has been compared with a series of numerical simulations, argued that AD in combination with magnetic braking during magnetized collapse leads to the formation of small disks, whose sizes ($\sim20$ au) are remarkably constant and only weakly depend on the initial rotational energy and magnetization of the cores (if $\beta$ and $\mu$ are not too large). This study thus suggests that magnetic self-regulation during the collapse may produce a rather narrow disk size distribution during the protostellar phase.   

At the current state of our knowledge, our PdBI results that suggest the paucity of large $r \simgt 60$ au disks during the Class\,0 phase may either point to 
\\(i) the extreme youth of these objects (which may not yet have developed a disk), 
\\(ii) an overestimate of the initial angular momentum of the cores in hydrodynamical collapse models, or
\\(iii) a magnetically regulated disk formation scenario.
\\The first possibility (i) is mostly ruled out by the presence of the large-scale outflows that are observed to be powered by every protostar in our sample (Podio et al. in prep). This testifies that accretion onto the central protostar has been proceeding for at least a few thousand years, similar to the typical timescale at which hydrodynamical models of the collapse of rotating dense cores predict large ($r \sim 100$ au) rotationally supported disks as a consequence of angular momentum conservation \citep{Goodwin04}. 
\\The possibility that rotation is overestimated in hydrodynamical models of protostellar collapse (ii) can be addressed since rotation (expressed as the ratio of rotational over gravitational energy $\beta$) has been observed to be $\beta\sim2\%-10\%$ in prestellar cores \citep{Goodman93,Caselli02a,Belloche13}.  
The typical angular momentum values observed in protostellar cores are on the order a few $10^{-3}$ km.s$^{-1}$.pc \citep{Belloche13,Yen15a}. A possible systematic over-estimate of the angular momentum has been reported from the analysis of numerical simulations of the collapse of protostellar cores \citep{Dib10}, although this was not confirmed by a recent numerical study \citep{Zhang18}.
Moreover, the rotation that is observed is usually interpreted as the value for a uniform core $\beta_{obs}=\frac{1}{3}\times\frac{R^3\Omega^2}{GM}\times \rm{sin(i),}$ with i the inclination angle, R and M the radius and mass of the core. Protostellar dense cores are centrally concentrated, however, which increases the gravitational energy and decreases the rotational energy, so that $\beta$ could actually be smaller (a factor of $\sim3$, see \citealt{Belloche02,Belloche13}). 
Hydrodynamical collapse, even when the cores have only a small fraction of the observed $\beta$, quickly develops much larger disks (e.g., Goodwin et al. 2004) than are observed in our sample because of angular momentum conservation if no mechanism is found to dissipate or redistribute the envelope angular momentum. 
For example, for a centrally condensed ($\rho\propto r^{-2}$) protostellar dense core of diameter 0.1 pc that is rotating uniformly with $\beta=0.02$, gravitational collapse would lead to the formation of a disk at $r_c \propto \beta \times R_{core}=200$ au when half the envelope mass has collapsed into a protostar-disk system (i.e., the end of the main-accretion phase). 
To produce a disk size that is consistent with the bulk of our observations, we would have to remove $>75\%$ of the angular momentum from the infalling material at scales $r>r_c$ in order to shrink the radius of the disk to $r_c \simlt 50$ au. 
Although it is unclear whether we have robust quantitative constraints on the angular momentum enclosed in star-forming cores, an overestimate of $\beta$ from observations by one order of magnitude, as would be requested to explain our observations, seems highly unlikely.
\\Magnetic torques acting in the envelope are, as far as we are aware, the only mechanism that is able to greatly reduce the transport of angular momentum during protostellar collapse \citep{LiPPVI}, as was shown by recent results that pointed toward a magnetically regulated disk formation scenario in the B335 protostar \citep{Maury18}.
Finally, if the magnetic field is a key player during the protostellar collapse that leads to the formation of solar-type stars, as suggested by our observations, we stress that identifying an embedded protostellar disk requires spectral line observations that show Keplerian rotation (such as L1527) before they are considered robust rotationally supported disks. Magnetized models of protostellar formation generically produce flattened inner envelope structures because the collapse is favored along the main direction of the magnetic field at core scale. Such disk-like structures (pseudo-disks, \citealt{Galli06}) are not rotationally supported but may be confused with disks in the absence of kinematical information.

\subsection{Multiplicity properties of Class 0 protostars}
\label{section:naturecont}

\subsubsection{In the CALYPSO sample}

The tentative nature of the multiple dust continuum components we detected in our maps is reported in Table~\ref{table:continuum-sources}. 
Only the sources that are detected at both frequencies (with separations $a>$ of the synthesized beam scale) are when possible claimed to be robust protostellar candidates (secondary protostar in the sixth column of Table \ref{table:continuum-sources}).
The PdBI maps have a synthesized beam at 231 GHz that can separate systems in the maps with separations larger than $\sim 60$ au in Taurus (three sources) and $\sim 90$ au in Perseus and Serpens South (nine sources). For L1157 and GF9-2, although the distance is subject to more uncertainties, we are able to probe systems down to separations $\sim 80-100$ au. We exclude the Serpens Main sources from the multiplicity analysis because the larger distance precludes us from probing systems closer than 160 au. The sample of 14 protostars we used for the multiplicity analysis allows us to detect companions down to separations 100 au.
The S/N of the 231 GHz dust continuum emission PdBI maps allows us to detect companions of fluxes that are comparable to the flux of the primary protostar in the low-luminosity Taurus sources, or 20 times lower than the flux of the primary protostar for the Perseus and Serpens South sources (except for IRAS2A, for which our high-resolution data are taken from the pilot study of \citealt{Maury10}, see Appendix \ref{section:appiras2a}). We are sensitive to multiples in an area around the targeted primary protostar of radius 1500 au in Taurus and 2800 au in Perseus and Serpens South. 
Of the 14 Class 0 protostars at distances $<260$ pc in our sample, 8 are single in our dust continuum maps at envelope scales (IRAM04191, L1521F, L1527, L1157, GF9-2, SerpS-MM22, SVS13B, and IRAS2A). However, in the case of IRAS2A, our sensitivity at the longest baselines is much lower than for the rest of the sample (see Appendix \ref{section:appiras2a}), and we do not detect the secondary continuum source  found with VLA \citet{Tobin15b} and ALMA observations (Maury et al. in prep.) at a separation 140 au. This means that 7 of the 14 closest CALYPSO protostars are single protostars at scales $100-2000$ au, within the nominal flux ratio limits given above. The secondary continuum sources detected in the remaining sources (L1448-2A, L1448-N, L1448-C, IRAS4A, IRAS4B, and SerpS-MM18) may be candidate protostellar companions, although in some cases, their nature has not been reliably assessed so far (see Appendix \ref{section:appindivsour} for details on individual sources). 
Assuming that all continuum sources detected around the primary sources are genuine protostellar companions, we find that about half of the CALYPSO protostars within 260\,pc are located in multiple systems, which leads to a multiplicity fraction $MF^{1mm}_{\rm Class0}=57\%$ (with the multiplicity fraction defined as $MF = \frac{B+T+Q+...}{S+B+T+Q+...}$). 
Because of the limited sample, excluding even one source would affect the overall multiplicity fraction by $\simgt 5\%,$  therefore we did not build distributions of multiplicity fractions with separations.

The multiplicity properties of SVS13B and IRAS4B can be discussed since there are no secondary sources detected within their individual envelopes, but they are associated with widely separated sources that are usually considered as individual protostars in a clustered environment ("separate envelope system" following the classification proposed by \citealt{Looney00}). It is unclear whether these sources, which are embedded in the same parsec-scale filamentary structure, will end up in a bound system similar to "ultra-wide pairs" \citep{Joncour17} since the future evolution of their proper motions cannot be extrapolated with the current data. We therefore consider that SVS13B and IRAS4B are single at envelope scales. 
When only the common-envelope systems are considered (excluding SVS13B and IRAS4B from the list of multiples), the Class\,0 multiplicity fraction in the CALYPSO sample drops to $MF^{1mm}_{\rm Class0}=43\%$.

Regarding the "close" multiple systems, we count three multiple systems at separations $a<210$ au: L1448-2Ab, whose nature remains to be confirmed, the secondary source in IRAS2A that we did not detect but that is seen with both VLA \citep{Tobin15b} and ALMA (Maury et al. in prep.), and L1448-NB2, which may itself be a binary according to \citet{Tobin16b}. Hence, in our sample we find that only $21\%$ of the protostars within 250 pc have candidate companions at scales $100<a<210$ au (detection limited to companions of similar flux for the Taurus sources). This paucity of multiple systems at $\sim$ 100 au scales has also been found by \citet{Tobin16a}. We suggest that it may be explained by the physical conditions in Class 0 envelopes at those scales, which limits the formation of massive $\simgt 100$ au disks and therefore the fragmentation of gravitationally unstable disks to form binaries at these $\sim 100$ au scales. The magnetized scenario, proposed in the previous section to explain the paucity of large disks, may explain the paucity of fragments at similar scales as well.

\subsubsection{Comparison with other works}

Large interferometric surveys of the molecular line emission in complete populations should be carried out to lift the current uncertainties on the multiplicity fraction of protostars, but unfortunately, no complete analysis that would allow unambiguously characterizing the nature of the multiple sources when they are detected is available in a large sample of protostars. 
When all continuum sources within 0.04 pc in Perseus are counted, as is done in the \citet{Tobin16a} VANDAM study, many separate-envelope systems are included: whether these will indeed end up as bound systems is debatable, especially because the multiplicity of Class\,I protostars at similar separations is found to be much lower in their sample (MF$=0.23$). We therefore focus on the candidate Class 0 multiple systems in the VANDAM sample that are separated by $50<a<5000$ au: the multiplicity fraction in VANDAM sources in this separation range is $MF^{8mm}_{\rm Class0}=45\%$. 
This fraction is slightly lower than but consistent with the $MF^{1mm}_{\rm Class0}=43-57\%$ that we obtain in the CALYPSO sample at separations 100-5000 au.

\subsubsection{Comparison to the multiplicity properties at later stages}

A large fraction of stars on the main sequence are observed within multiple systems: the frequency of multiple systems is $MF_{0.7-1.3\,M_\odot}^{\rm MS}=44\%$ for solar-type stars \citep[see][for a review]{DucheneReview13}. 
T-Tauri stars and Class II objects have a higher multiplicity fraction than their main-sequence descendants \citep{Kraus11}. 
For the direct progenitors of T-Tauri stars, Class\,I YSOs, the multiplicity properties are less clear since these systems are still partly embedded and therefore should be observed in the near- to mid-infrared where spatial resolution is usually lower, and more confusion arises from various types of circumstellar emission. 
Studies of the multiplicity of Class I protostars in the infrared suggest a lower companion frequency (from $\sim47\% \pm8\%$ at separations 14-1400 au to $\sim26\%$ at separations 110-1400 au and $\sim16\%$ at separations 300-1400 au, \citealt{Duchene04,Duchene07}) than in Class II YSOs. This might be due to the narrower range of separations where these studies are complete (between 50 and 200 au). 
The infrared study by \citet{Connelley09} found an MF$^{IR}_{\rm ClassI}=44\%$ at separations 50-25000 au, and a distribution of Class\,I system separations consistent with a flat distribution.
The fact that multiple systems are observed in Class\,I YSOs suggests that fragmentation processes have taken place, which means that they would have occurred during the earliest stages of star formation (prestellar cores and Class~0 protostars).

We compared our results in the CALYPSO Class~0 protostars with infrared studies of the multiplicity of Class\,I protostars.
Our Class\,0 sample is well matched to the scales probed in the "restricted sample" of \citet{Connelley09}, which consists of 32 Class\,I YSOs at $d<500$ pc. The authors discarded binary companions with a projected  separation smaller than 50 au for completeness arguments. 
Sixteen of the 32 Class\,I protostars they observed have a companion, either wide or close, at separations 50-5000 au (i.e., $50\%$ of the sample). 
Our multiplicity statistics in the CALYPSO sample (7 single and 8 multiples) at the Class\,0 stage, although on an admittedly small sample, is not significantly different from the restricted sample of Connelley et al. (2009).
It has been suggested \citep{Reipurth00,Sadavoy17} that a large portion of Class\,0 protostars form in non-hierarchical  multiple systems that dynamically decay during the Class~0  phase, in which case the overall multiplicity fraction of Class\,0 protostars should be higher than that of Class I protostars. 
Although our conclusions should be viewed as preliminary at this stage because of the small sample and inhomogeneous distances of our sources, our results do not seem to be consistent with this scenario, at least at the scales common in the CALYPSO sample and the Connelley sample since the overall multiplicity is found to be very similar in both samples.

\section{Conclusions}

In the framework of the CALYPSO survey, we have observed the dust continuum emission in a sample of 16 Class 0 protostars with synthesized beams $\sim0.4\arcsec$ using the Plateau de Bure interferometer at IRAM (50 to 150 au at the distances of our sources).
We performed an analysis of the dust continuum visibility data using models of Plummer envelopes and explored the range of parameters for envelope properties. Six of these sixteen protostars cannot be satisfactorily described by a single circularly symmetric envelope model over the whole range of spatial scales (40-1000 au) sampled by our observations at either 1.3\,mm or 3\,mm. We thus modeled the visibility profiles including an additional Gaussian component for all sources in our sample in order to test whether adding a disk-like component would explain the properties of some of the CALYPSO protostars at long baselines. 
The main results of our study are listed below.
   \begin{enumerate}
      \item For 11 of the 16 protostellar profiles we analyzed, the inclusion of disk-like components improves the description of the 231\,GHz continuum visibility profiles at the longest baselines we probed.
      \item However, only four protostars in our sample require a candidate disk structure resolved with radii $>60$ au to reproduce their dust continuum visibility profiles (IRAS4B, L1527, Serpens South MM22 and Serpens Main SMM4). Two other protostars (L1448-C, and GF9-2) are better reproduced when an additional disk-like component is included; they are marginally resolved by our data at radii $\simlt 60$ au. In one close multiple system (L1448-NB1/2),  an additional circumbinary structure with radius $\sim 200$ au may be present, while the two individual
protostars do not show evidence of individual disks resolved by our observations.
      \item Four out of the 16 protostars in the CALYPSO sample are likely to harbor large, well-resolved individual protostellar disks at radii $> 60$ au. Our observations suggest that while most Class 0 protostars show evidence of embedded disks, most ($>75\%$) Class 0 continuum disks are small. When we combined all the recent 0.8--1mm interferometric observations of Class\,0 protostars that probed the bulk of the envelope emission down to radii $\sim 50$ au, only 5 to 7 out of 26 Class 0 protostars, that is $< 28\%$, show either confirmed or candidate large disks at radii $\simgt$ 60 au. This upper limit on the occurrence of large disks at the Class 0 stage confirms our earlier results \citep{Maury10} and suggests that Class 0 disks are small on average.
      \item From a theoretical point of view, if Class 0 protostars contain similar rotational energy as currently estimated in prestellar cores, only magnetized models of protostellar collapse can reproduce such a large population of small disks during the main accretion phase.
      \item The multiplicity fraction in the CALYPSO sample is found to be $\sim 43-57\% \pm 5\%$ at scales 100--5000 au, which is slightly larger than but in general agreement with previous studies of Class 0 multiplicity. 
      \\

Our results suggest that the formation of disks and multiple systems during the Class~0 phase could occur at smaller scales than predicted by hydrodynamical models of rotating protostellar collapse. However, we stress that confirming the properties of the embedded protostellar structures requires additional spectral line analysis that either traces rotationally supported motions to robustly identify disk components, or confirms the protostellar nature of the systems at $\sim 50$ au scales.

   \end{enumerate}

\begin{acknowledgements}
      We thank the IRAM staff for their support in carrying out the CALYPSO observations, and the anonymous referee for a detailed reading that significantly helped us to improve the paper. We thank Kees Dullemond, Ralf Klessen, Sandrine Bottinelli, Benoit Commercon, and Ralf Launhardt for their participation to the preparation of the CALYPSO project. This work has benefited from the support of the European Research Council under the European Union’s Seventh Framework Programme (Advanced Grant ORISTARS with grant agreement no. 291294 and Starting Grant MagneticYSOs with grant agreement no. 679937). 
\end{acknowledgements}


\bibliographystyle{aa}
\bibliography{Bibliographie}

\clearpage
\onecolumn

\begin{center}
\tablefirsthead{%
\hline
\hfill & \hfill & \hfill & \hfill & \hfill & \hfill & \hfill & \hfill  & \hfill & \hfill \\  
\hfill & \hfill & {Plummer} & {Plummer} & {Plummer} & {Inner} & {Gaussian} & {Gaussian} & {Reduced} & {PG better} \\ 
{Source}        & {Model} & {radius}    & {index}       & {flux} & {radius}         & {FWHM}        & {flux}        & {$\chi^2$} & {than Pl ?} \\
\hfill & \hfill & {($R_{\rm{out}}$)} & {(p+q)} & \hfill & {($R\inside$)} & \hfill & \hfill & {}  & {}  \\  
\hfill                  & {}    & {($\arcsec$)}                 & \hfill                         & {(mJy)}  & {($\arcsec$)}                      & {($\arcsec$)}           &       {(mJy)}                 &  \hfill & \hfill  \\ 
{\tiny{[1]}} & {\tiny{[2]}} & {\tiny{[3]}} & {\tiny{[4]}} & {\tiny{[5]}} & {\tiny{[6]}} & {\tiny{[7]}} & {\tiny{[8]}} & {\tiny{[9]}} & {\tiny{[10]}} \\ 

\hline}
\tablehead{%
\hline
\hfill & \hfill & \hfill & \hfill & \hfill & \hfill & \hfill & \hfill  & \hfill & \hfill \\  
\hfill & \hfill & {Plummer} & {Plummer} & {Plummer} & {Inner} & {Gaussian} & {Gaussian} & {Reduced} & {Better} \\ 
{Source}        & {Freq.} & {radius}    & {index}       & {flux} & {radius}         & {FWHM}        & {flux}        & {$\chi^2$} & {than Pl ?} \\
\hfill & \hfill & {($R_{\rm{out}}$)} & {(p+q)} & \hfill & {($R\inside$)} & \hfill & \hfill & {}  & {}  \\  
\hfill                  & {(GHz)}       & {($\arcsec$)}                 & \hfill                  & {(mJy)}  & {($\arcsec$)}                      & {($\arcsec$)}           &       {(mJy)} &  \hfill & \hfill  \\ 
        {[1]} & {[2]} & {[3]} & {[4]} & {[5]} & {[6]} & {[7]} & {[8]} & {[9]} & {[10]} \\ 
\hline}
\tabletail{%
\hline
\multicolumn{10}{|r|}{\small\sl continues on next page}\\
\hline}
\tablelasttail{\hline}
\tablecaption{Parameters of the best-fit Plummer envelope models (Pl models) and Plummer+Gaussian models (PG models) that best reproduce the observed dust continuum emission radial profiles.}
\begin{supertabular}{|c|c|cccc|cc|c|c|}
\label{table:popg-fits}
\hfill & \hfill & \hfill & \hfill & \hfill & \hfill & \hfill & \hfill  & \hfill & \hfill  \\  
{ L1448-2A}     & Pl231 & 18 $\pm$ 4            & 2.3 $\pm$ 0.1 & 527 $\pm$ 30      & $0.01\pm0.01$ &  --                           & --                            &       1.0              & -- \\
{ }                     & Pl94          & 25 $\pm$ 5            & 2.2 $\pm$ 0.2     & 58 $\pm$ 5    & $0.01\pm0.01$ &  --                           & --                              &       1.0             & -- \\
{}                      & PG231 & 20 $\pm$ 10           & 2.2 $\pm$ 0.1 & 526$\pm$ 100    & $0.09\pm0.01$ & $<$ 0.15              & 12 $\pm$ 3            &       1.0              & no \\
{}                      & PG94          & 22 $\pm$ 2            & 2.0 $\pm$ 0.1     & 60 $\pm$ 30   & $0.01\pm0.01$ &  $<$ 0.3                      & 2.2 $\pm$ 0.5   &       0.9             & no \\
\hfill & \hfill & \hfill & \hfill & \hfill & \hfill & \hfill & \hfill & \hfill & \hfill   \\  
{ L1448-NB1}    & Pl231 & 25 $\pm$ 5            & 2.2 $\pm$ 0.2 & 3373 $\pm$ 200 & $0.01\pm0.01$     &  --                           & --                            &       3.0                 & --     \\
{ }                     & Pl94          & 19 $\pm$ 3            & 2.7 $\pm$ 0.3     & 226 $\pm$ 110 & $0.2\pm0.1$ &  --                             & --                              &       7.0             & -- \\
{ }                     & PG231 & 23 $\pm$ 3            & 2.2 $\pm$ 0.2 & 3176 $\pm$ 380 & $0.06\pm0.02$  &  $<$ 0.15             & 38 $\pm$ 11           &       3.2                 & no     \\
{ }                     & PG94          & 23 $\pm$ 3            & 2.7 $\pm$ 0.2     & 243 $\pm$ 18  & $0.94\pm0.1$ &  0.94 $\pm$ 0.1        & 39 $\pm$ 4               &       5.0             & no \\
\hfill & \hfill & \hfill & \hfill & \hfill & \hfill & \hfill & \hfill & \hfill & \hfill   \\  
{ L1448-C}      & Pl231 & 12 $\pm$ 3            & 2.5 $\pm$ 0.2 & 660 $\pm$ 40      & $0.01\pm0.01$ &  --                           & --                            &       6.8                 & --     \\
{ }                     & Pl94          & 17 $\pm$ 7            & 2.7 $\pm$ 0.2     & 53 $\pm$ 10   & $0.01\pm0.01$ &  --                           & --                              &       9               & -- \\
{ }                     & PG231 & 14 $\pm$ 4            & 1.7 $\pm$ 0.2 & 860 $\pm$ 70    & $0.14\pm0.05$ &  0.16 $\pm$ 0.05      & 130 $\pm$ 5           &       0.7                 & yes \\
{ }                     & PG94          & 14 $\pm$ 1            & 1.4 $\pm$ 0.2     & 79 $\pm$ 2    & $0.05\pm0.03$ &  $<$ 0.3                      & 18 $\pm$ 1              &       0.1             &  yes \\
\hfill & \hfill & \hfill & \hfill & \hfill & \hfill & \hfill & \hfill & \hfill & \hfill   \\  
{ IRAS2A1}      & Pl231 & 5.7$\pm$ 1            & 2.5 $\pm$ 0.2 & 600 $\pm$ 50      & $0.01\pm0.01$ & --                            & --                            &       1.3                 & --     \\
{ }                     & Pl94          & 10 $\pm$ 5            & 2.6 $\pm$ 0.2     & 65 $\pm$ 2    & $0.01\pm0.01$ &  --                           & --                              &       0.9             &  -- \\
{ }     & PG231 & 7.7 $\pm$ 1                           & 2.5 $\pm$ 0.2 & 600 $\pm$ 40    & $0.05\pm0.02$ & $<$ 0.15              & 52 $\pm$ 5    &       1.3                 & no     \\
{ }                     & PG94          & 10 $\pm$ 5            & 2.4 $\pm$ 0.2     & 56 $\pm$ 2    &       $0.06\pm0.02$ &  $<$ 0.3                        & 9 $\pm$ 1               &       0.5             &  yes \\
\hfill & \hfill & \hfill & \hfill & \hfill & \hfill & \hfill & \hfill & \hfill & \hfill  \\ 
 { SVS13B}      & Pl231 & 14 $\pm$ 7            & 2.9  $\pm$ 0.2        & 446 $\pm$ 15 & $0.06\pm0.02$    & --                            & --                            &         2.5             & --    \\
 { }                    & Pl94          & 6.7 $\pm$ 2           & 2.6 $\pm$ 0.2     & 52 $\pm$ 4    &       $0.01\pm0.01$ & --                              &  --                             &       0.45            & --     \\
 { }                    & PG231 & 9 $\pm$ 3             & 2.5  $\pm$ 0.3        & 636 $\pm$ 88    & $0.2\pm0.1$ & 0.19 $\pm$ 0.1  & 80 $\pm$ 7            &         4.6             & yes\tablefootmark{$\star$}    \\
 { }                    & PG94          & 10 $\pm$ 3            & 2.4 $\pm$ 0.2     & 58 $\pm$ 4    &       $0.1\pm0.05$ & $<$ 0.3                  &  10 $\pm$ 1             &       0.6             & no     \\
 \hfill & \hfill & \hfill & \hfill & \hfill & \hfill & \hfill & \hfill & \hfill & \hfill   \\ 
{ IRAS4A1}      & Pl231 &       3.7$\pm$0.5             & 2.48$\pm$ 0.1 & 3489 $\pm$ 106  & $0.1\pm0.05$  &       --                      & --                            &       1.6              & -- \\
{ }                     & Pl94          &       5.4 $\pm$ 0.8           & 2.8 $\pm$ 0.2   & 363 $\pm$ 12  & $0.07\pm0.02$ &       --                      & --                              &       2.2             & -- \\
{ }                     & PG231 &       3.8$\pm$ 0.6    & 2.69 $\pm$ 0.2        & 3155 $\pm$ 60   & $0.4\pm0.1$ &         0.32 $\pm$ 0.1  & 348 $\pm$ 25  &         1.7             & no \\
{ }                     & PG94          &       4.5$\pm$ 0.3            & 2.3 $\pm$ 0.1   & 363 $\pm$ 30  & $0.5\pm0.1$ & 0.5 $\pm$ 0.1   & 130 $\pm$ 9               &       4.3             & no \\
\hfill & \hfill & \hfill & \hfill & \hfill & \hfill & \hfill & \hfill & \hfill & \hfill   \\ 
{ IRAS4B}       & Pl231 & 4 $\pm$ 2             & 2.9 $\pm$ 0.4 & 1448 $\pm$ 40      & $0.14\pm0.1$ & --                             &  --                           & 3.3                     & -- \\
{ }                     & Pl94          & 3.9 $\pm$ 0.4         & 2.9 $\pm$ 0.4     & 141 $\pm$ 6   & $0.14\pm0.05$ & --                            &  --                             & 0.8                   & -- \\
{ }                     & PG231 & 3.8 $\pm$ 1           & 2.9 $\pm$ 0.4 & 841 $\pm$ 80    & $1.0\pm0.2$ & 0.53 $\pm$ 0.1  &  645 $\pm$ 35 &       2.5              & yes \\
{ }                     & PG94          & 7.8 $\pm$ 0.5         & 2.9 $\pm$ 0.4     & 131 $\pm$ 11  & $0.3\pm0.1$ & 0.29 $\pm$ 0.15 &  24 $\pm$ 7           &       1.8              & no \\
\hfill & \hfill & \hfill & \hfill & \hfill & \hfill & \hfill & \hfill & \hfill & \hfill  \\  
{ IRAM04191}    & Pl231 & 27$\pm$2              & 1.6 $\pm$ 0.2         & 410 $\pm$ 60 & $0.01\pm0.01$    & --                            & --                            & 2.37            & --            \\
{ }                     & Pl94          & 29$\pm$2              & 1.4 $\pm$ 0.3     & 44 $\pm$ 9    & $0.01\pm0.01$ & --                            & --                              & 0.86                  & --    \\
{ }                     & PG231 & 28$\pm$2              & 1.4 $\pm$ 0.3         & 539 $\pm$ 10    & $0.2\pm0.3$ & 0.17 $\pm$ 0.1  & 3.6 $\pm$ 1           & 0.84            & yes           \\
{ }                     & PG94          & 29$\pm$4              & 1.54 $\pm$ 0.2     & 32.8 $\pm$ 9  & $1.28\pm1$    & $<$ 0.3                       & 0.2 $\pm$ 0.1           & 1.0                   & no    \\
\hfill & \hfill & \hfill & \hfill & \hfill & \hfill & \hfill & \hfill & \hfill & \hfill  \\  
{ L1521F}       & Pl231 & 32$\pm$2              & 1.7 $\pm$ 0.2         & 1100 $\pm$ 100  & $3.0\pm3.0$ & --                              & --                            & 2.4             & --            \\
{ }                     & Pl94          & 37$\pm$8              & 2.0 $\pm$ 0.2     & 53 $\pm$ 30   &       $3.0\pm2.0$ & --                                & --                              & 1.35                  & --    \\
{ }                     & PG231 & 32$\pm$3              & 1.6 $\pm$ 0.4         & 1100 $\pm$ 100  & $2.6\pm0.8$ & $0.13\pm0.1$    & 1.3 $\pm$ 0.4 & 0.79          & yes             \\
{ }                     & PG94          & 39$\pm$10             & 1.65 $\pm$ 0.5     & 53 $\pm$ 30   & $1.8\pm0.8$   & $0.16\pm0.2$                  & <0.5 $\pm$ 0.1  & 5.0                   & no    \\
\hfill & \hfill & \hfill & \hfill & \hfill & \hfill & \hfill & \hfill & \hfill & \hfill  \\  
{ L1527}        & Pl231 & 35$\pm$5              & 2.57 $\pm$ 0.2        & 1080 $\pm$ 100  & $0.01\pm0.01$ & --                            & --                            & 19                      & --            \\
{ }             & Pl94          & 35$\pm$10             & 2.6 $\pm$ 0.2         & 85 $\pm$ 20     & $0.01\pm0.01$ & --                            & --                            & 4                       & --    \\
{ }             & PG231 & 28$\pm$4              & 1.68 $\pm$ 0.4        & 1275 $\pm$ 320          & $0.4\pm0.2$ & 0.4 $\pm$ 0.1   & 215 $\pm$ 14          & 2.9             & yes           \\
{ }             & PG94          & 35$\pm$10             & 1.78 $\pm$ 0.3         & 85 $\pm$ 20   & $0.28\pm1.0$  & $0.3\pm0.1$                   & 23 $\pm$ 1              & 0.6                   & yes   \\
\hfill & \hfill & \hfill & \hfill & \hfill & \hfill & \hfill & \hfill & \hfill & \hfill  \\  
{ SerpM-S68N}           & Pl231         & 15$\pm$2              & 2.28$\pm$0.1                 & 800$\pm$80    & $0.01\pm0.01$ & --                    & --              & 1.67  & --                     \\
{ }                             & Pl94          & 22$\pm$5              & 2.56$\pm$0.2            & 35$\pm$10     & $0.03\pm0.03$ & --                    & --              & 1.2   & --             \\
{ }                             & PG231         & 13.9$\pm$3            & 2.1$\pm$0.2     & 800$\pm$80    & $0.1\pm0.1$   & 0.11$\pm$0.1          & 28$\pm$ 11              & 2.1   & no                     \\
{ }                             & PG94          & 15$\pm$2              & 2.08$\pm$0.2            & 35$\pm$10             & $0.5\pm0.1$ & 0.5$\pm$0.1                     & 6.5$\pm$ 1              & 3.4           & no             \\
\hfill & \hfill & \hfill & \hfill & \hfill & \hfill & \hfill & \hfill & \hfill & \hfill  \\  
{ SerpM-SMM4}   & Pl231         & 25$\pm$5              & 2.8$\pm$0.4           & 1700$\pm$200    & $0.06\pm0.05 $ & --                                   & --                      & 2.7                   & --                     \\
{ }                             & Pl94          & 15$\pm$4              & 2.8$\pm$0.4             & 105$\pm$6     & $0.01\pm0.01$ & --                                    & --                      & 3.7                   & --             \\
{ }                             & PG231         & 22$\pm$5              & 1.8$\pm$0.5             & 1305$\pm$500  & $0.70\pm0.1$ & 0.70$\pm$0.2           & 595$\pm$ 35             & 0.14  & yes                    \\
{ }                             & PG94          & 25$\pm$5              & 1.6$\pm$0.3             & 103$\pm$20    & $0.6\pm0.1$   & 0.62$\pm$0.1          & 61$\pm$ 3               & 0.39          & yes            \\
\hfill & \hfill & \hfill & \hfill & \hfill & \hfill & \hfill & \hfill & \hfill & \hfill  \\  
{ SerpS-MM18}   & Pl231         & 16$\pm$2              & 2.24$\pm$0.2          & 2208$\pm$190    & $0.01\pm0.01$ & --                                    & --                              & 1.8   & --                     \\
{ }                             & Pl94          & 20$\pm$5              & 2.4$\pm$0.2             & 114$\pm$20    & $0.015\pm0.05$        & --                                    & --                              & 0.87  & --     \\
{ }                             & PG231         & 15.5$\pm$6            & 2.17$\pm$0.2            & 2327$\pm$55   & $0.13\pm0.07$ & 0.128$\pm$0.08                & 76$\pm$ 4               & 0.68  & yes                    \\
{ }                             & PG94          & 11$\pm$2              & 2.24$\pm$0.2            & 114$\pm$20    & $0.04\pm0.02$ & < 0.1$\pm$ 0.1                & < 1$\pm$ 0.5    & 2.78          & no                     \\
\hfill & \hfill & \hfill & \hfill & \hfill & \hfill & \hfill & \hfill & \hfill & \hfill  \\  
{ SerpS-MM22}   & Pl231         & 11$\pm$2              & 2.56$\pm$0.2          & 135$\pm$30      & $0.01\pm0.01$ & --                                    & --                              & 1.8   & --                     \\
{ }                             & Pl94          & 25$\pm$5              & 2.7$\pm$0.3             & 10$\pm$5      & $0.01\pm0.01$ & --                                    & --                              & 1.16          & --     \\
{ }                             & PG231         & 10$\pm$4              & 2.0$\pm$0.3             & 148$\pm$9     & $0.26\pm0.08$ & 0.25$\pm$0.08                 & 31$\pm$ 4               & 1.1   & yes                    \\
{ }                             & PG94          & 19$\pm$3              & 1.98$\pm$0.3            & 10$\pm$5      & $0.31\pm$ 0.07        & $0.31\pm$ 0.07            & 3.2$\pm$ 0.5          & 0.59          & yes                    \\
\hfill & \hfill & \hfill & \hfill & \hfill & \hfill & \hfill & \hfill & \hfill & \hfill  \\  
{ L1157}                        & Pl231                 & 12$\pm$10             & 2.68$\pm$0.2            & 520$\pm$170    & 0.01$\pm$0.01                & --                                      & --                            & 1.29    & --                     \\
{ }                             & Pl94                  & 16$\pm$5              & 2.65$\pm$0.2            & 61$\pm$10             & 0.01$\pm$0.01 & --                                    & --                              & 1.4           & --     \\
{ }                             & PG231                 & 6.8$\pm$2             & 2.5$\pm$0.2             & 494$\pm$35    & 0.05$\pm$0.04 & <0.1$\pm$0.1          & 56$\pm$6                & 0.76          & yes                    \\
{ }                             & PG94                  & 8.2$\pm$2             & 2.24$\pm$0.3            & 61$\pm$10             & 0.05$\pm$0.04 & <0.3$\pm$ 1                       & 9$\pm$ 1              & 2.5           & no                     \\
\hfill & \hfill & \hfill & \hfill & \hfill & \hfill & \hfill & \hfill & \hfill & \hfill  \\  
{ GF9-2}        & Pl231                 & 40$\pm$7              &  2.36$\pm$0.2         &  179$\pm$20 & $0.01\pm0.01$     & --                                    & --                              & 5             & --             \\
{ }             & Pl94                  & 40$\pm$10             &  2.2$\pm$0.2          &  27$\pm$7       & $0.01\pm0.01$ & --                                    & --                              & 1.8           & --                     \\
{ }             & PG231                 & 37$\pm$3              &  1.67$\pm$0.3         &  407$\pm$80 &   $0.18\pm0.05$ & 0.18 $\pm$ 0.05         & 11.8 $\pm$ 2          & 0.55            & yes            \\
{ }             & PG94                  & 34$\pm$5              &  1.71$\pm$0.3         &  39$\pm$20      & $0.15\pm0.07$ & <0.3$\pm$0.1          & 1.3$\pm$0.5           & 0.45            & yes                    \\
\hfill & \hfill & \hfill & \hfill & \hfill & \hfill & \hfill & \hfill & \hfill & \hfill  \\  
\hline
\end{supertabular}\\
\end{center}
\tablefoot{The first column shows the name of the primary protostar. Column 2 reports the type of model used (Pl or PG plus the frequency of the modeled visibility profile): for each source, the first two lines report the parameters of the best Plummer envelope (Pl) models and the two following lines report the parameters of the best Plummer+Gaussian (PG) models.
Columns 3 to 8 report parameters of the best-fit model: the size of the Plummer envelope (Col. 3), the value of the (p+q) brightness radial distribution index (Col. 4), the total flux of the Plummer envelope component emission at the considered frequency (Col. 5), the FWHM of the Gaussian component (Col. 6), the inner radius (Col. 7), and the flux of the Gaussian component (Col. 8). Column 9 reports the reduced $\chi^2$ value associated with this best-fit model, and Col. 10 indicates whether the result of the F-test suggests that the PG model is a better model than the Pl model. \\
\tablefoottext{$\star$} {For SVS13B, although the reduced chi square of the PG model is not statistically better than the one obtained for the Pl model, the more reasonable value of the $p+q$ parameter makes the PG model a satisfactory model at 231 GHz as well. The fact that the visibilities at 231 GHz might need an additional unresolved component to be properly reproduced while the 94 GHz visibility profile is fine without this might suggest that the additional component might be due to an optically thick component at 231 GHz that is not seen at 94 GHz and is embedded in the otherwise optically thin envelope.}}


\begin{appendix}

\section{Details of CALYPSO observations}
\label{section:appobs}

\begin{table*}[!h]
\caption{Fields observed with CALYPSO, grouped by shared observing tracks.}
\label{table:observations}
\centering
\begin{tabular}{c|cc|ccc}
\hline
 {\bf{Field}\tablefootmark{1}} & \multicolumn{2}{c|}{Phase center} & \multicolumn{3}{c}{Central frequency of WideX observations} \\
 {} & {$\alpha_{2000}$} & {$\delta_{2000}$} & 231 GHz & 219 GHz & 94 GHz \\
\hline
\multicolumn{6}{|c|}{Track1} \\ 
\hline 
\hfill & \hfill & \hfill & \multicolumn{3}{c}{Observation dates} \\ 
{ L1448-2A} & 03:25:22.4 & +30:45:12 & 2010 Jan.23-24 & 2010 Dec.11-12  & 2010 Nov.02-03  \\
{ L1448-C} & 03:25:38.9 & $+$30:44:05 & 2010 Nov.18-19  & 2011 Jan.24-25  & 2011 Jan.27-28  \\
{ L1521F} & 04:28:38.9 & $+$26:51:36 & 2012 Nov.15-16  & 2011 Apr.05-13 - & \\
\hfill & \hfill & \hfill & \hfill & \hfill & \hfill \\
\hline
\multicolumn{6}{|c|}{Track2} \\ 
\hline
\hfill & \hfill & \hfill & \multicolumn{3}{c}{Observation dates} \\ 
{ SVS13B} & 03:29:03.7 & +31:15:52 & 2011 Feb.05 & 2011 Nov.13-14  & 2011 Nov.11-12  \\
{ IRAS4A} & 03:29:10.5 & +31:13:31 & 2011 Oct.15 & 2012 Feb.04 & 2011 Nov.18-19-20 \\
{ IRAS4B} & 03:29:11.9 & +31:13:08 & 2012 Feb.12-13  & 2013 Jan.24 & 2012 Feb.06-07 \\
{ L1527}  & 04:39:53.9 & +26:03:10 & 2012 Nov.07-08 & 2013 Feb.16 & \\
\hfill & \hfill & \hfill & \hfill & \hfill & \hfill \\
\hline
\multicolumn{6}{|c|}{Track3} \\ 
\hline
\hfill & \hfill & \hfill & \multicolumn{3}{c}{Observation dates} \\ 
{ L1448-N} & 03:25:36.3 & +30:45:15 & 2008.Feb.09\tablefootmark{$^{\star}$}  & 2010 Dec.04-05 & 2010 Nov.06-07 \\
{ IRAS2A}  & 03:28:55.6 & +31:14:37 & 2010 Nov.07-08-21-22 & 2010 Dec.04-05 & 2010 Nov.06-07 \\
{ IRAM04191} & 04:21:56.9 & +15:29:46 & 2011 Jan.25 / Feb.01  & 2011 Jan.29-30 & \\
\hfill & \hfill & \hfill & \hfill & \hfill & \hfill \\
\hline
\multicolumn{6}{|c|}{Track4} \\ 
\hline
\hfill & \hfill & \hfill & \multicolumn{3}{c}{Observation dates} \\ 
{SerpM-S68N} & 18:29:48.10 & +01:16:43.6 & 2011 Feb.08  & 2011 Nov.13-14-16 & 2011 Nov.09-11-12  \\
{SerpM-SMM4} & 18:29:56.70 & +01:13:15.0 & 2011 Mar.25  & 2012 Feb.04-21-25 & 2012 Feb.12     \\
{SerpS-MM18}  & 18:30:03.86 & -02:03:04.9 & 2012 Feb.02  & 2013 Feb.16       & 2012 Mar.01  \\
{SerpS-MM22}  & 18:30:12.34 & -02:06:52.4 & 2012 Nov.16  &        &        \\
\hfill & \hfill & \hfill & \hfill & \hfill & \hfill \\
\hline
\multicolumn{6}{|c|}{Track5} \\ 
\hline
\hfill & \hfill & \hfill & \multicolumn{3}{c}{Observation dates} \\ 
{L1157} & 20:39:06.3 & +68:02:15 & 2011 Feb.04-06-07  & 2011 Nov.08-10-11-12-13 & 2011 Nov.07 \\
{GF9-2} & 20:51:30.1 & +60:18:39 & 2011 Mar.20 & 2012 Feb.11-17-18 & 2012 Feb.08 \\
\hfill & \hfill  & \hfill & 2012 Oct.30 & & 2013 Jan.12 \\
\hfill & \hfill & \hfill & \hfill & \hfill & \hfill \\
\hline
\end{tabular}
\tablefoottext{1}{The region mapped by a given field is limited by the PdBI primary beam: 20$\arcsec$ at 231GHz, 23$\arcsec$ at 219GHz, and 54$\arcsec$ at 94~GHz.} \tablefoottext{$^{\star}$}{Identifies observations from the pilot observing program R068 \citep{Maury10}}
\end{table*}

\clearpage
\section{CALYPSO dust continuum maps}
\label{section:appmaps}




\begin{figure*}[!h]
\centering
\includegraphics[width=0.71\textwidth]
{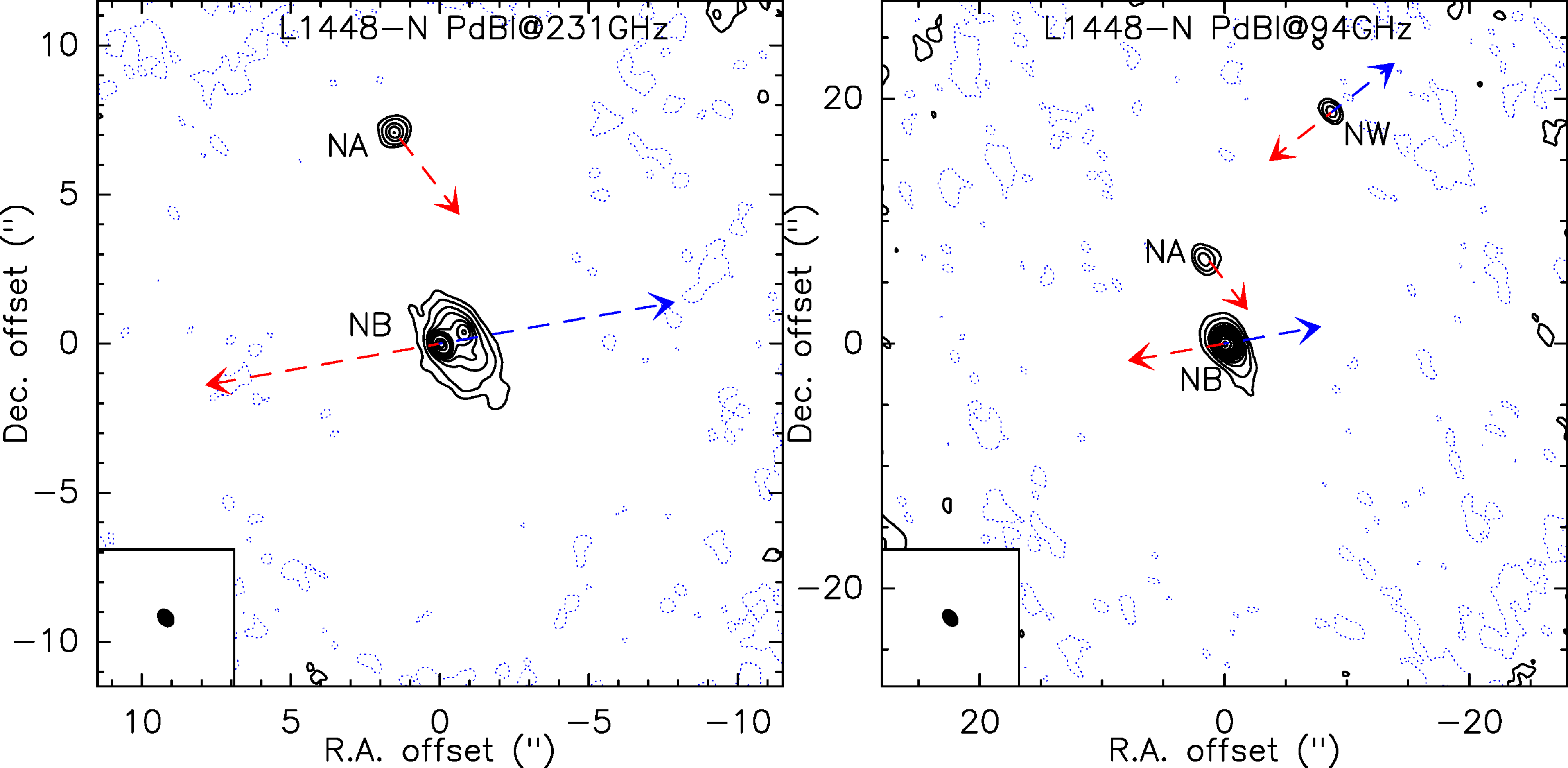}
\caption{\label{fig:l1448n_maps} 
1.3~mm (231 GHz) and 3.3~mm (94 GHz) PdBI dust continuum emission maps of L1448-N. The ellipses in the bottom left corner show the respective synthesized beam sizes. The contours are levels of -3$\sigma$ (dashed) and 5$\sigma,$ then in steps of 10$\sigma$ from 10$\sigma$ to 100$\sigma$, and finally in steps of 20$\sigma$ beyond (rms noise computed in the map before primary beam correction as reported in Table \ref{table:continuumobs}). The maps have been corrected for primary beam attenuation. The blue and red arrows show the direction of the protostellar jet(s) associated with the millimeter sources in the field, see Table \ref{table:sample} for further details.
}
\end{figure*}
\begin{figure*}[!h]
\centering
\includegraphics[trim={0 0 0 0},clip,width=0.71\textwidth]
{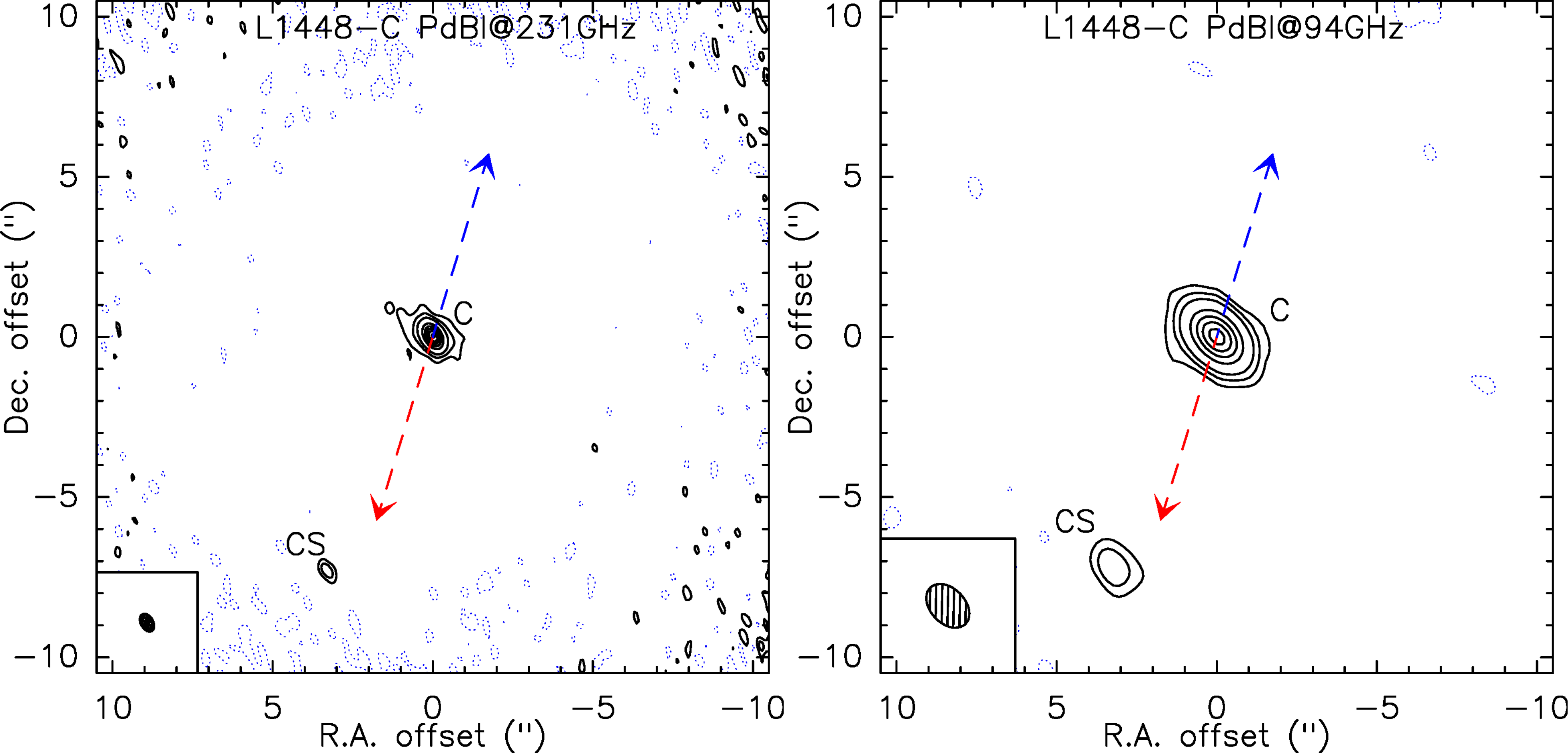}
\caption{\label{fig:l1448c_maps} 
Same as Fig. \ref{fig:l1448n_maps} for L1448-C. The contours show 
levels of -3$\sigma$ (dashed), 5$\sigma$,  and 10$\sigma$  and then use 
steps of 40$\sigma$ from 20$\sigma$ on. 
}
\end{figure*}
\begin{figure*}[!h]
\centering
\includegraphics[trim={0 0 0 0},clip,width=0.71\textwidth]
{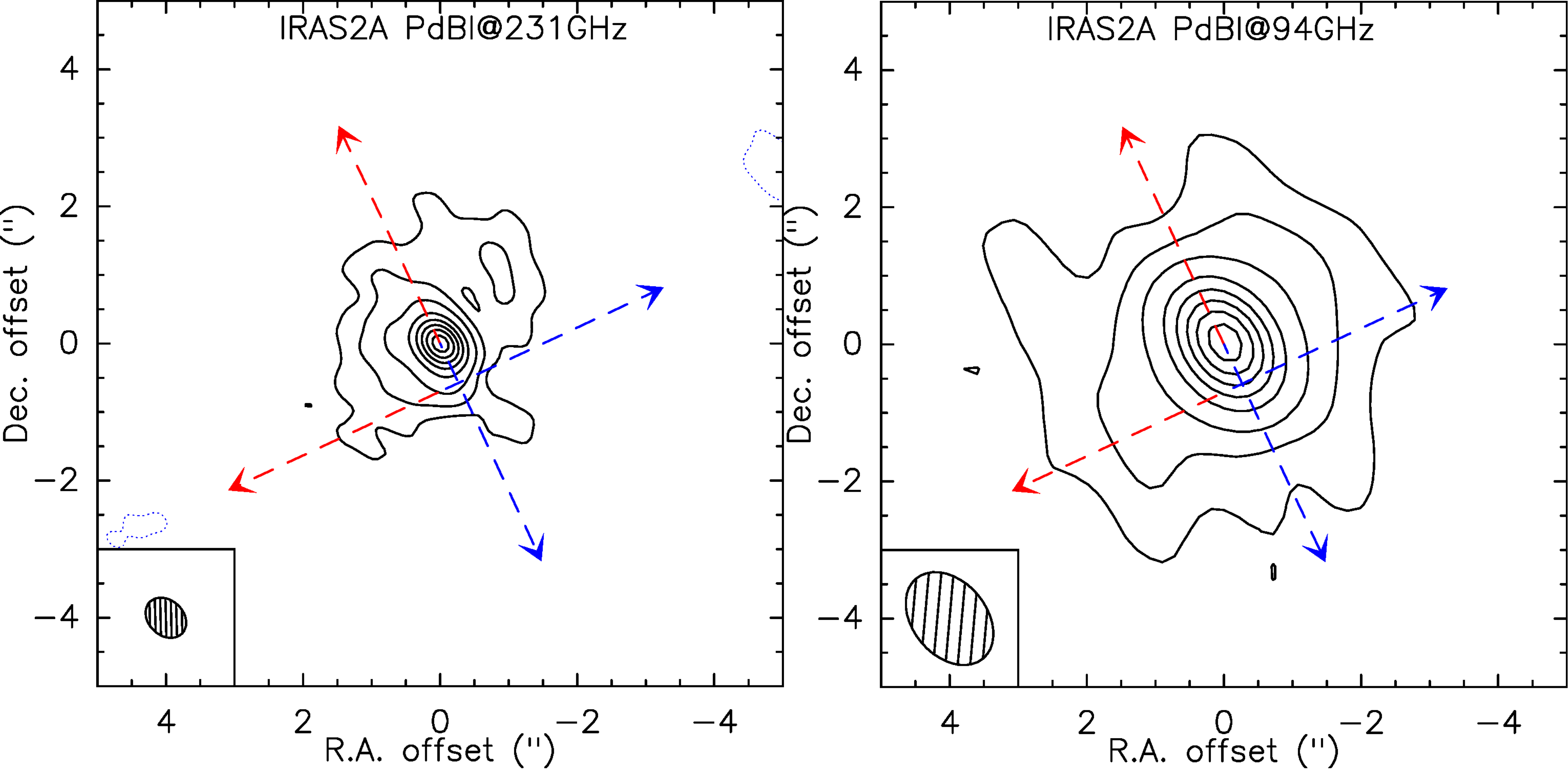}
\caption{\label{fig:iras2a_maps} 
Same as Fig. \ref{fig:l1448n_maps} for IRAS2A. The contours show 
levels of -3$\sigma$ (dashed), 5$\sigma$, and 10$\sigma$  and then use 
steps of 20$\sigma$ from 20$\sigma$ on.
}
\end{figure*}
\begin{figure*}[!h]
\centering
\includegraphics[trim={0 0 0 0},clip,width=0.73\textwidth]
{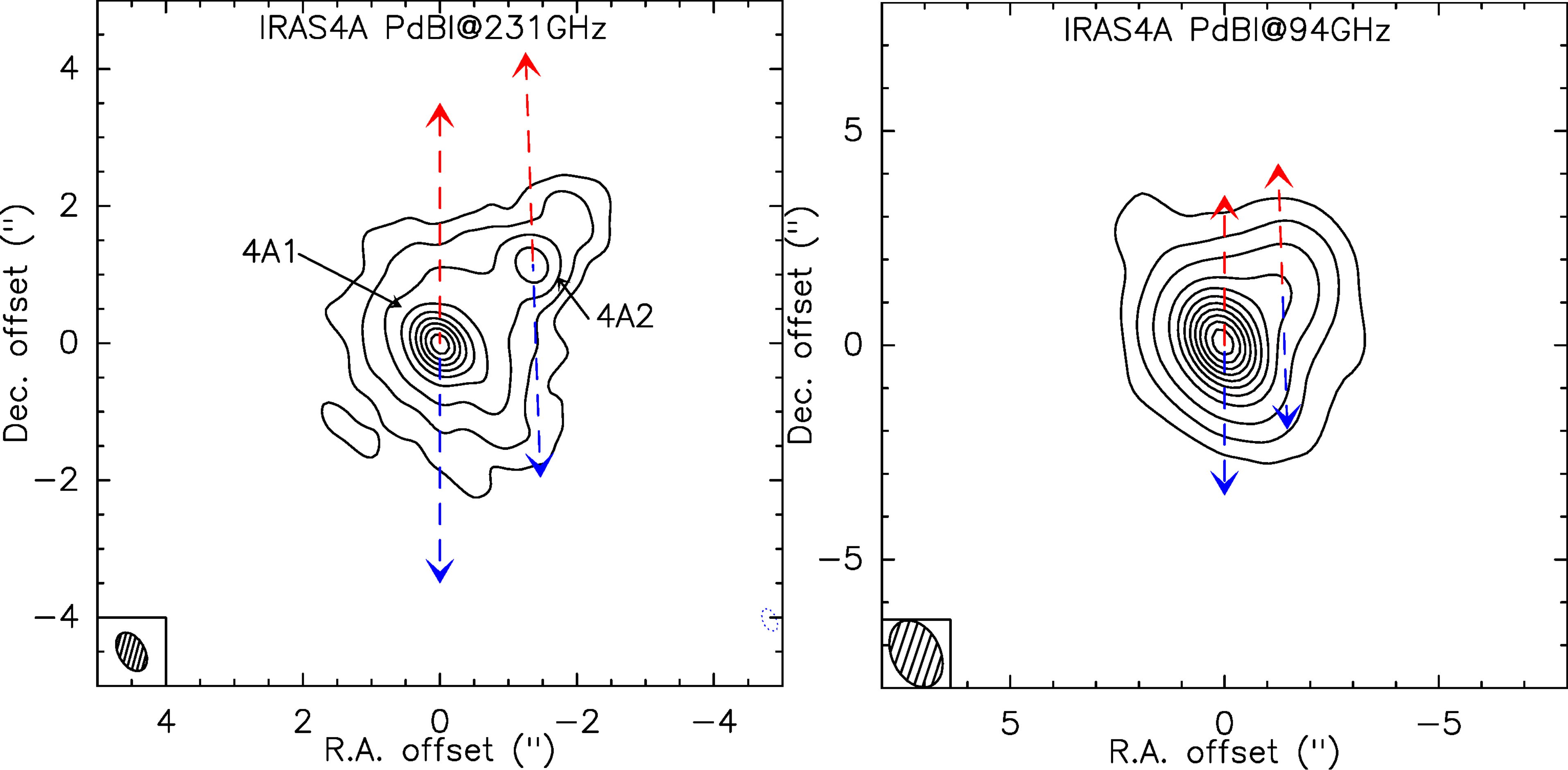}
\caption{\label{fig:iras4a_maps} 
Same as Fig. \ref{fig:l1448n_maps} for IRAS4A. The contours show 
levels of -3$\sigma$ (dashed), 5$\sigma$, and 10$\sigma$ and then use steps 
of 20$\sigma$ from 20$\sigma$ on. IRAS4B is detected to the southwest in the 94 GHz map, 
outside the area shown here, so that the structure of IRAS4A is well 
distinguished in the figure.
}
\end{figure*}
\begin{figure*}[!h]
\centering
\includegraphics[trim={0 0 0 0},clip,width=0.73\textwidth]
{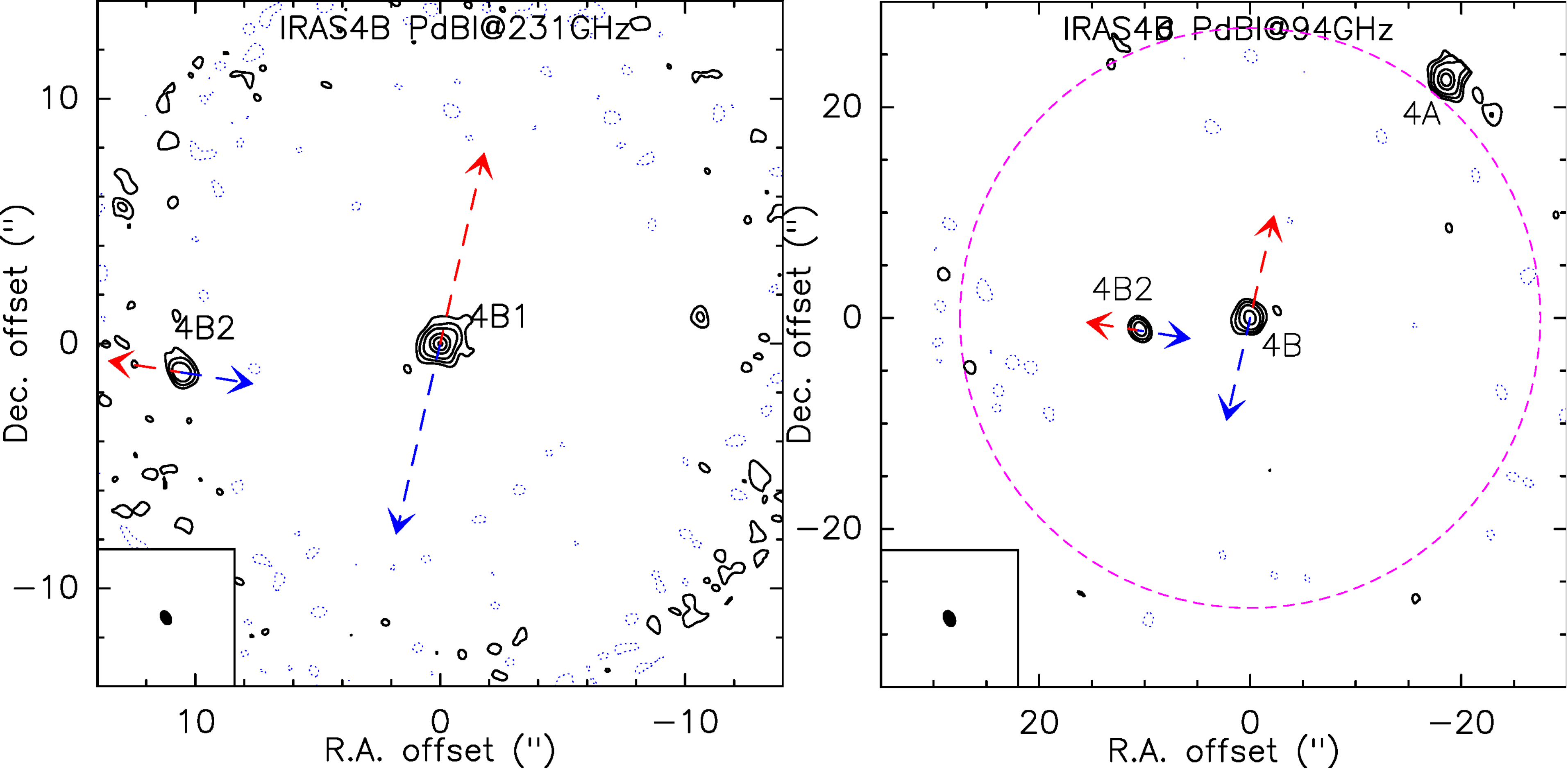}
\caption{\label{fig:iras4b_maps} 
Same as Fig. \ref{fig:l1448n_maps} for IRAS4B. The contours show 
levels of -3$\sigma$ (dashed), 5$\sigma$, and 10$\sigma$ and then use steps 
of 30$\sigma$ from 20$\sigma$ on. The flux density in the maps has been corrected for primary beam attenuation (the primary beam FWHM is shown as a pink dashed circle in the 94 GHz map).
}
\end{figure*}
\begin{figure*}[!h]
\centering
\includegraphics[trim={0 0 0 0},clip,width=0.73\textwidth]
{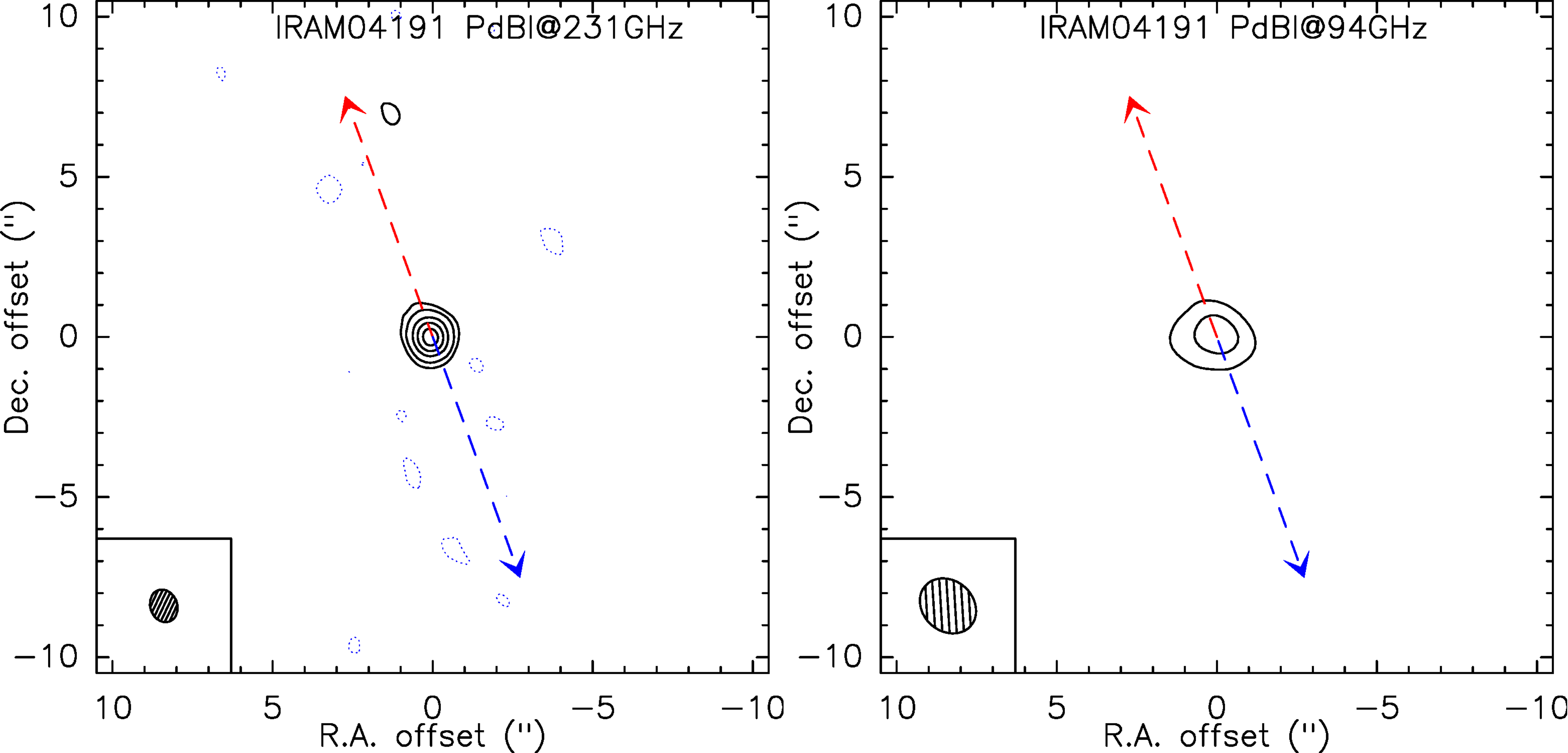}
\caption{\label{fig:iram04191_maps} 
Same as Fig. \ref{fig:l1448n_maps} for IRAM04191. The contours show 
levels of -3$\sigma$ and 5$\sigma$ and then use steps of 10$\sigma$ from 
10$\sigma$ on.
}
\end{figure*}
\begin{figure*}[!h]
\centering
\includegraphics[trim={0 0 0 0},clip,width=0.75\textwidth]
{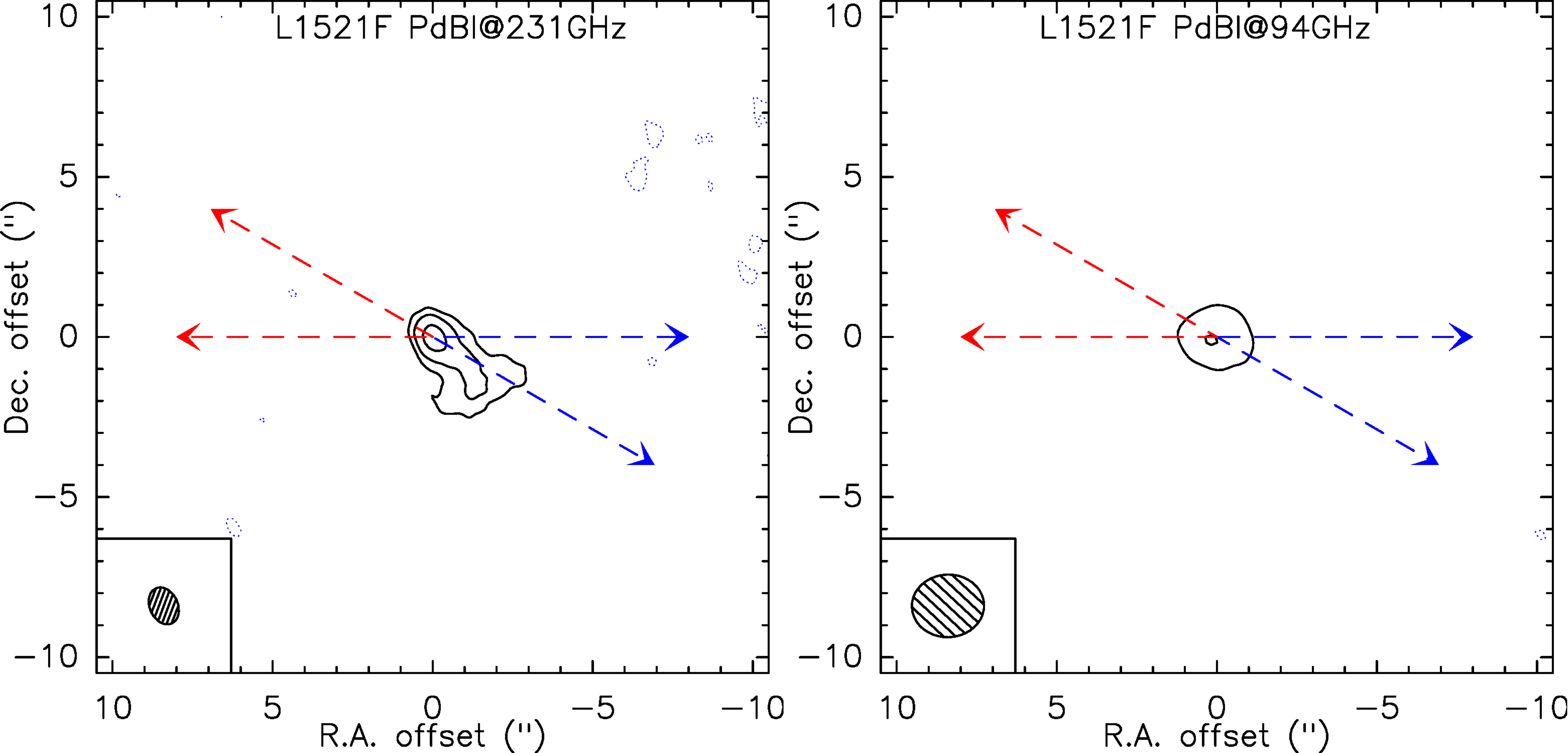}
\caption{\label{fig:l1521f_maps} 
Same as Fig. \ref{fig:l14482a_maps} for L1521F. The contours show 
levels of -3$\sigma$ and  5$\sigma$ and then use steps of 10$\sigma$ 
from 10$\sigma$ on.
}
\end{figure*}
\begin{figure*}[!h]
\centering
\includegraphics[trim={0 0 0 0},clip,width=0.75\textwidth]
{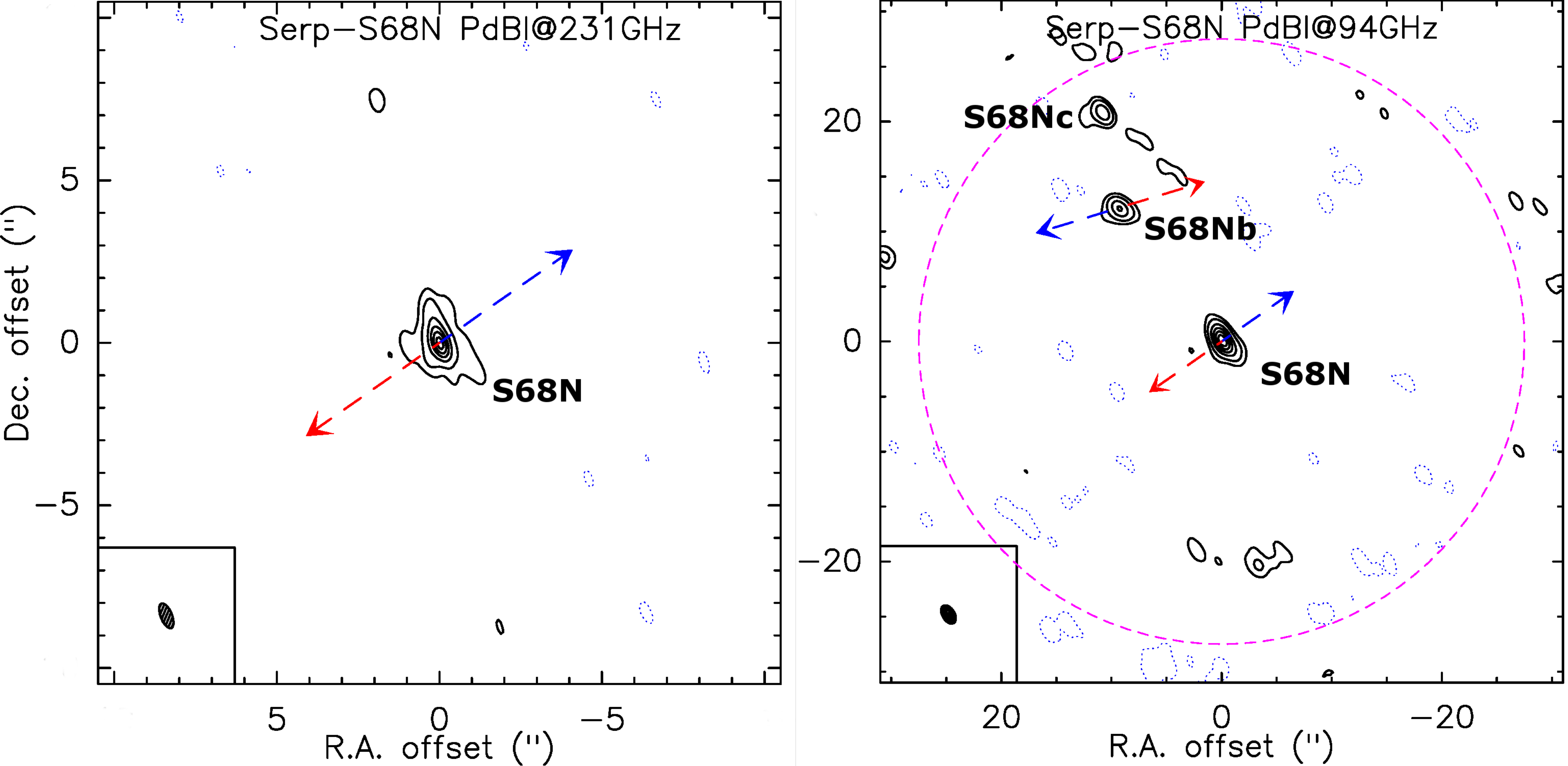}
\caption{\label{fig:serps68n_maps} 
Same as Fig. \ref{fig:l1448n_maps} for SerpM-S68N. The contours show 
levels of -3$\sigma$ and 5$\sigma$ and then use steps of 10$\sigma$ 
from 10$\sigma$ on. The dashed pink contour in the 94 GHz continuum 
emission map shows the PdBI primary beam. 
}
\end{figure*}
\begin{figure*}[!h]
\centering
\includegraphics[trim={0 0 0 0},clip,width=0.75\textwidth]
{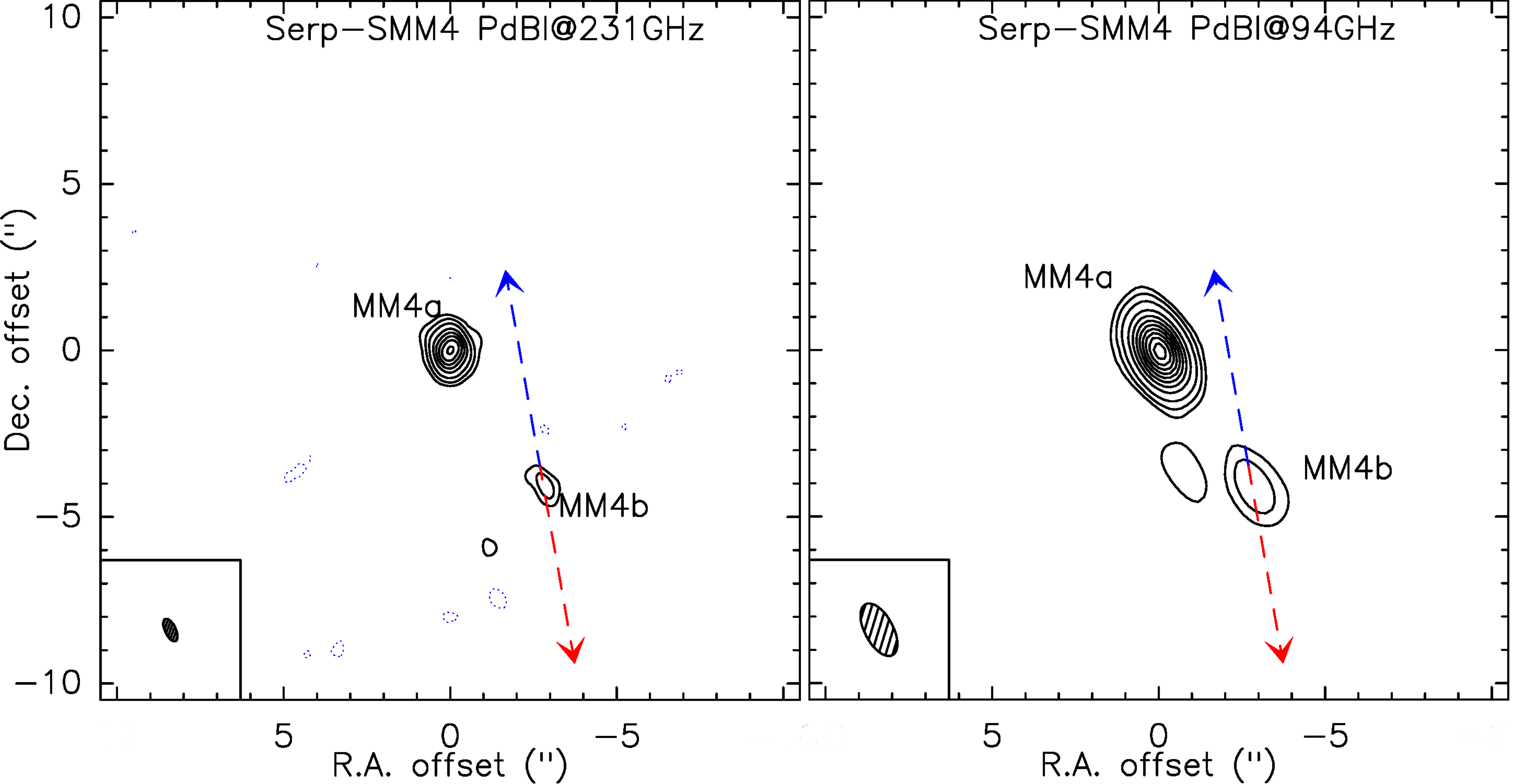}
\caption{\label{fig:serpsmm4_maps} 
Same as Fig. \ref{fig:l1448n_maps} for SerpM-SMM4. The contours 
show levels of -3$\sigma$, 5$\sigma,$ and 10$\sigma$ and then use 
steps of 20$\sigma$ from 20$\sigma$ on.
}
\end{figure*}
\begin{figure*}[!h]
\centering
\includegraphics[trim={0 0 0 0.1},clip,width=0.73\textwidth]
{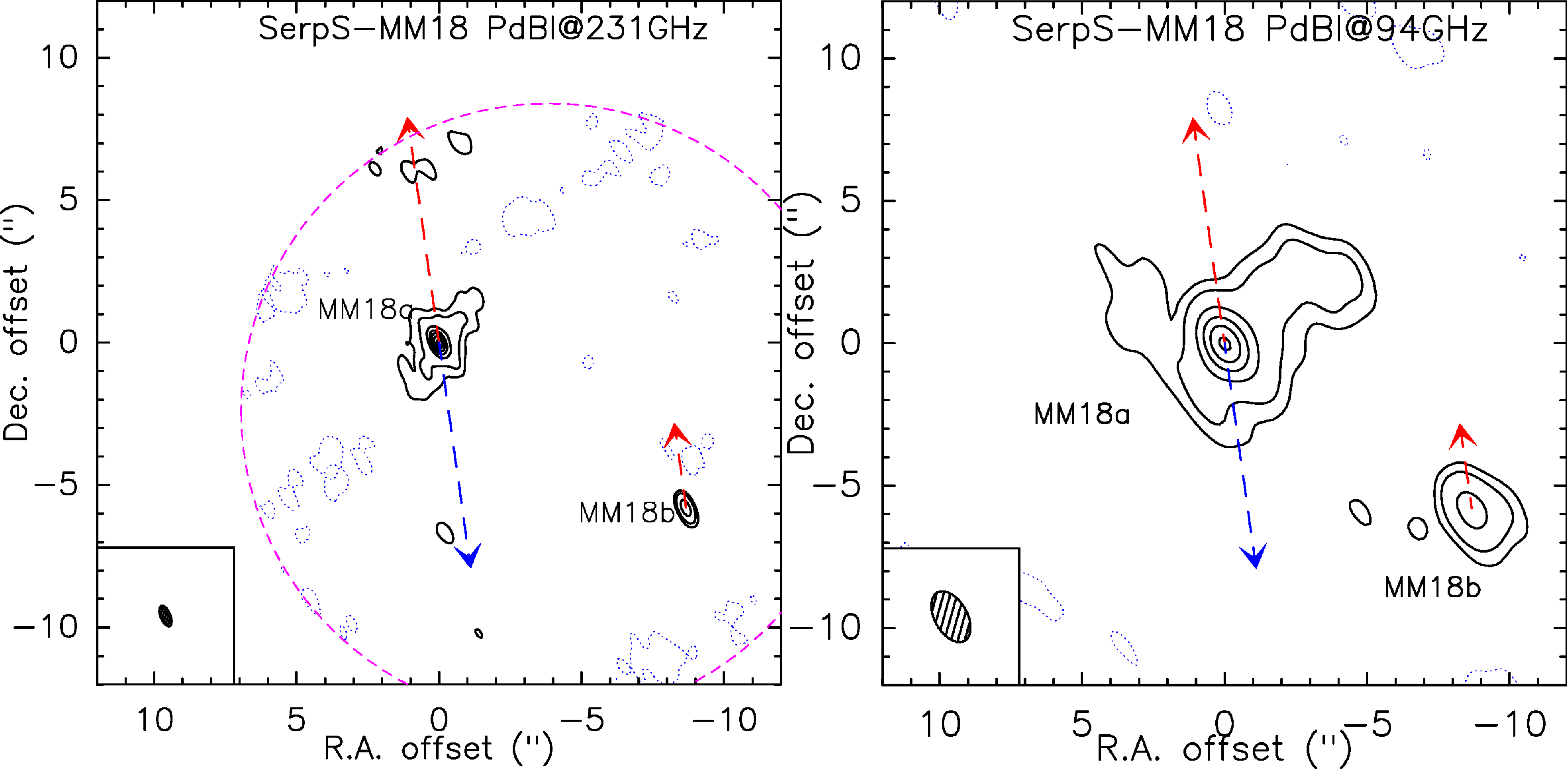}
\caption{\label{fig:serpsmm18_maps} 
Same as Fig. \ref{fig:l1448n_maps} for SerpS-MM18. The contours 
show levels of -3$\sigma$ and 5$\sigma$ and then use steps of 20$\sigma$ 
from 10$\sigma$ on. The dashed pink contour shows the PdBI primary 
beam.
}
\end{figure*}
\begin{figure*}[!h]
\centering
\includegraphics[trim={0 0 0 0.1},clip,width=0.73\textwidth]
{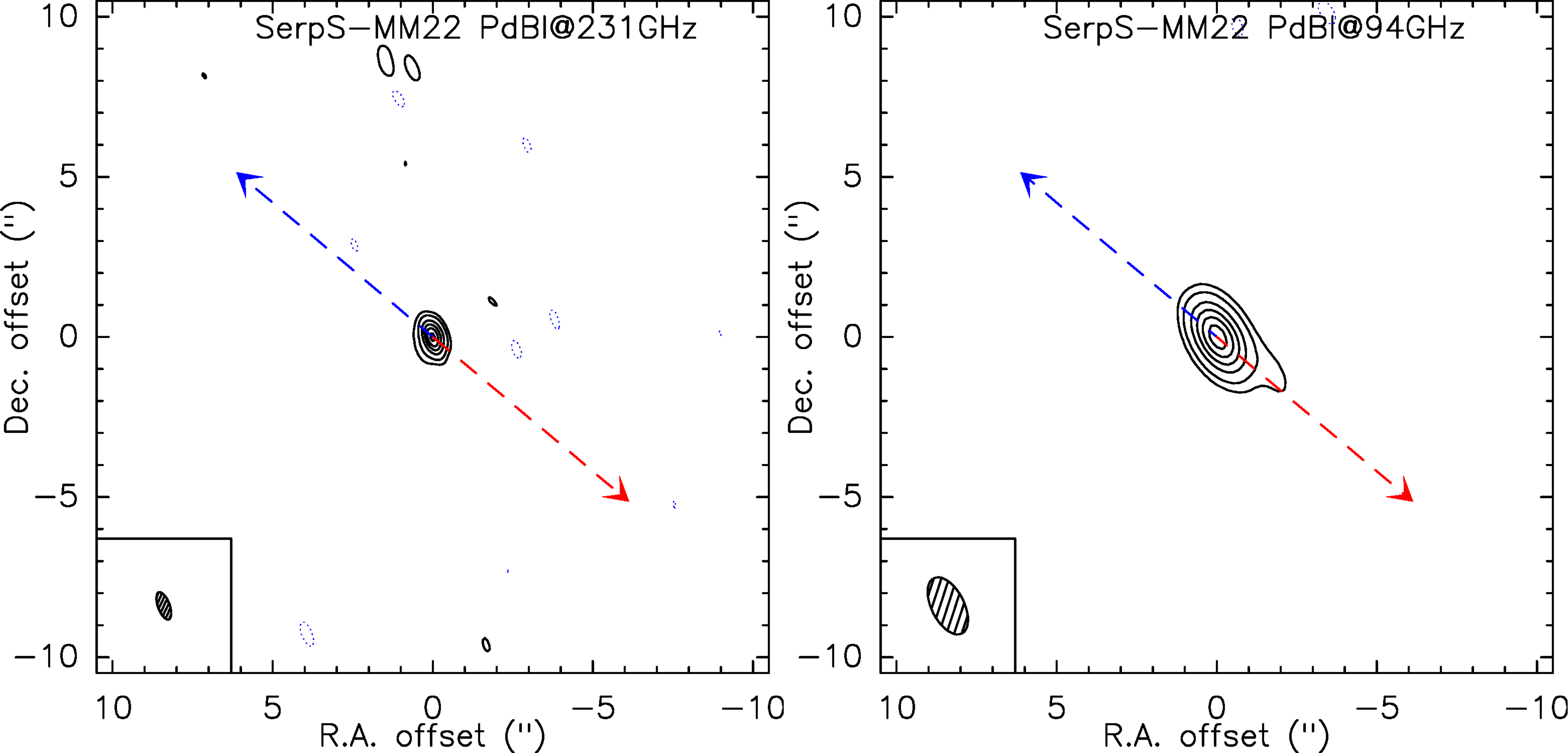}
\caption{\label{fig:serpsmm22_maps} 
Same as Fig. \ref{fig:l1448n_maps} for SerpS-MM22. The contours show 
levels of -3$\sigma$ and 5$\sigma$ and then use steps of 10$\sigma$ 
from 10$\sigma$ on.
}
\end{figure*}
\begin{figure*}[!h]
\centering
\includegraphics[trim={0 0 0 0.1},clip,width=0.73\textwidth]
{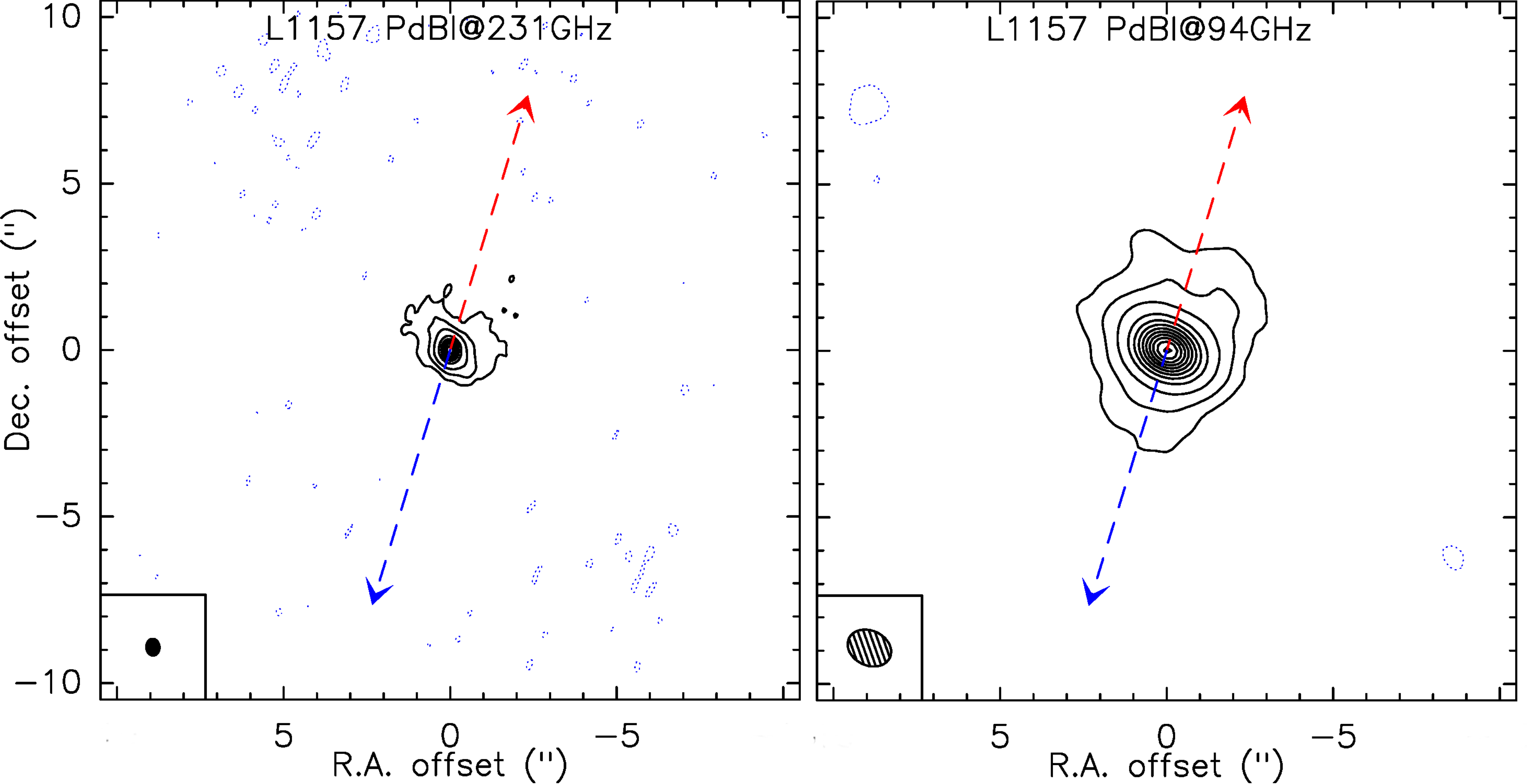}
\caption{\label{fig:l1157_maps} 
Same as Fig. \ref{fig:l1448n_maps} for L1157. The contours 
show levels of -3$\sigma $ and  5$\sigma$ and then use steps of 
30$\sigma$ from 20$\sigma$ on.}
\end{figure*}
\begin{figure*}[!h]
\centering
\includegraphics[trim={0 0 0 0.1},clip,width=0.75\textwidth]
{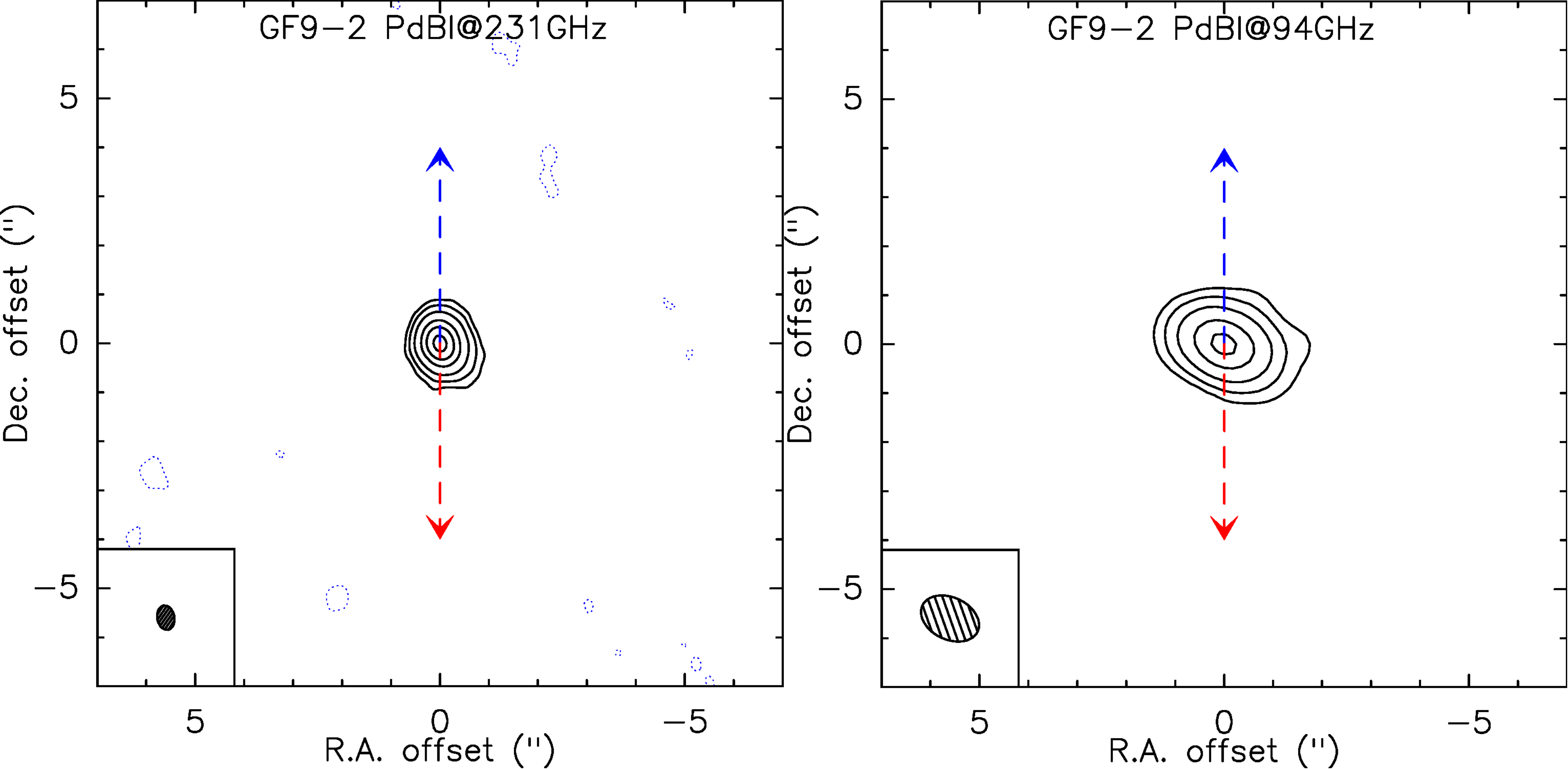}
\caption{\label{fig:gf92_maps} 
Same as Fig. \ref{fig:l1448n_maps} for GF9-2. The contours show 
levels of -3$\sigma$, 5$\sigma$, and 10$\sigma$ and then use steps of 
20$\sigma$ from 20$\sigma$ on.
}
\end{figure*}

\twocolumn
\clearpage
\section{Comments on individual sources}
\label{section:appindivsour}

\subsection{L1448-2A}
\label{section:appl14482a}

\begin{figure}[!h]
\centering
\includegraphics[trim={0.5cm 2cm 2cm 2cm},clip,width=0.99\linewidth]{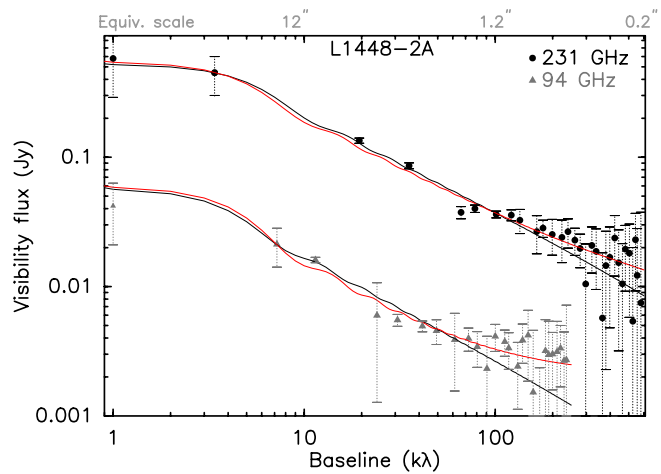}
\caption{Visibility profiles of the 231 GHz (black dots) and 94 GHz (gray triangles) continuum emission from L1448-2A. In each visibility curve, we show the best-fit Plummer-only model (Pl, black curve) and the best-fit Plummer+Gauss (PG, red curve) model. In the case of L1448-2A, the F-test suggests that the PG models are not statistically better than the Plummer-only (Pl) models, although the presence of an unresolved (radius $< 0.15\arcsec$) disk-like component cannot be robustly excluded.
}
\label{fig:l1448-2a_bothmodels}
\end{figure}

\begin{itemize}
\item{\it{Single-dish constraints}}\\
We used the 268 GHz Bolocam fluxes reported in \citet{Enoch06}, and subtracted the 15 mJy from L1448-2Ab, which was removed from the PdBI visibilities. We used the dual-frequency PdBI spectral index at $20 k\lambda$ (see Table \ref{table:continuum-pdbi-sd}) to extrapolate the flux from 268 GHz to 231 GHz.
The uncertainties on the 231 GHz extrapolated fluxes are $\pm 30\%$. The same method was followed to extrapolate the 94 GHz envelope fluxes (integrated flux and peak flux), with estimated uncertainties of $\pm 40\%$.

\item{\it{Multiplicity}}\\ 
L1448-2A is the only millimeter source detected in the field, see Fig. \ref{fig:l14482a_maps}, but an extension is also detected west of the millimeter peak position. The coordinates of this secondary source, L1448-2Ab, are reported in Table \ref{table:continuum-sources}.
L1448-2Ab was detected with the VLA \citep{Tobin16a} with a peak flux density 0.151 Jy at 8mm, but was never previously detected at higher frequencies, including in the 1.3 mm CARMA map at $0.3\arcsec$ resolution published by \citet{Tobin15a}. Its location in the equatorial plane of L1448-2A (as estimated from the jet position angle from L1448-2A), detection in independent datasets at different millimeter wavelengths, and 1.3\,mm peak flux half of the flux toward the primary protostar make it a robust protostellar companion candidate. 
Moreover, we tentatively detect two distinct jets from L1448-2A in our CALYPSO maps of the high-velocity $^{12}$CO(2-1) emission (blue and red jets are misaligned; Podio et al. in prep.).
The exact nature of L1448-2Ab needs to be further characterized with observations at higher angular resolution, but we consider it a candidate protostellar companion in our multiplicity analysis.

\item{\it{Candidate disk}}\\ 
The visibility profiles at 231\,GHz and 94\,GHz are shown in Figures \ref{fig:l14482a_popgmodels} and \ref{fig:l1448-2a_bothmodels} together with the best Plummer-only (Pl) and Plummer+Gauss (PG) models.
When all the dust continuum visibilities are used, the best-fit model to reproduce the continuum emission visibility profiles of L1448-2A is the Plummer-only model for both frequencies (see main text and Table \ref{table:popg-fits}). The best-fit PG model for the 231 GHz visibility profile includes a 12 mJy Gaussian source with an unresolved FWHM $<0.15\arcsec$. The upper limits on the parameters for this Gaussian component are reported as upper limits on the disk properties in Table \ref{table:disks-properties}. 
Similarly, the best-fit PG model that reproduces the 94 GHz visibilities includes the minimum (size and flux) Gaussian component allowed in the fitting procedure (see Table \ref{table:popg-fits}).
When only the equatorial visibilities are used (i.e., only the visibilities at PA $-80^{\circ}\pm30^{\circ}$ are selected), we find that the best-fit model is still the Plummer-only model (see models Pleq and PGeq in Table \ref{table:l14482a}).  
It is therefore clear from our analysis of the CALYPSO data that the continuum structure around L1448-2A traces the inner part of the envelope, and no resolved continuum disk-like emission is detected at scales 50-500 au.
\end{itemize}

\subsection{L1448-NB}
\label{section:appl1448n}

\begin{itemize}
\item{\it{Single dish constraints}}\\
We used the data from IRAM-30m MAMBO observations reported in \citet{Motte01a}, and subtracted the fluxes from the sources that were removed from the PdBI visibilities, that is, L1448-NA and the secondary western source NB2.
To extrapolate the MAMBO 243 GHz fluxes to our observing frequencies, we used the large-scale spectral index computed directly from our PdBI dual-frequency visibility amplitudes at 20 k$\lambda$ (see Table \ref{table:continuum-pdbi-sd}).
While the integrated flux over 4200 au is 2.1 Jy in \citet{Motte01a} (MA01), we would expect a total envelope flux 1.79 Jy in a $14\arcsec$ radius (the distance adopted in MA01 was 300 pc). However, the modeled $R_{out}$ of 10000 au in \citet{Motte01a} for L1448-NB suggests a larger radius ($20\arcsec$ at 300 pc) than the radius in which they report this integrated flux. From their map, we computed the flux in an area of radius $20\arcsec$, obtaining an integrated flux 3.7 Jy, which translates into a $3.24$ Jy total envelope flux  at 231 GHz.
Similarly, we used the MA01 peak flux to complete our visibility profile at 9 k$\lambda$ (for a $20\arcsec$ radius source in a $11\arcsec$ beam).
The estimated uncertainties on the 231 GHz extrapolated fluxes are $\pm 30\%$. The same method was followed to extrapolate the 94 GHz envelope fluxes (integrated and peak fluxes), and the estimated uncertainties are $\pm 40\%$.

\item{\it{Multiplicity}}\\
L1448-NB is the strongest millimeter source in the field, see Fig. \ref{fig:l1448n_maps}. 
The secondary source L1448-NA, located 1500 au away, was previously detected at both infrared and millimeter wavelengths and is classified as a Class I protostar \citep{Ciardi03,Kwon06}. It drives an outflow whose redshifted lobe was detected by \citet{Lee-K15} at a P.A. $218^{\circ}$. 
The source L1448-NW was also previously detected in the BIMA and SMA surveys \citet{Looney00,Lee-K15}, although it is sometimes referred to as L1448-IRS3C: it is associated with two 8mm sources, might not be gravitationally bound to the NA/NB system, and seems to drive an outflow \citep{Lee-K15}, but its nature is not precisely determined.

At smaller scales, NB is resolved into two components, NB1 and NB2. NB1 is the strongest millimeter source, and the western secondary component L1448-NB2 was previously detected in the 1.3\,mm dust continuum emission with the PdBI and ALMA maps \citep{Maury15IAUGA, Tobin16b} and with the VLA at 8mm \citep{Lee-K15}. It is tentatively resolved into two components called IRS3B-a and IRS3B-b with ALMA and VLA. Its location along the axis of the collimated jet driven by L1448-NB1 (projected onto the plane of the sky) and its flat spectral index (possibly due to free-free emission) would naively suggest that this source is due to interaction between the jet and the surrounding material. However, the presence of a small-scale structure (at scales 200--300 au, see our modeling of the visibility profile for this source described below) surrounding NB1 and NB2 can be used to argue that NB2 does not trace a jet feature, but possibly a protostellar companion in a circumbinary disk (this point is further discussed below). A robust assessment of the nature of this millimeter continuum component will require analyzing a larger set of data, including molecular lines and tracers of protostellar nature, which is beyond the scope of this paper. To build robust estimates of the upper-limit MF from our sample, we consider it a candidate protostellar companion in our multiplicity analysis.

\item{{\it{Candidate disk}}}\\

\begin{itemize}
\item{Centered on L1448-NB1:}
Our analysis is based on the visibilities that only contain the NB1 primary protostar, after removing the secondary sources (NA and NB2 in the 231 GHz data, and also NW in the 94 GHz map: as an illustrating example, Figure \ref{fig:l1448nb1-onlymain} shows the maps we obtained after removing these components, to be compared to the maps shown in Fig. \ref{fig:l1448n_maps}).
We show that adding a Gaussian structure to the Plummer envelope model centered on NB1 does not improve the modeling of the 231~GHz visibility profile (see Table \ref{table:popg-fits} and Figure \ref{fig:l1448nb1_bothmodels}).
Regarding the 94 GHz visibility profile centered on NB1, our analysis suggests that an additional Gaussian component improves the modeling: the F value computed between the best-fit Pl and PG models is 15, while the critical value (corresponding to a probability 0.3\% that the Plummer+Gaussian model is better only because it includes two more free parameters) is 8 in this case. 
None of the models for the 94\,GHz visibility profile is considered satisfactory, however, since none of their reduced $\chi^2$ values drop below 3. 
The discrepancy between the 231 GHz and 94 GHz models might either arise because the 94 GHz observations do not separate the two millimeter continuum components L1448-NB1/NB2 and detect the additional component associated with L1448-NB2 as an additional Gaussian, while we could remove L1448-NB2 from the 231GHz observations that do separate the two components.
It might also suggest that an additional Gaussian component is indeed present in the system, but is not centered on the main continuum source L1448-NB1. If it is slightly shifted toward the western source L1448-NB2, this shift would not greatly affect the 94 GHz visibility profile, which is insensitive to asymmetries at scales $<1\arcsec$, while our analysis of the 231 GHz visibility profile centered on the main millimeter source L1448-NB1 would see this circumbinary structure as an asymmetry. Strengthening this hypothesis, the 231 GHz PdBI visibility profile shows oscillations in the circular bins in the 231 GHz visibility profile at baselines > $100 k\lambda$. This suggests that there are scales at which the spatial distribution of the emission is not circularly symmetric around the chosen phase center, here L1448-NB1. However, even if we use the equatorial plane visibilities, we still find that the best-fit models are the Plummer-only models.

\begin{figure}
\begin{center}
\includegraphics[trim={0cm 0cm 0cm 0cm},clip,width=0.97\linewidth]{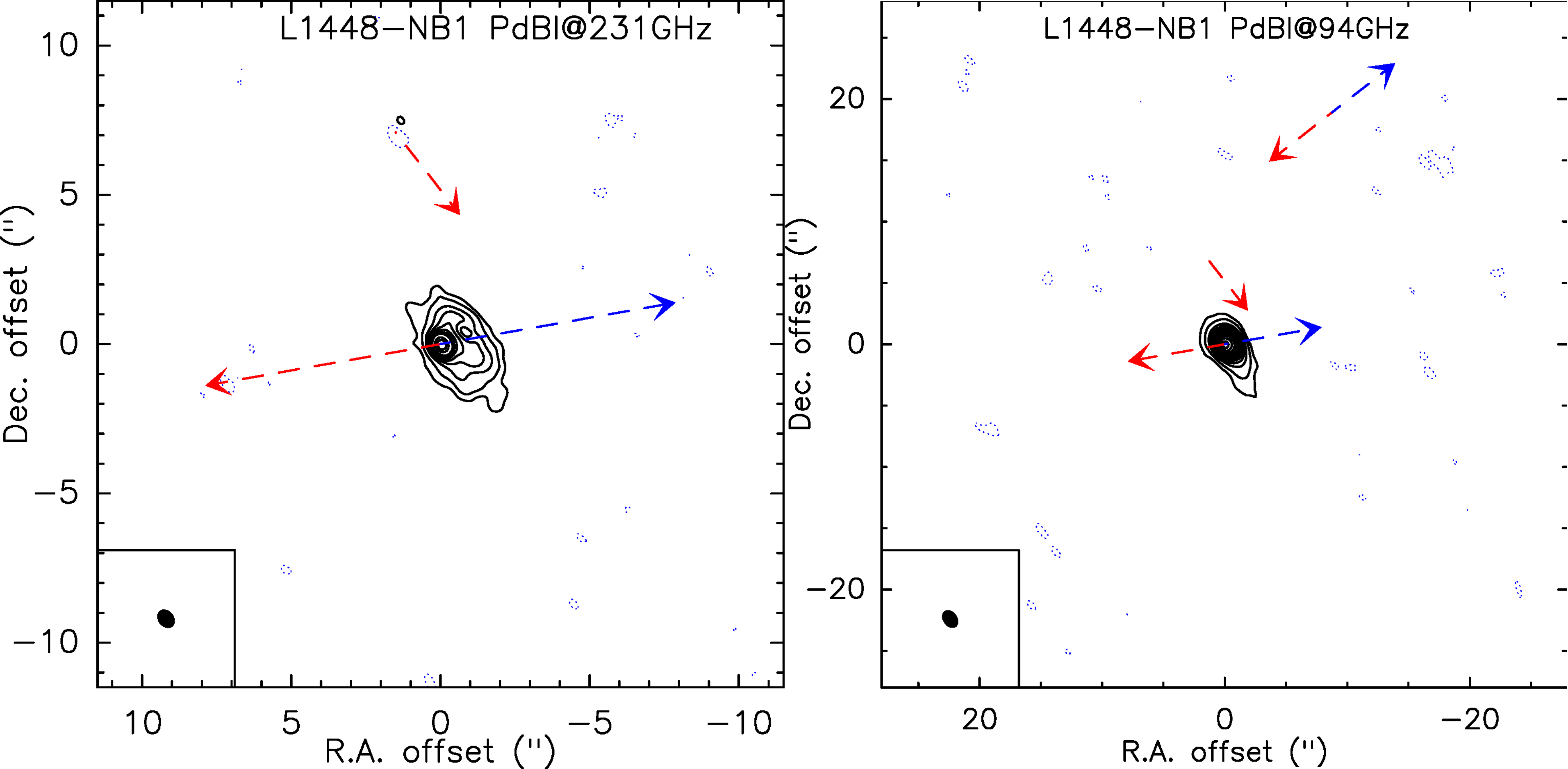}
\caption{1.3~mm (231 GHz) and 3.3~mm (94 GHz) PdBI dust continuum emission maps of L1448-NB1 after subtracting the secondary components. The contours are the same as in Figure \ref{fig:l1448n_maps}.
 \label{fig:l1448nb1-onlymain}
 }
 \end{center}
 \end{figure}

%
%
\item{Centered on L1448-NB2:}
Recent ALMA observations by \cite{Tobin16b} have shown a continuum structure that resembles a disk structure. The authors suggested that it might be centered on L1448-NB2 (also called IRS3B-a in their paper): they argued that although NB2 is not the strongest millimeter source in the system, its mass dominates, and hence the Keplerian motions are centered around the NB2 secondary source. 
Following these results, we wished to test the hypothesis of an additional disk component centered on L1448-NB2 that would be missed by our analysis, which is centered on the main millimeter source (NB1).
We shifted the phases of the PdBI visibilities so that they were centered on the L1448-NB2 component and modeled the continuum visibilities toward L1448-NB2. To do this, we had to remove the brightest millimeter source L1448-NB1 from the visibility tables because otherwise its flux dominated (we subtracted a model point-like source with a flux of the peak flux of L1448-NB1 at the position of L1448-NB1). Centering the Plummer envelope model on NB2 improves the minimization, but for both the 94 GHz and the 231 GHz data, the best-fit models are the Plummer-only models (see models Pl and PG in Table \ref{table:l1448nb2}).
The dust continuum emission in L1448-NB is not circularly symmetric: if we only use the equatorial plane visibilities, however, we find that the best-fit models are still the Plummer-only models.

%
%
\item{Centered at the barycenter:}
As an ultimate test of our modeling, we tested the possibility that a circumbinary disk-like additional structure might surround the two main millimeter sources L1448-NB1 and NB2. We removed the point-like contributions from the three protostars L1448-NA, L1448-NB1, and L1448-NB2, which produced profiles of the visibility amplitudes of the remaining underlying structure that we phase-shifted to be centered in between the two millimeter sources. 
Overall, the continuum emission around the L1448-NB1/NB2 system is only satisfactorily modeled at 94 GHz by either a Plummer envelope or an envelope with an additional Gaussian component (see Table \ref{table:l1448nbcent}). The PG model is better in this case, however, and points toward the presence of a $\sim 1.1\arcsec$ (260 au radius) additional structure with a flux of 50 mJy at 94\,GHz, which centered in between the NB1 and NB2 protostars.
The strong asymmetries of the continuum emission at 231 GHz when the two strongest sources in the map are excluded preclude us from performing a more advanced modeling of the PdBI continuum visibilities to further explore the nature of the circumbinary structure that emits in the millimeter continuum. 

\end{itemize}

\begin{figure}[!h]
\centering
\includegraphics[trim={0.5cm 2cm 2cm 2cm},clip,width=0.97\linewidth]{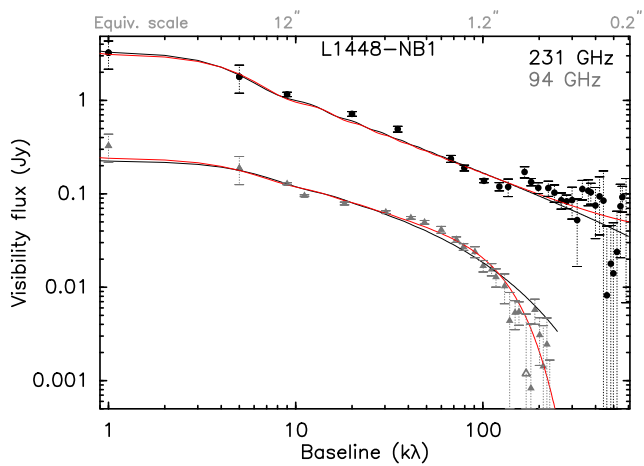}
\caption{Visibility profiles of the 231 GHz (black dots) and 94 GHz (gray triangles) continuum emission from L1448-NB1. Open symbols are used when the visibility real part has a negative value (since absolute values are shown in the log-log plot). In each visibility curve, we show both the best Plummer-only model (Pl, black curve) and the best Plummer+Gauss (PG, red curve) model. In the case of L1448-NB1, the F-test suggests that the PG model for the 231 GHz profile does not perform statistically better than the Plummer-only (Pl) model (see Table \ref{table:popg-fits} for more information on the two models). For the 94\,GHz profile, the PG model that includes a $0.94\arcsec$ Gaussian component performs better than the Pl model, although none of the two models does satisfactorily reproduce the visibility profile (reduced chi square $> 3$ in both cases).
}
\label{fig:l1448nb1_bothmodels}
\end{figure}

We conclude that while the individual protostars NB1 and NB2 do not harbor individual disk-like structures that can be resolved with our PdBI observations, they might both be embedded within a candidate circumbinary disk-like structure at scales $\sim 200$ au. The nature of this additional structure needs to be investigated further, especially since its disk nature, suggested in \cite{Tobin16b}, is questioned because it does not seem to be peaking on any of the two bright millimeter sources (that are assumed to contain most of the mass), and no indication that the material in this structure is rotationally supported has been found so far (a clear velocity gradient is detected in CALYPSO observations by Maret et al. in prep and Gaudel et al. in prep, but it is not well reproduced by Keplerian rotation). It is possible that the structure traces tidal arms that are created by a differential gravitational potential due to relative motions of the multiple components.

\end{itemize}

\subsection{L1448-C}
\label{section:appl1448c}

\begin{figure}[!h]
\centering
\includegraphics[trim={0.5cm 2cm 2cm 2cm},clip,width=0.99\linewidth]{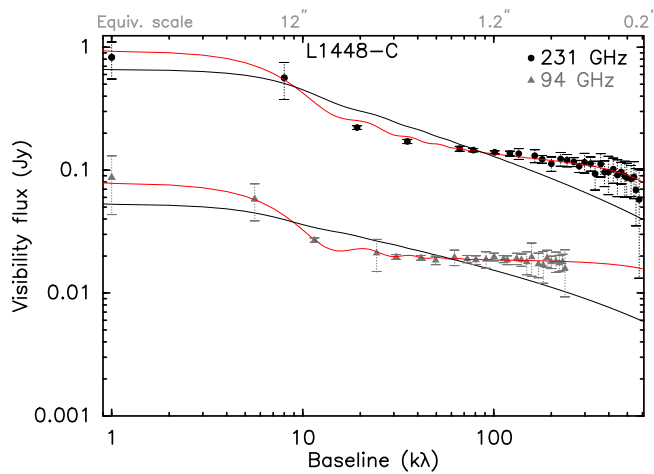}
\caption{Visibility profiles of the 231 GHz (black dots) and 94 GHz (gray triangles) continuum emission from L1448-C. In each visibility curve, we show both the best Plummer-only model (Pl, black curve) and the best Plummer+Gauss (PG, red curve) model. The PG model that includes a marginally resolved $0.16\arcsec$-FWHM Gaussian component reproduces the 231 GHz profile better than the Plummer-only (Pl) model (see Table \ref{table:popg-fits} for more information on the two models). Moreover, the PG model that includes an unresolved additional Gaussian component reproduces the 94 GHz visibility profile better.}
\label{fig:l1448c_bothmodels}
\end{figure}

\begin{itemize}
\item{\it{Single-dish constraints}}\\
From their IRAM-30m  MAMBO observations, \citet{Motte01a} found the following fluxes for the L1448-C envelope: F$_{\rm{peak}}=620$ mJy and F$_{\rm{int}}=910$ mJy in a 4200 au radius (assuming d=300 pc, which translates into a $14\arcsec$ radius on the map).
We scaled down the IRAM-30m fluxes, which are obtained at an observing frequency 243 GHz, to 94\,GHz and 231\,GHz using the PdBI spectral index at short baselines (see Table  \ref{table:continuum-pdbi-sd}).
We removed one additional component from the PdBI visibilities: L1448-CS, but it is out of the single-dish beam, therefore we did not remove its flux from the extrapolated flux. For the 231 GHz profile, we used an integrated flux $(910*(231/240)^{2.5}) = 827$ mJy, and a peak flux $F_{\rm{peak}}$563 mJy, at 8 k$\lambda$. We used a $20\%$ uncertainty on both extrapolated fluxes (this sets the upper- and lower-limit values that the total flux parameter is allowed to take in the fitting procedure).
At 94 GHz, the peak flux extrapolated from \citet{Motte01a} is consistent with the \citet{Looney03} BIMA flux at short baselines (65 mJy at 2 k$\lambda$). 
We used a $40\%$ uncertainty on both the integrated flux and the peak flux we used to model the 94 GHz profile.

\item{\it{Multiplicity}}\\ 
Our CALYPSO dust continuum emission maps are shown in Fig. \ref{fig:l1448c_maps}. The primary protostar is well detected at the center of the field, and we detect continuum emission associated with L1448-CS, $8\arcsec$ (2000 au) southeast, at both frequencies.
This source has previously been detected at millimeter wavelengths \citep{Jorgensen07a,Maury10,Hirano10}, and is associated with a mid-infrared source seen with {\it{Spitzer}} \citep{Jorgensen06}. 
This southern source is brighter than L1448-C in the mid-infrared, but much weaker in the millimeter and submillimeter bands: hence it is probably a more evolved source (Class I or older), with a separate envelope.

\item{\it{Candidate disk}}\\
The 231\,GHz visibility profiles and models for L1448-C are shown in Figure \ref{fig:l1448c_bothmodels}.
When all the dust continuum visibilities are used, the best-fit model to reproduce the continuum emission visibility profiles of L1448-C is the Plummer+Gaussian model for both frequencies (see Table \ref{table:popg-fits}). The best-fit PG model for the 231 GHz visibility profile includes a 130 mJy Gaussian source with an FWHM $0.16\arcsec$, which is marginally resolved by our observations. The parameters for this Gaussian component are reported as the candidate disk properties in Table \ref{table:disks-properties}. 
Similarly, the best-fit PG model to reproduce the 94 GHz visibilities includes an unresolved (FWHM$<0.3\arcsec$) Gaussian component with a flux of 18 mJy (see Table \ref{table:popg-fits}).
When only the equatorial visibilities are used (in a direction orthogonal to the jet axis position angle, i.e., selecting only the visibilities at PA $-107^{\circ}\pm30^{\circ}$ for L1448-C), we find that the best-fit model to reproduce the 231 GHz profile is still the Plummer+Gaussian model with a size and flux similar to the bestfit -model value using all visibilities (see models Pleq and PGeq in Table \ref{table:l1448c}). 
Hence, our analysis of the CALYPSO data suggests that a candidate disk is detected in L1448-C, and that it is marginally resolved at radii $\sim 40-50$ au.

\end{itemize}

\subsection{NGC1333 IRAS2A}
\label{section:appiras2a}

\begin{figure}[!h]
\centering
\includegraphics[trim={0.5cm 2cm 2cm 2cm},clip,width=0.99\linewidth]{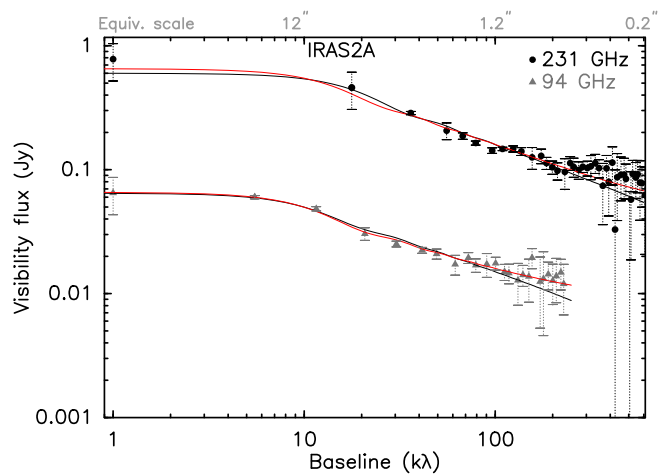}
\caption{Same as Fig. \ref{fig:l1448c_bothmodels} for IRAS2A. The Plummer-only model is the best-fit model to reproduce the 231 GHz profile. The PG model, which includes an unresolved additional Gaussian component, performs marginally better than the Pl model for the 94 GHz visibility profile.
}
\label{fig:iras2a_bothmodels}
\end{figure}

\begin{itemize}
\item{\it{Single-dish constraints}}\\
In their MAMBO observations, \citet{Motte01a} found a peak flux 875 mJy/beam in the $11\arcsec$ beam of the IRAM-30m, and they claimed that the source is almost unresolved.
Hence, we used this peak flux as the envelope-integrated flux for IRAS2A.
Using the 20 k$\lambda$ spectral index from our dual-frequency continuum visibilities (see Table \ref{table:continuum-pdbi-sd}), we scaled this value down to our frequencies to obtain the total envelope fluxes at 231\,GHz and 94\,GHz. We left the source size quite loose in our fitting and allowed values for $R_{\rm{out}}\sim6\pm4\arcsec$ since it is unresolved by the single-dish observations. The uncertainty on the total flux was set to $30\%$.

\item{\it{Multiplicity}}\\ 
In our CALYPSO maps, shown in Fig. \ref{fig:iras2a_maps}, a single protostar is detected. The two continuum sources IRAS2A2 and IRAS2A3, reported in \citet{Maury14,Codella14a}, are now shown to originate from an envelope structure that is detected by the interferometer: our improved dataset after self-calibration (which improved the rms noise and the imaging fidelity) allows us to recover the lower surface brightness emission that surrounds them, and they are found to reconnect with the main source envelope emission. This suggests that, similarly to what was proposed in \citet{Santangelo15} for IRAS4A, these sources represent the dust continuum emission from envelope structures that are detected by the interferometer when only a restricted range of spatial scales is sampled. They are therefore no longer considered as robust compact continuum emission components. Recent VLA \citep{Tobin15a} and ALMA observations (not yet published, Maury et al. in prep) of IRAS2A have revealed the presence of a secondary millimeter dust continuum source located $0.4\arcsec$ from the main millimeter source. Although the nature of this secondary component remains to be investigated, it is likely that IRAS2A is a close ($\leq 100-200$ au) binary system (two separate jets are detected, Podio \& CALYPSO, in prep.). This source is not separated from the main millimeter source in our PdBI maps, although a slight extension is detected at its location. Our highest angular resolution data for IRAS2A (configuration A of the PdBi at 231 GHz) was obtained as part of the pilot R068 project \citep{Maury10}: the lower sensitivity of these earlier observations (obtained before the installation of the WideX correlator) explains the non-detection of a separated compact source at the secondary position. To build robust estimates of the upper-limit MF from our sample, and considering that a secondary source is detected with ALMA and the VLA, we count IRAS2A as a binary system in our multiplicity analysis.

\item{\it{Candidate disk}}\\
The 231\,GHz visibility profiles and models are shown in Figure \ref{fig:iras2a_bothmodels} for IRAS2A.
When all the dust continuum visibilities are used, the best-fit model to reproduce the continuum emission visibility profiles of IRAS2A is the Plummer-only model for the 231 GHz profile (see main text and Table \ref{table:popg-fits}). The best-fit PG model for the 231 GHz visibility profile that satisfactorily reproduces the visibility profile but is not better than the Plummer-only model includes a 52 mJy Gaussian source that is unresolved with an FWHM $0.01\arcsec$ (the smallest size allowed in our fitting procedure). The upper limits on the parameters for this Gaussian component are reported as upper limits on the disk properties in Table \ref{table:disks-properties}. 
The PG model performs marginally better than the Pl model at reproducing the 94 GHz visibilities (a comparison of the two models produces an F-value of 11, while the critical value of F is 8. Above this, the probability that the PG model is better only by chance is $<0.3\%$ ). It includes an unresolved Gaussian with a flux of 9 mJy (see Table \ref{table:popg-fits}).
When only the equatorial visibilities are used (in a direction orthogonal to the jet axis position angle, i.e., selecting only the visibilities at PA $-65^{\circ}\pm30^{\circ}$), we find that the best-fit model is still the Plummer-only model for the 231 GHz profile (see models Pleq and PGeq in Table \ref{table:iras2a}). 
Our analysis of the CALYPSO data therefore suggests that no resolved disk-like emission is detected at scales 50-500 au.

We also tested the effect of removing a point source from the flux of the secondary source that is resolved with the ALMA 1.3 mm observations (14 mJy at 230 GHz, Maury et al. in prep) from the PdBI visibilities and shifting the phase center to the position of the main millimeter source at (03:28:55.569;31:14:36.952) to model the 231 GHz PdBI visibilities anew. For the sake of clarity and brevity, we do not report this model here, but the best-fit model is still the Plummer-only model, with very similar parameters as were found as best fits for the whole visibility dataset, which is reported in Table \ref{table:popg-fits}.
\end{itemize}

\subsection{SVS13B}
\label{section:appsvs13b}

\begin{figure}[!h]
\centering
\includegraphics[trim={0.5cm 2cm 2cm 2cm},clip,width=0.99\linewidth]{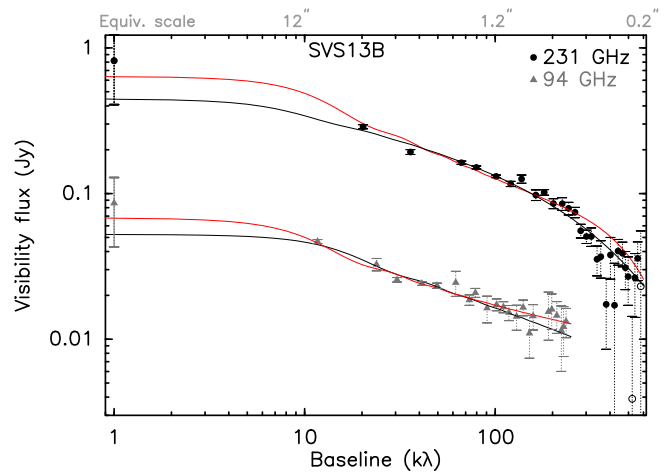}
\caption{Same as Fig. \ref{fig:l1448c_bothmodels} for SVS13B. The Plummer-only models are the best-fit models to reproduce the two profiles, although at 231 GHz, a more reasonable $(p+q)$ index is found in the PG model that includes a $0.19\arcsec$ FWHM Gaussian component (flux 80 mJy).
}
\label{fig:svs13b_bothmodels}
\end{figure}

\begin{itemize}
\item{\it{Single-dish constraints}}\\
Fluxes obtained using the MAMBO bolometer array on the IRAM-30m telescope are reported in \citet{Chini97}: F$_{\rm{peak}}=900$ mJy and F$_{\rm{int}}=1180$ mJy. However, \citet{Lefloch98} reported a peak flux for SVS13B of only 320 mJy/beam, with the same instrument and telescope.
We used the \citet{Chini97} peak flux as the envelope integrated flux in a source the size of the MAMBO beam ($11\arcsec$-FWHM), and we used the \citet{Lefloch98} value as the lower limit allowed for the total envelope flux for the minimization. We used the peak flux as an integrated flux in the MAMBO beam because contamination from the surrounding filament, and more especially, from the nearby SVS13A, precluded obtaining robust integrated fluxes for SVS13B alone at radii larger than $6\arcsec$.
We scale these fluxes down using the 20 k$\lambda$ spectral index from our dual-frequency continuum visibilities (see Table \ref{table:continuum-pdbi-sd}).
We also let the source size quite loose in the minimization procedure, allowing radii from 4$\arcsec$ to $14\arcsec$.
The uncertainty on the total envelope flux was set to $40\%$.

\item{\it{Multiplicity}}\\ 
SVS13B is in a large-scale multiple system (separate envelopes) with SVS13A, a Class I protostar located 3500 au away, and SVS13C, whose nature is as yet undetermined, and which lies 4500 au away. These three sources are detected in our 94 GHz dust continuum emission map shown in Fig. \ref{fig:svs13b_maps}. A fourth source, VLA3 detected previously at centimeter wavelengths with the VLA by \citet{Rodriguez99}, is also detected in both our maps (although not at the 10$\sigma$ level in our 231 GHz data). A millimeter counterpart to VLA3 was also previously tentatively detected in the BIMA maps of \citet{Looney00} as an extension from the SVS13A source at 110 GHz (they named it A2). The primary Class 0 protostar in the field, SVS13B, does not exhibit any sign of further multiplicity down to the scales of $\sim50$ au that are probed by our data. A single protostellar jet from SVS13B is detected in our CALYPSO maps of molecular line emission at a P.A. $+167^{\circ}$ (Podio \& CALYPSO in prep.).

\item{\it{Candidate disk}}\\
When all the dust continuum visibilities are used, the best-fit model to reproduce the continuum emission visibility profiles of SVS13B is the Plummer-only model for both frequencies (see main text and Table \ref{table:popg-fits}). However, the best Plummer-only model to reproduce the 231 GHz visibility profile is found to have $p+q=2.9$ which seems excessively high. Although we do not exclude that such a value might be physically possible, we stress that including a Gaussian component allows reducing this slope index to $p+q=2.5$, which is a more standard value that is expected in protostellar envelopes ( \citealt{Looney03} also found $p+q=2.4$ at short $<40$k$\lambda$ baselines in their 3mm BIMA data). This best-fit PG model for the 231 GHz visibility profile includes a 80 mJy Gaussian source with an FWHM $0.19\arcsec$: although this model is not statistically better than the Plummer-only model, we have to keep these facts in mind about SVS13B. In Table \ref{table:disks-properties} we report these values as upper limits for the candidate disk component. 
The F-test shows that the best-fit model to reproduce the 94 GHz visibilities is also the Plummer-only model (see Table \ref{table:popg-fits}, Fig. \ref{fig:svs13b_popgmodels} and Fig. \ref{fig:svs13b_bothmodels}), with a satisfactorily reduced $\chi^2$ value that is only slightly lower than that of the Plummer+Gaussian model, but has two free parameters less.
When only the equatorial visibilities are used (in a direction orthogonal to the jet axis position angle, i.e., selecting only the visibilities at PA $-103^{\circ}\pm30^{\circ}$ from the jet axis found in the CALYPSO data, Podio et al. in prep), we find that the best-fit model is still the Plummer-only model at 231 GHz with a high index $p+q=2.9$ (see models Pleq and PGeq in Table \ref{table:svs13b}). 
Our analysis of the CALYPSO data therefore suggests that the continuum emission from SVS13B is better reproduced by an envelope model down to scales 50 au, but we cannot exclude the presence of a $\simlt60$ au candidate disk which addition produces more reasonable $p+q$ values.

\end{itemize}

\subsection{NGC1333 IRAS4A1}
\label{section:appiras4a}

\begin{figure}[!h]
\centering
\includegraphics[trim={0.5cm 2cm 2cm 2cm},clip,width=0.99\linewidth]{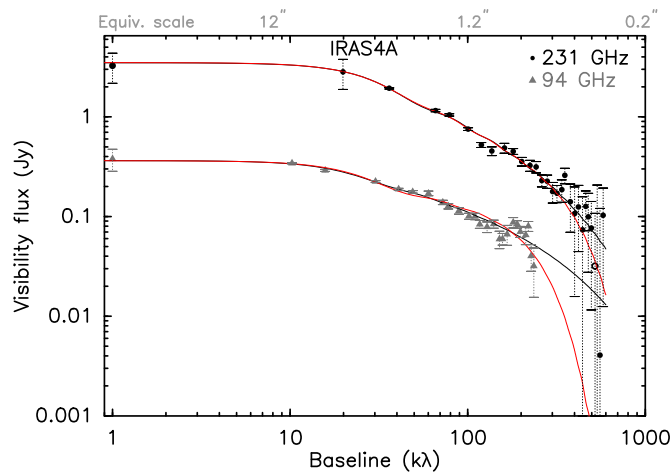}
\caption{Same as Fig. \ref{fig:l1448c_bothmodels} for IRAS4A1. The Plummer-only models are the best-fit models to reproduce the two visibility profiles.
}
\label{fig:iras4a_bothmodels}
\end{figure}

\begin{itemize}
\item{\it{Single-dish constraints}}\\
\citet{Motte01a} reported that the NGC1333 IRAS4A envelope is unresolved in their IRAM-30m observations ($12\arcsec$-FWHM beam). The peak flux is F$_{\rm{peak}}=4.1$ Jy in the IRAM-30m beam, which we used as the total envelope flux in a $6\arcsec$ source. We removed the contribution from IRAS4A2 and rescaled the MAMBO flux that was obtained at a central frequency of 243\,GHz to obtain extrapolated fluxes at both 94\,GHz and 231\,GHz (using the spectral index at large scales from our dual-frequency PdBI data, see Table \ref{table:continuum-pdbi-sd}). We used a $20\%$ uncertainty on the single-dish flux, and let the outer envelope radius vary between $3\arcsec$ and $9\arcsec$.

\item{\it{Multiplicity}}\\ 
The IRAS4A system is resolved by our CALYPSO observations: see Fig. \ref{fig:iras4a_maps}. The secondary protostar IRAS4A2 was extensively discussed by \citet{Santangelo15} and drives its own high-velocity jet. The continuum emission associated with the component IRAS4A3 reported in \citet{Santangelo15} disappears when robust weighting is used. This confirms the suggestion made in \citet{Santangelo15} that this traces a structure of dust continuum emission that is produced by the outflow interaction with the envelope, and is not a compact component associated with a true protostellar source.

\item{\it{Candidate disk}}\\
The models and visibility profiles of IRAS4A1 are shown in Figure \ref{fig:iras4a_bothmodels}.
When all the dust continuum visibilities are used, the best-fit model to reproduce the continuum emission visibility profiles of IRAS4A1 is the Plummer-only model for both the 231 GHz and 94 GHz visibility profiles (see main text and Table \ref{table:popg-fits}). The best-fit PG model for the 231 GHz visibility profile includes a 348 mJy Gaussian source with an FWHM $0.3\arcsec$, but does not perform better than the Plummer-only model (slightly larger reduced $\chi^2$, but two additional free parameters). The upper limits on the parameters for this Gaussian component are reported as upper limits on the disk properties in Table \ref{table:disks-properties}. 
When only the equatorial visibilities are used (in the direction of the C$^{17}$O(3--2) velocity gradient that is shown in the SMA map by \citet{Ching16} at the $4\arcsec$ core scale, i.e., selecting only the visibilities at PA $-62^{\circ}\pm30^{\circ}$), we find that the best-fit model is still the Plummer-only model for the 231 GHz profile (see models Pleq and PGeq in Table \ref{table:iras2a}). 
It is therefore clear from our analysis of the CALYPSO data that the continuum structure around NGC1333 IRAS4A1 can be well described with the inner part of the envelope: no resolved disk-like emission is detected at scales 50-500 au around A1.
Finally, we note that the brightness temperature obtained from the 231 GHz flux density of IRAS4A1 is 44K in the $\sim0.5\arcsec$ beam: this is about the temperature expected from $T_{dust} = 38\times L_{int}^{0.2}\times(r/100\rm{au})^{-0.4} \sim 60$K at $r\sim 100$ au. This suggests that the dust continuum emission might be partially optically thick at scales $\text{of about}$ the size of the synthesized beam (30-80) au. Using the optically thin 3.2\,mm flux density in the $\sim1\arcsec$ beam, we deduce a column density $0.1-4\times10^{26}$ cm$^{-2}$ (using $\kappa_{3.2mm}$=0.017 from the spectral index computed at 20 k$\lambda$ between the 1.3mm and 3.2mm visibility amplitudes).  We acknowledge that an optically thick disk could show a shallow flux decline at long baselines that might be confused with the inner part of an optically thin envelope. In our sample, the 231 GHz dust continuum emission is optically thin at all scales probed in all sources except for IRAS4A: the possible confusion caused by the partial optical thickness of the 231 GHz emission in IRAS4A is mitigated by the fact that (i) both visibility profiles at 231 GHz and 94 GHz show a steep decline at long baselines and (ii) the 94\,GHz dust continuum emission is optically thin and is well reproduced by an envelope model down to scales $50$au. We therefore argue that despite the partial optical thickness of the 231 GHz emission in IRAS4A1, our non-detection of a candidate disk structure in IRAS4A1 is robust.

We note that the neighboring IRAS4A2 probably includes a small candidate disk structure.  When we removed the A2 component to produce clean visibilities for A1 (with the aim of modeling the A1 structure), we realized that our PdBI data are best modeled when a circular disk component of diameter $0.7\arcsec$ and flux 430 mJy is used at both 231 GHz and 221 GHz independently for A2: this suggests that a disk-like component of radius $0.3-0.5\arcsec$ is associated with IRAS4A2. This is in agreement with observations by \citet{Choi07, Choi10}, who showed maps of the blueshifted and redshifted emission of the NH$_{3}$(3,3) line around IRAS4A2, which they interpreted as tracing the rotation of a disk at PA $\sim109^{\circ}$, while no such rotation is detected around IRAS4A1

\end{itemize} 

\subsection{NGC1333 IRAS4B1}
\label{section:appiras4b}

\begin{figure}[!h]
\centering
\includegraphics[trim={0.5cm 2cm 2cm 2cm},clip,width=0.99\linewidth]{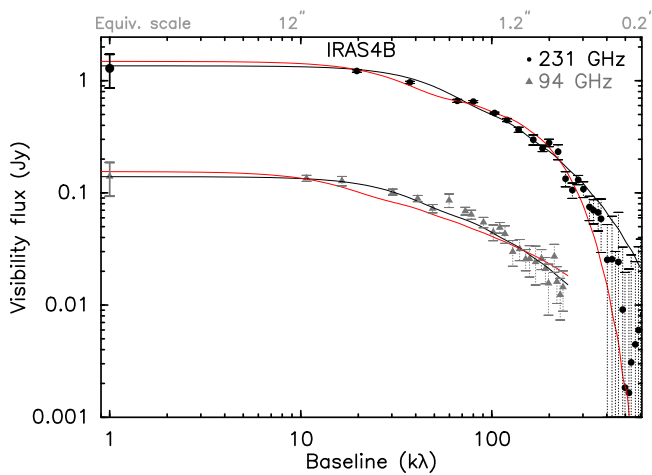}
\caption{Same as Fig. \ref{fig:l1448c_bothmodels} for IRAS4B1. The PG model that includes a $0.53\arcsec$ Gaussian source (flux 645 mJy) reproduces the 231\,GHz visibility profile better, while the 94\,GHz visibility profile is satisfactorily reproduced by the Plummer-only model.
}
\label{fig:iras4b_bothmodels}
\end{figure}

\begin{itemize}
\item{\it{Single-dish constraints}}\\
The IRAM-30m observations of the protostellar envelope with the MAMBO camera at 243 GHz \citep{Motte01a} suggest that the IRAS4B envelope is unresolved in the IRAM-30m $11\arcsec$ beam, with a peak flux 1470 mJy/beam. We scaled this flux density down from 243 GHz to 231 GHz and 94\,GHz using the PdBI spectral index at 20 k$\lambda$ (see Table \ref{table:continuum-pdbi-sd}). However, the 94\,GHz extrapolated flux is lower than the flux obtained with our PdBI observations at 10 k$\lambda$, so that we used the \citet{Looney00} BIMA flux at 108\,GHz (200 mJy at 2-8 k$\lambda$) to extrapolate an envelope flux at 94 GHz. We also note that these authors find a best-fit envelope size of $\sim$2000 au using a distance of 350pc (i.e., $6\arcsec$ radius). This confirms that the envelope is mostly unresolved in the IRAM-30m beam. We recovered the entire single-dish flux at 20 k$\lambda$ in our observations at 231 GHz, which suggests that the envelope of IRAS4B is very compact and is fully probed by the CALYPSO observations
We used a $40\%$ uncertainty on the single-dish fluxes, and let the envelope outer radius vary between $3\arcsec$ and $9\arcsec$.

\item{\it{Multiplicity}}\\
The secondary source IRAS4B2 (also called IRAS 4BE, IRAS 4B$^{\prime}$, IRAS4C, and IRAS 4BII, see \citealt{Looney00}) is detected $10\arcsec$ east (see Fig. \ref{fig:iras4b_maps}). It is outside of the IRAS4B1 envelope.
Our PdBI CALYPSO data do not detect COM emission in this secondary source \citep{DeSimone17}, and IRAS4B2 is not detected at 70$\mu$m in the Herschel HGBS maps \citep{Sadavoy14}. This suggests that it is either a very low luminosity source or a very young protostar (candidate first hydrostatic core). The SMA MASSES survey (Lee et al. 2016) and CARMA TADPOL survey (Hull et al. 2014) have recently suggested that some high-velocity $^{12}$CO emission could trace outflowing gas associated with this secondary millimeter source.
Moreover, our CALYPSO observations detect high-velocity blue-shifted SO emission originating from IRAS4B2 (Podio \& CALYPSO collaboration, in prep) at P.A. $\sim -99^{\circ}$. Since B2 is not detected in the infrared, it cannot currently be characterized well enough to firmly establish a robust protostellar nature. To build robust estimates of the upper-limit MF from our sample, we consider that IRAS4B is a separate-envelope system in our multiplicity analysis.

\item{\it{Candidate disk}}\\
When all the dust continuum visibilities are used, the best-fit model to reproduce the 231 GHz continuum emission visibility profiles of IRAS4B1 is the Plummer+Gaussian model (see main text and Table \ref{table:popg-fits}). It includes a 645 mJy Gaussian source with an FWHM $0.53\arcsec$. Although the modeling is not satisfactory because the power-law index (p+q) is unrealistically high (2.9), we report the parameters for this Gaussian component as the size and flux of the candidate disk in IRAS4B in Table \ref{table:disks-properties}. 
The PG and Pl models can both reproduce the 94 GHz visibilities, but they also feature an unrealistically high power-law index (2.9, see Table \ref{table:popg-fits}).
When only the equatorial visibilities are used (in a direction orthogonal to the jet axis position angle, i.e., selecting only the visibilities at PA $-103^{\circ}\pm30^{\circ}$ for IRAS4B1), we find the Plummer+Gaussian model performs slightly better at reproducing the 231\,GHz visibility profile, with a Gaussian size and flux similar to the best-fit model values found using all visibilities (see models Pleq and PGeq in Table \ref{table:iras4b}). 
We stress that the envelope of IRAS4B is very compact at both frequencies, and its spatial extent seems fully probed by our CALYPSO observations (see Figure \ref{fig:iras4b_bothmodels}). The IRAS4B visibility profiles are very flat, especially at 94 GHz: such a steep slope has previously been noted in the BIMA observations by \citet{Looney03} ($(p+q) \sim 2.8$ at 108 GHz between 5 and 90 k$\lambda$).
We also find it striking that half of the envelope dust continuum emission flux is included in a Gaussian-like structure: when a dust temperature of 50K is used, such a 231 GHz flux translates into a very high ''disk'' mass of $0.2-0.4\msol$ depending on the assumptions made on the dust emissivity. This raises questions on the nature of the dust continuum emission in IRAS4B, and we stress that were it a candidate disk, it should show some evident rotational signature that currently is not observed at $\sim$arcsecond scales \citep{Yen15b}.
We conclude that our analysis of the CALYPSO data suggests that a candidate disk-like structure might be detected at radii $\sim 120$ au in IRAS4B1, but our current modeling does not allow us to robustly conclude on the nature of the dust continuum emission recovered by the PdBI in this source.

\end{itemize} 

\subsection{IRAM04191}
\label{section:appiram04191}

\begin{figure}[!h]
\centering
\includegraphics[trim={0.5cm 2cm 2cm 2cm},clip,width=0.99\linewidth]{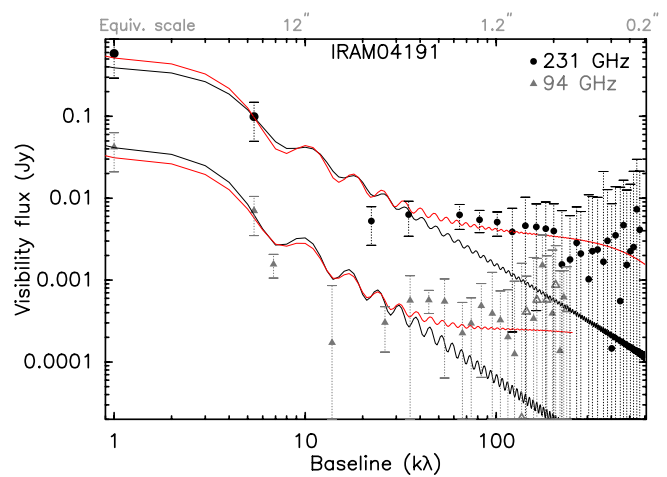}
\caption{Same as Fig. \ref{fig:l1448c_bothmodels} for IRAM04191. Open symbols are used when the visibility real part has a negative value (since absolute values are shown in the log-log plot). The red Plummer+Gaussian (PG) model that includes an unresolved $0.17\arcsec$-FWHM Gaussian component (see Table \ref{table:popg-fits}) is only statistically better than the black Plummer-only (Pl) model at 231 GHz. The 94 GHz profile can be well modeled with a Plummer-only model. The oscillations of the models are due to the Hankel transform of the power-law envelope model with a low index (p+q=1.4).
}
\label{fig:iram04191_bothmodels}
\end{figure}

\begin{itemize}
\item{\it{Single-dish constraints}}\\
In the IRAM-30m MAMBO observations by \citet{Motte01a} at 240 GHz, the envelope is resolved with an integrated flux 650 mJy in a 4200 au radius. The peak flux is 110 mJy in the $11\arcsec$ IRAM-30m beam. 
We scaled these fluxes down from 243 GHz to 94\,GHz and 231\,GHz using the PdBI spectral index at 20 k$\lambda$ (see Table \ref{table:continuum-pdbi-sd}).
We used a $20\%$ uncertainty on the single-dish fluxes, and let the envelope outer radius vary between $20\arcsec$ and $40\arcsec$.

\item{\it{Multiplicity}}\\
The secondary source reported by \citet{Chen12} is not detected in our dust continuum maps, and is probably due to a deconvolution artifact in their maps: IRAM04191 is single at all scales probed with the CALYPSO data (50-3000 au: see Fig. \ref{fig:iram04191_maps}).

\item{\it{Candidate disk}}\\
The low S/N of the binned visibilities from our PdBI dataset for IRAM04191 makes it difficult to robustly model them in detail.
However, the envelope intensity radial distribution of IRAM04191 is quite well characterized at scales $>2\arcsec$ (see, e.g., \citealt{Motte01a, Belloche04} ), and it is possible to extrapolate the outer envelope properties to establish constraints on the maximum disk-like component (i.e., maximum radius and flux) that can be added while reproducing the dust continuum interferometric fluxes obtained in IRAM04191.
We show that the PG model that includes an unresolved Gaussian component (FWHM $\sim 24$ au) is better than the Pl model at 231 GHz (see Table \ref{table:iram04191}), while the 94 GHz profile can be well modeled with a Plummer-only model.
The flat profile of the 231 GHz dust continuum emission observed at baselines $>300 k\lambda$ strongly suggest that an unresolved candidate disk is responsible for the dust continuum emission at scales $<50$au, but our limited S/N cannot provide a robust characterization of the protostellar disk size in IRAM04191 (all error bars on the fluxes overlap in the baseline range 300-600 k$\lambda$).
Although the current CALYPSO observations do not allow us to firmly establish the size of the disk-like component in IRAM04191, the 231 GHz visibility curve (see Fig. \ref{fig:iram04191_bothmodels}) shows that we should be able to detect an additional disk-like component that would have a half-maximum flux of 4 mJy at 300 k$\lambda$: we report this maximum size and flux for the additional disk-like component, for example, $<57$ au and flux $<4 \pm 1$ mJy, in Table \ref{table:disks-properties} and report the disk candidate as unresolved at these scales in Table \ref{table:disks-properties}. 

\end{itemize} 

\subsection{L1521F}
\label{section:appl1521f}

\begin{figure}[!h]
\centering
\includegraphics[trim={0.5cm 2cm 2cm 2cm},clip,width=0.99\linewidth]{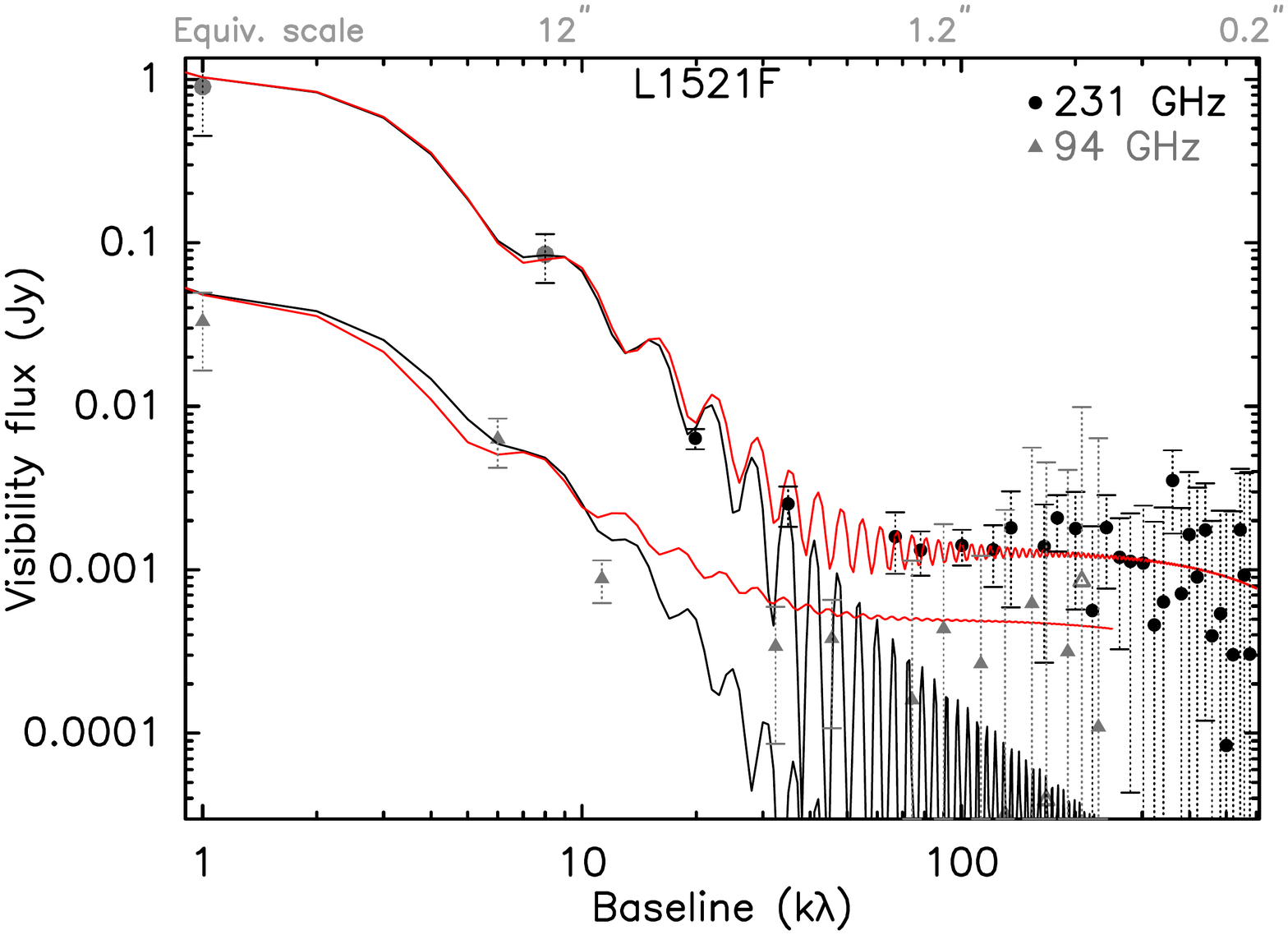}
\caption{Same as Fig. \ref{fig:l1448c_bothmodels} for L1521F. Open symbols are used when the visibility real part has a negative value (since absolute values are shown in the log-log plot). The red Plummer+Gaussian (PG) model that includes an unresolved $0.13\arcsec$ FWHM Gaussian component (see Table \ref{table:popg-fits}) is statistically better than the black Plummer-only (Pl) model at 231 GHz. The 94 GHz profile can be well modeled with a Plummer-only model.
}
\label{fig:l1521f_bothmodels}
\end{figure}

\begin{itemize}
\item{\it{Single-dish constraints}}\\
In the IRAM-30m MAMBO observations by \citet{Tokuda16} at 240 GHz, the envelope is resolved with an integrated flux 1.0 Jy in a $30\arcsec$ radius. The peak flux is $\sim 90$ mJy in the $11\arcsec$ IRAM-30m beam \citep{Crapsi04}. 
We scaled these fluxes down from the native 243 GHz observing frequency to 94\,GHz and 231\,GHz using the PdBI spectral index at 20 k$\lambda$ (see Table \ref{table:continuum-pdbi-sd}).
We used a $30\%$ uncertainty on the single-dish fluxes, and let the envelope outer radius vary between $20\arcsec$ and $40\arcsec$.

\item{\it{Multiplicity}}\\
In our dust continuum maps, L1521F is a single source (see Fig. \ref{fig:l1521f_maps}).
We detect a southeast extension in the 231\,GHz map (see Fig. \ref{fig:l1521f_maps}, also called MMS2 in \citealt{Tokuda16}), however: this emission cannot be modeled with a compact component from our dust continuum emission visibilities, and it does not have an infrared counterpart.
We therefore suggest that MMS2, lying in the outflow cavity, is probably a dust emission feature from a structured cavity.

\item{\it{Candidate disk}}\\
The low S/N of the binned visibilities from our PdBI dataset for L1521F make it difficult to robustly model them in detail, but it is possible to extrapolate the outer envelope properties to establish constraints on the maximum disk-like component (i.e., maximum radius and flux) that can be added while reproducing the dust continuum interferometric fluxes obtained in L1521F.
The 231 GHZ profile is best modeled using a PG model that includes an unresolved Gaussian component (FWHM $0.13\arcsec$, see Fig. \ref{fig:l1521f_bothmodels}): this suggests that a candidate disk of radius $\sim <20$ au and flux $\sim 1.3\pm0.4$ mJy may be detected in L1521F.
However, as in the case of IRAM04191, we are limited by the S/N $<3$ at long ($>300$ k$\lambda$) baselines for L1521F, so that the size of the Gaussian component that can be added to the Plummer envelope model is subject to high uncertainties. 
To remain robust in our characterization of a potential protostellar disk in L1521F, we report in Table \ref{table:disks-properties} that a candidate disk-like component might contribute a flux 1.3 mJy at scales $< 57$ au ($300$ k$\lambda$). We also note that 0.87mm ALMA observations by \citet{Tokuda17} reported the detection of a small 10 au disk in L1521F, which is consistent with our upper-limit size and flux at 1.3\,mm.

\end{itemize} 

\subsection{L1527}
\label{section:appl1527}

\begin{figure}[!h]
\centering
\includegraphics[trim={0.5cm 2cm 2cm 2cm},clip,width=0.99\linewidth]{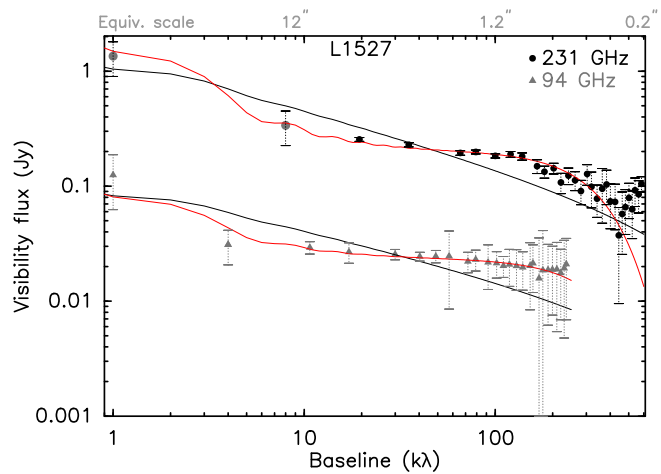}
\caption{Same as Fig. \ref{fig:l1448c_bothmodels} for L1527. The two visibility profiles are better reproduced with models that include an additional Gaussian source whose FWHM is $0.4\arcsec$ at 231 GHz. 
}
\label{fig:l1527_bothmodels}
\end{figure}

\begin{itemize}
\item{\it{Single-dish constraints}}\\
In the IRAM-30m observations by \citet{Motte01a} with MAMBO at 240 GHz, the L1527 envelope is resolved with an integrated flux 1500 mJy in a 4200 au radius. The peak flux is 375 mJy in the $11\arcsec$ IRAM-30m beam. 
We scaled these fluxes down from 243 GHz to 94\,GHz and 231\,GHz using the PdBI spectral index at 20 k$\lambda$ (see Table \ref{table:continuum-pdbi-sd}).
We used a $20\%$ uncertainty on the extrapolated single-dish fluxes, and let the envelope outer radius vary between $25\arcsec$ and $35\arcsec$.

\item{\it{Multiplicity}}\\
In our dust continuum maps, L1527 is a single source (see Fig. \ref{fig:l1527_maps}).

\item{\it{Candidate disk}}\\
The models and the visibility profiles of L1527 are shown in Figure \ref{fig:l1527_popgmodels} in the text and in Figure \ref{fig:l1527_bothmodels} here.
When all the dust continuum visibilities are used, the best-fit model to reproduce the continuum emission visibility profiles of L1527 is the Plummer+Gaussian model for both frequencies (see main text and Table \ref{table:popg-fits}). The best-fit PG model for the 231 GHz visibility profile includes a 215 mJy Gaussian source with FWHM $0.4\arcsec$. The parameters for this Gaussian component are reported as the candidate disk properties in Table \ref{table:disks-properties}. 
Similarly, the best-fit PG model to reproduce the 94 GHz visibilities includes a marginally resolved Gaussian component with FWHM$=0.3\arcsec$ and flux 23 mJy (see Table \ref{table:popg-fits}).
When only the equatorial visibilities are used (in a direction orthogonal to the jet axis position angle, i.e., selecting only the visibilities at PA $19^{\circ}\pm30^{\circ}$ for L1527), we find that the Plummer+Gaussian model performs better at reproducing the 231\,GHz visibility profile, with a Gaussian size and flux similar to the best-fit model values found using all visibilities (see models Pleq and PGeq in Table \ref{table:l1527}). 
Our analysis of the CALYPSO data therefore suggests that a candidate disk-like structure is detected at radii $\sim 50-70$ au in L1527.

\end{itemize} 

\subsection{Serpens Main S68N}
\label{section:appserpms68n}

\begin{itemize}
\item{\it{Single-dish constraints}}\\
SerpM-S68N \citep{McMullin94} is also known as SMM 9 \citep{Casali93} or Ser-emb 8 \citep{Enoch11}, and is located in the Serpens Main cluster.  
Two protostars are located within the FWHM of the Bolocam core Ser-emb 8 associated with S68N, but the envelope associated with S68N can be clearly identified in the IRAM-30m observations by \citet{Kaas04} at better angular resolution. Based on the MAMBO map, we estimate an integrated flux 1030 mJy in a $15\arcsec$ radius area, and a peak flux 550 mJy in the $11\arcsec$ IRAM-30m beam.
We scaled these fluxes down from the MAMBO central frequency of 243\,GHz to 94\,GHz and 231\,GHz using the PdBI spectral index at 20 k$\lambda$ (see Table \ref{table:continuum-pdbi-sd}).
We used a $15\%$ uncertainty on the extrapolated single-dish fluxes, and let the envelope outer radius vary between $8\arcsec$ and $22\arcsec$.

\item{\it{Multiplicity}}\\
SerpM-S68N (S68N in the following) is single in the 231\,GHz PdBI primary beam area ($21\arcsec$ FWHM, see Fig. \ref{fig:serps68n_maps}).
Our PdBI 94\,GHz map indicates a compact source at the position of S68N, and two additional sources $\sim 12\arcsec$ and $\sim 20\arcsec$ to the northeast. This is consistent with the CARMA 230\,GHz sources found by \citet{Enoch11}. 
While the region is crowded and it is difficult to robustly associate sources at different wavelengths with different resolutions, it seems that the secondary source S68Nb has no {\it{Spitzer}} MIPS counterpart at 24$\mu$m, while S68Nc has an associated {\it{Spitzer}} IRAC source at 8$\mu$m \citep{Enoch09,Harvey07}: it is classified as a Class I protostar.
The three sources (S68N, S68Nb, and S68Nc) are embedded in a common parsec-scale filamentary structure seen in the single-dish maps of the dust continuum emission, but they have separate envelopes.

\item{\it{Candidate disk}}\\
When all the dust continuum visibilities are used, the Plummer model is as good a model as the PG model to reproduce the continuum emission visibility profiles of S68N at both frequencies (see main text and Table \ref{table:popg-fits}). The best-fit PG model for the 231 GHz visibility profile includes an unresolved Gaussian source with an FWHM $0.11\arcsec$ and flux 28 mJy. The parameters for this Gaussian component are reported as the upper-limit values for any disk in S68N in Table \ref{table:disks-properties}, although our observations do not hint at the presence of a disk in S68N. 
When only the equatorial visibilities are used (in a direction orthogonal to the jet axis position angle, i.e., selecting only the visibilities at PA $45^{\circ}\pm30^{\circ}$ for S68N), we find that the Plummer model performs as well as the Plummer+Gaussian model to reproduce the 231\,GHz visibility profile, with a Gaussian size and flux similar to the best-fit model values found using all visibilities (see models Pleq and PGeq in Table \ref{table:l1527}). 
Our analysis of the CALYPSO data therefore suggests that no disk-like structure is resolved in S68N at our spatial resolution, and any disk can only be present at radii $< 50$ au.

\end{itemize} 

\begin{figure}[!h]
\centering
\includegraphics[trim={0.5cm 2cm 2cm 2cm},clip,width=0.99\linewidth]{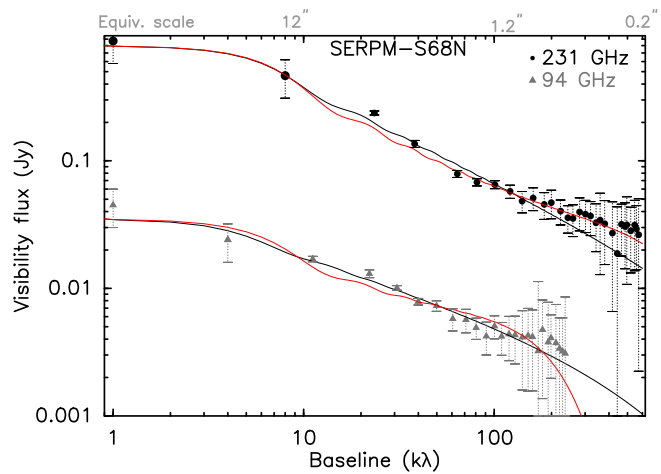}
\caption{Same as Fig. \ref{fig:l1448c_bothmodels} for S68N. The Plummer-only models are the best-fit models to reproduce both visibility profiles.
}
\label{fig:s68n_bothmodels}
\end{figure}

\subsection{Serpens Main SMM4}
\label{section:appserpmsmm4}

\begin{itemize}
\item{\it{Single-dish constraints}}\\
The envelope associated with SerpM-SMM4 (SMM4 in the following) can be clearly identified in the IRAM-30m observations by \citet{Kaas04}: based on the MAMBO map, we estimate an integrated flux 2350 mJy in a $20\arcsec$ radius area (background subtracted), and a peak flux 1000 mJy in the $11\arcsec$ IRAM-30m beam. We removed the contribution from MM4b that is included in the $11\arcsec$ beam, and scaled these fluxes down from the MAMBO central frequency of 243\,GHz to 94\,GHz and 231\,GHz (using the PdBI spectral index at 20 k$\lambda$, see Table \ref{table:continuum-pdbi-sd}).
We used a $20\%$ uncertainty on the extrapolated single-dish fluxes, and let the envelope outer radius vary between $15\arcsec$ and $25\arcsec$.

\item{\it{Multiplicity}}\\
In both our dust continuum maps at 231\,GHz and 94\,GHz, we detect a secondary component, SMM4b, about $7\arcsec$ southwest of the strongest millimeter source (see Fig. \ref{fig:serpsmm4_maps}).
This is the first time that this secondary component is detected, and its nature is therefore unclear. However, it seems that the Herschel/PACS emission peaks toward SMM4b and not toward the main protostar. Moreover, a water maser was detected at (18:29:56.51, 01:13:11.6, equ. J2000) with the VLA, see \citet{Furuya03}: this position is closer to the position of MM4b than to the position of MM4a. Finally, the methanol outflow detected in \citet{Kristensen10} was not found to peak on the millimeter emission peak reported in the literature. Based on these pieces of evidence, we suggest that SMM4b is probably the driving source of the jet or outflow, while the more quiescent SMM4a dominates the millimeter dust continuum emission. A more detailed analysis of the jet properties and chemical content for these two sources will be proposed by Podio \& CALYPSO (in prep.) and Belloche \& CALYPSO (in prep.).

\item{\it{Candidate disk}}\\
When all the dust continuum visibilities are used, the best-fit model to reproduce the continuum emission visibility profiles of SMM4a is the Plummer+Gauss model for both frequencies (see main text and Table \ref{table:popg-fits}). The best-fit PG model for the 231 GHz visibility profile includes a Gaussian source with an FWHM $0.7\arcsec$ and flux 595 mJy. The parameters for this Gaussian component are reported as the values for the properties of the candidate disk in SMM4a in Table \ref{table:disks-properties}. 
Hence, our analysis of the CALYPSO data suggests that continuum emission that is not accounted for by circular-symmetric Plummer-like envelope models may trace a disk-like structure in SMM4a that is resolved with our observations with a radius $\sim 300$ au. If this continuum emission indeed traces a disk structure, the disk flux at 231 GHz accounts for $\sim30\%$ of the total envelope flux from single-dish observations (obtained within a $20\arcsec$ radius), which is very unusual for Class 0 protostars. Based on the internal luminosity of SMM4a in the HGBS (2\lsol, see Table 1), we would expect a dust temperature $\sim20-30 $K at the $200\,k\lambda$ scale, turning the 231 GHz disk flux into a disk mass > 0.3\msol. Such a high disk mass is surprising: a more complete analysis of SMM4a with 2D modeling of interferometric data that cover a larger spatial dynamic range is needed to better understand the nature of this source.

\end{itemize} 

\begin{figure}[!h]
\centering
\includegraphics[trim={0.5cm 2cm 2cm 2cm},clip,width=0.99\linewidth]{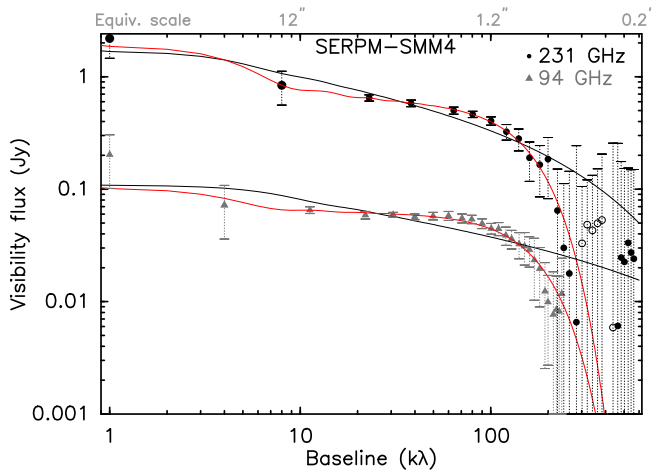}
\caption{Same as Fig. \ref{fig:l1448c_bothmodels} for SMM4. Open symbols are used when the visibility real part has a negative value (since absolute values are shown in the log-log plot). The PG models are the best-fit models to reproduce the two CALYPSO visibility profiles: at 231 GHz, the additional Gaussian component has an FWHM $0.7\arcsec$ and flux 595 mJy.
}
\label{fig:smm4_bothmodels}
\end{figure}

\subsection{Serpens South MM18}
\label{section:appserpsmm18}

\begin{itemize}
\item{\it{Single-dish constraints}}\\
The envelope associated with Serpens South MM18 (SerpS-MM18 in the following)  has been mapped with MAMBO on the IRAM-30m telescope \citep{Maury11}: based on the MAMBO map, we estimate an integrated flux 2505 mJy in a $18\arcsec$ FWHM source, and a peak flux 1376 mJy in the $11\arcsec$ IRAM-30m beam. We removed the contribution from MM18b and scaled these fluxes down from the MAMBO central frequency of 243\,GHz to 94\,GHz and 231\,GHz (using the PdBI spectral index at 20 k$\lambda$, see Table \ref{table:continuum-pdbi-sd}).
We used a $20\%$ uncertainty on the extrapolated single-dish fluxes, and let the envelope outer radius vary between $10\arcsec$ and $20\arcsec$.

\item{\it{Multiplicity}}\\ 
SerpS-MM18 is separated into two sources in our CALYPSO maps: MM18a the primary protostar (dominating the millimeter continuum emission) and a secondary  source MM18b, weaker and more compact, $10\arcsec$ to the southwest (see Fig. \ref{fig:serpsmm18_maps}). At a distance of 250 pc for Serpens South, these two sources are separated by 2600 au. We note that \citet{Ortiz15} argued that Serpens Main and W40  are at a same distance, about 430 pc, based on their VLA parallax measurements of seven sources in both clouds. While W40 and Serpens South seem to belong to the same extinction wall \citep{Straizys03}, it is not yet clear at which distance the Serpens South filament is located \citep{Konyves15}. For consistency with previous studies, we use here the value of 250 pc but acknowledge that the distance might be twice larger, and the physical separation between MM18a and MM18b could thus be up to 5000 au. MM18b was originally detected in \citet{Maury11}, then with CARMA by \citet{Plunkett13} at 3mm (CARMA-6) and with the VLA by Kern et al. (2016), who classified this source as a Class~I (source VLA$\_$13). More recently, it was detected with ALMA in Band 6 (\citealt{Plunkett18}, source serps33). 
MM18a drives a collimated jet (P.A. $+188^{\circ}$ for the blue lobe, Podio et al. in prep.), while MM18b is associated with outflowing gas that follows a cavity with a rather large opening angle (P.A. of the redshifted emission $\sim +8^{\circ}$, from our CALYPSO data).

\item{\it{Candidate disk}}\\ 
When all the dust continuum visibilities are used, the best-fit model to reproduce the 231\,GHz continuum emission visibility profile of SerpS-MM18a is the Plummer model that includes a Gaussian (see main text and Table \ref{table:popg-fits}), with an FWHM $0.13\arcsec$ and flux 76 mJy. At 94\,GHz, the Plummer-only model satisfactorily reproduces the visibility profile. Our modeling therefore suggests that a marginally resolved candidate disk might be detected in SerpS-MM18a, and the parameters of the Gaussian component that are included in the best-fit PG model for the 231\,GHz profile are reported as the upper-limit disk size and disk flux in Table \ref{table:disks-properties}. 

\end{itemize}

\begin{figure}[!h]
\centering
\includegraphics[trim={0.5cm 2cm 2cm 2cm},clip,width=0.99\linewidth]{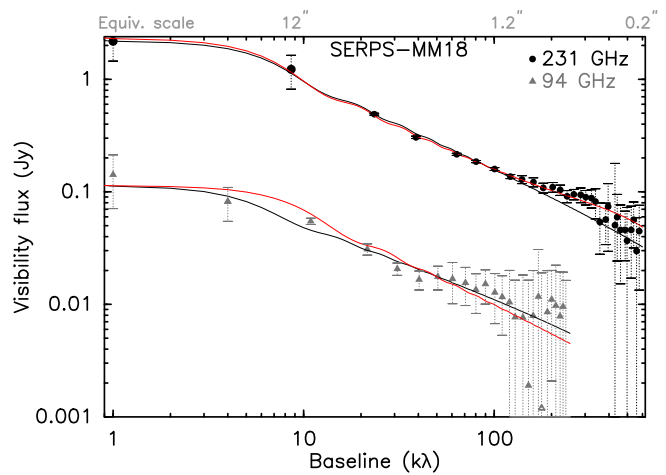}
\caption{Same as Fig. \ref{fig:l1448c_bothmodels} for SerpS-MM18. The PG model is the best-fit model to reproduce the 231 GHz visibility profile: it includes a marginally resolved additional Gaussian component with an FWHM $0.13\arcsec$ (flux 76 mJy). The 94 GHz profile can be well modeled with a Plummer-only envelope model.
}
\label{fig:ssmm18_bothmodels}
\end{figure}

\subsection{Serpens South MM22}
\label{section:appserpsmm22}

\begin{itemize}
\item{\it{Single-dish constraints}}\\
The envelope associated with Serpens South MM22 (SerpS-MM22 in the following) has been mapped with the MAMBO camera on the IRAM-30m \citep{Maury11}: based on the MAMBO map, we estimate an integrated flux 261 mJy in a $20\arcsec$ FWHM, and a peak flux 129 mJy in the $11\arcsec$ IRAM-30m beam. We scaled these fluxes down from the MAMBO central frequency of 243\,GHz to 94\,GHz and 231\,GHz (using the PdBI spectral index at 20 k$\lambda$, see Table \ref{table:continuum-pdbi-sd}).
We used a $40\%$ uncertainty on the extrapolated single-dish fluxes, and let the envelope outer radius vary between $10\arcsec$ and $25\arcsec$.

\item{\it{Multiplicity}}\\ 
SerpS-MM22 is single in our dust continuum maps (see Fig. \ref{fig:serpsmm22_maps}).

\item{\it{Candidate disk}}\\ 
When all the dust continuum visibilities are used, the Pl and PG models are satisfactory for both frequencies (see main text and Table \ref{table:popg-fits}), but the Plummer+Gauss models are statistically better (at 231\,GHz, the F value is 19, compared to a critical value of 10 above which the probability of a better minimization that is only due to the use of a model containing two more free parameters is $<0.3\%$). The PG model at 231\,GHz includes a Gaussian source with an FWHM $0.25\arcsec$ and flux 31 mJy.
At 94\,GHz, the PG model includes a marginally resolved Gaussian component similar to the one found in the 231\,GHz visibility profile (FWHM $0.31\arcsec$ and flux 3.2 mJy). The parameters of the Gaussian component included in the best-fit model for the 231\,GHz profile are reported as properties of the candidate disk in SerpS-MM22, in Table \ref{table:disks-properties}. 
Hence, our analysis suggests that a disk-like structure might be present in SerpS-MM22 at radii $\sim 62$ au. However, we stress that the low S/N\ of our CALYPSO visibility profiles for this low-luminosity source precludes us from concluding firmly: additional deeper observations of SerpS-MM22 are needed to fully characterize this candidate disk structure.

\end{itemize}

\begin{figure}[!h]
\centering
\includegraphics[trim={0.5cm 2cm 2cm 2cm},clip,width=0.99\linewidth]{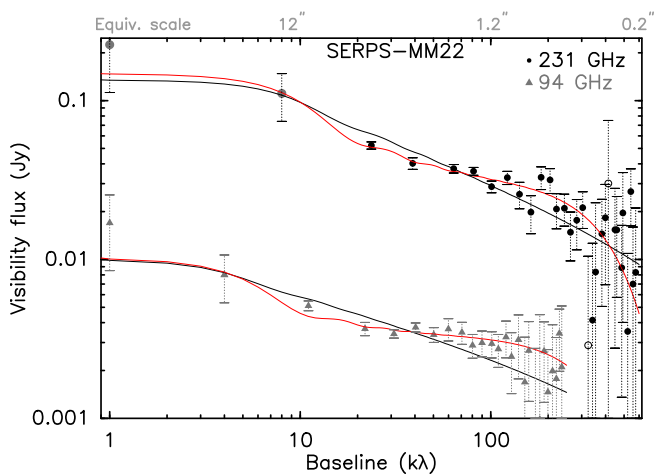}
\caption{Same as Fig. \ref{fig:l1448c_bothmodels} for SerpS-MM22. Open symbols are used when the visibility real part has a negative value (since absolute values are shown in the log-log plot). The red Plummer+Gaussian (PG) models are the best-fit models for the two CALYPSO visibility curves. The PG model at 231 GHz includes a $0.25\arcsec$ FWHM Gaussian component (see Table \ref{table:popg-fits}).
}
\label{fig:ssmm22_bothmodels}
\end{figure}

\subsection{L1157}
\label{section:appl1157}

\begin{itemize}
\item{\it{Single-dish constraints}}\\
\citet{Motte01a} IRAM 30m MAMBO observations at 240 GHz show an unresolved envelope with a peak flux 630 mJy/beam in the IRAM-30m beam $11\arcsec$ beam.
We used the dual-frequency PdBI spectral index at $20 k\lambda$ to extrapolate the flux from 240 GHz to 231 GHz.
The extrapolated flux is 569 mJy: since the envelope is unresolved in the IRAM-30m $11\arcsec$ beam, we used the IRAM-30m peak flux as envelope-integrated flux, and constrained the core angular size at $8\arcsec-16\arcsec$.
The uncertainties on the 231 GHz extrapolated fluxes are $\pm 30\%$. 
For the 94 GHz envelope fluxes, we used an integrated flux extrapolated from the shortest baseline CARMA flux in \citet{Kwon15}, 70 mJy, and a peak flux extrapolated from \citet{Motte01a} flux at 240\,GHz, 53 mJy at 4k$\lambda$, with estimated uncertainties $\pm 40\%$.

\item{\it{Multiplicity}}\\ 
We find no evidence of multiplicity in L1157 down to the $0.4\arcsec$ scales probed by our longest baselines. \citet{Kwon15} suggested that multiple jets originate from the L1157 inner envelope, but the detection of a unique resolved high-velocity jet with CALYPSO observations \citep{Podio16} rather suggests L1157 is indeed a single protostar at scales $70-1000$ au.

\item{\it{Candidate disk}}\\ 
The models that reproduce the visibility profile for L1157 are shown in Figure \ref{fig:l1157_popgmodels}.
When all the dust continuum visibilities are used, the best-fit model to reproduce the continuum emission visibility profiles of L1157 is the PG model for the 231\,GHz data (see main text and Table \ref{table:popg-fits}): it includes an unresolved $56\pm6$ mJy Gaussian source. The F-value obtained when the best-fit Pl and best-fit PG model are compared is 9 for the 231\,GHz dataset, which is slightly above the critical value (the critical value of F above which the probability of a coincidental better minimization due to two additional parameters is 8). 
The parameters for this Gaussian component are reported in Table \ref{table:disks-properties}. 
The best-fit model to reproduce the 94 GHz visibilities is the Pl model. 

When only the equatorial visibilities are used (in a direction orthogonal to the jet axis position angle, i.e., selecting only the visibilities at PA $+73^{\circ}\pm30^{\circ}$), we find that the PG model performs slightly better than the Pl model as well (see models Pleq and PGeq in Table \ref{table:l1157}).  
Our analysis of the CALYPSO data therefore suggests that the continuum structure around L1157 is dominated by emission from the inner part of the envelope, and an unresolved candidate disk only contributes at scales $<50$ au to the long-baseline 231\,GHz fluxes.
\end{itemize}

\begin{figure}[!h]
\centering
\includegraphics[trim={0.5cm 2cm 2cm 2cm},clip,width=0.99\linewidth]{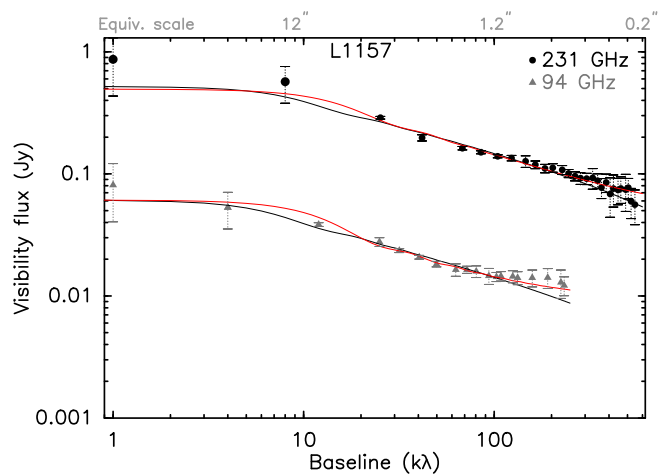}
\caption{Same as Fig. \ref{fig:l1157_bothmodels} for L1157. The red Plummer+Gaussian (PG) model that includes an unresolved Gaussian component (see Table \ref{table:popg-fits} for more informations on both models) is statistically better than the black Plummer-only (Pl) model at 231 GHz. At 94 GHz, the best-fit model is the Plummer-envelope model.
}
\label{fig:l1157_bothmodels}
\end{figure}

\subsection{GF9-2}
\label{section:appgf92}

\begin{figure}[!h]
\centering
\includegraphics[trim={0.5cm 2cm 2cm 2cm},clip,width=0.99\linewidth]{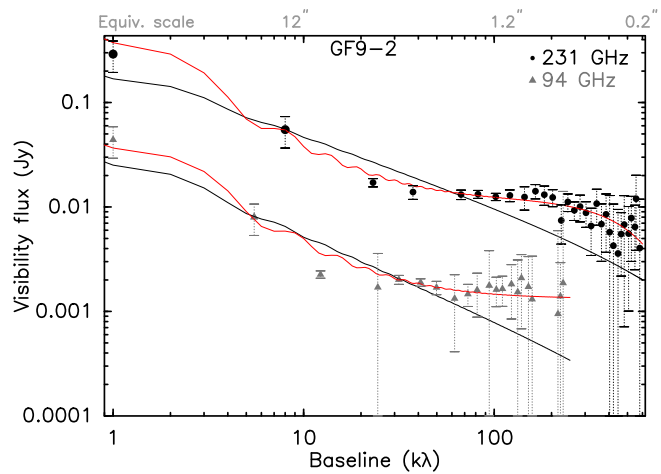}
\caption{Same as Fig. \ref{fig:l1448c_bothmodels} for GF9-2. The 231\,GHz visibility profile is shown with black dots and the 94\,GHz visibility profile with gray triangles. The red Plummer+Gaussian (PG) model that includes a Gaussian component ($0.18\arcsec$ FWHM for the 231\,GHz and unresolved at 94\,GHz) is statistically better than the black Plummer-only (Pl) model. See Tables \ref{table:popg-fits} and \ref{table:gf92} for more information on the models.
}
\label{fig:gf92_bothmodels}
\end{figure}

\begin{itemize}
\item{\it{Single-dish constraints}}\\
The envelope associated with GF9-2 is quite unconstrained, and this core has been little studied so far. It is also known as L1082C \citep{Bontemps96a,Caselli02c}. Located at 200 pc, it has been shown to be associated with an infrared source \citep{Furuya06, Furuya14}, and a maser was detected to be associated with the core \citep{Furuya01}. 
In Helmut Wiesemeyer's PhD thesis \citep{Wiesemeyer97}, where the NH$_3$ emission maps are used as a temperature probe at an arcminute scale, the envelope mass is estimated to be $\sim0.3$\msol.
We note that a virial analysis that uses the C$^{18}$O line width at similar scales suggests an envelope mass 1\msol, while \citet{Furuya06} found an envelope mass $0.6 \msol$ based on the dust continuum maps at 350 \mic within a 5400 au area. 

We used the IRAM-30m observations by \citet{Wiesemeyer97} with the MPIfR bolometer to constrain the large-scale envelope properties. Based on these data, we estimate an integrated flux 315 mJy in an area $20\arcsec$ in radius, and a peak flux 60 mJy/beam in a $12\arcsec$ beam.
We scaled these fluxes down from the original central frequency of 243\,GHz to 94\,GHz and 231\,GHz (using the PdBI spectral index at 20 k$\lambda$, see Table \ref{table:continuum-pdbi-sd}).
We used a $40\%$ uncertainty on the extrapolated single-dish fluxes, and let the envelope outer radius vary between $20\arcsec$ and $40\arcsec$.

\item{\it{Multiplicity}}\\ 
We find  no evidence of multiplicity in GF9-2 at the scales probed by our CALYPSO observations (50 to 5000 au).

\item{\it{Candidate disk}}\\ 
For the two visibility profiles at 231\,GHz and 94\,GHz, models that include an additional Gaussian component are statistically better at reproducing the intensity radial distribution.
Therefore we conclude that the GF9-2 core seems to include a small-scale disk-like component whose size is $\sim 36\pm10$au and whose flux is 12 mJy at 231\,GHz. 
We stress that the fluxes in the two visibility profiles obtained with the PdBI are surprisingly flat (see Fig. \ref{fig:gf92_bothmodels}), even at the shortest baselines we probed (10 k$\lambda$ at 94 GHz): no clear envelope-like profile is detected in our data even at the relatively large scales we probe ($2-10\arcsec$). The envelope emission might either be completely filtered out at these scales, or GF9-2 might be a compact object that is embedded in a large-scale filamentary structure that is detected with the single-dish observations. The OVRO 3mm continuum emission was found to be shifted to the east with respect to the peak of the N$_2$H$^{+}$ core, which led \citet{Furuya06} to propose that the millimeter continuum emission was dominated by a protostellar object while the molecular core traces a younger object that might be prestellar in nature. Here the peaks of our PdBI dust continuum emission coincide with the OVRO 3mm continuum peak position, which trace the same object. 
The possibility that GF9-2 is a very young object (e.g., first hydrostatic core, see discussion about this possibility in \citealt{Furuya14}) seems to be ruled out by our detection of a jet that originates from the continuum source (Podio et al. in prep), but the current observations do not allow us to firmly establish the nature of the PdBI dust continuum source we detect.

Our modeling relies on the large-scale envelope parameters that are obtained solely from single-dish observations since emission from the envelope, if any, is not detected in our PdBI data. Because for GF9-2 the quality of the IRAM-30m data is significantly poorer than in other single-dish data that we used for the remaining sources in the sample, we had to use quite loose constraints on the single-dish fluxes that trace the outer scales in our modeling: we stress that the exact properties of this candidate disk in the GF9-2 core need to be confirmed.

\end{itemize}

\newpage


\begin{table*}
\caption{L1448-2A: overview of the modeling results}             
\label{table:l14482a}      
\centering                          
\begin{tabular}{c c c c c c c c c}        
\hline\hline                 
Model & DoF & $R_{\rm{out}}$ & $R_{\rm{i}}$ & $p+q$ & $F_{\rm{Tot}}$ & $\Theta_{\rm{Gauss}}$ & $F_{\rm{Gauss}}$ & $\chi^{2}_{\rm{red}}$  \\ 
 &  & ($\arcsec$) & ($\arcsec$) &  & (mJy) & ($\arcsec$) & (mJy) &   \\     
\hline                       
\multicolumn{9}{c}{{\bf{231 GHz}}} \\
   Pl & 27 & $18\pm4$ & 0.01$^{\ddagger}$ & $2.27\pm0.1$ & $527\pm30$ & - & - & 1.0 \\      
   PG & 25 & $20\pm10$ & $0.09\pm0.01$ & $2.22\pm0.1$ & $538\pm100$ & $0.06\pm0.2$ & $12\pm3$ & 1.0  \\
   PGf & 29 & 18\fp  & 0.01\fp & $2.27$\fp & 527\fp & 0.01$^{\ddagger}$ & 1$^{\ddagger}$ & 0.9 \\ 
   PGt & 26 & $18\pm5$ & $0.1\pm0.1$ & $2.0\pm0.1$ & $426\pm39$ & $0.1\pm0.1$ & 42 & 16 \\ 
   Pleq & 14 & 15$^{\ddagger}$ & 0.01$^{\ddagger}$ & $2.28\pm0.1$ & $509\pm15$ & - & - & 0.37 \\
   PGeq & 12 & $20\pm1$ & $0.05\pm0.01$ & $2.3\pm0.2$ & $555\pm41$ & 0.01$^{\ddagger}$ & $11\pm4$ & 0.67 \\
   \hline                        
\multicolumn{9}{c}{{\bf{94 GHz}}} \\
   Pl & 22 & 25\lul &  0.01$^{\ddagger}$ & $2.21\pm0.2$ & $58\pm5$ & - & - & 1.0 \\      
   PG & 20 & $22\pm2$ & 0.01$^{\ddagger}$ & $2.0\pm0.1$ & 60\lul & 0.01$^{\ddagger}$ & $2.19\pm0.5$ & 0.9  \\
\hline                                   
\end{tabular}
\tablefoot{Best-fit parameters for all models with which we attempted to reproduce the dust continuum visibilities profile of L1448-2A. Column 1 reports the model identifier, Col. 2 the number of degrees of freedom we used for the modeling, Col. 3 the outer radius of the truncated Plummer envelope model, Col. 4 the inner radius, Col. 5 the value of the (p+q) brightness radial distribution index, Col. 6 the total flux in the model (sum of the Plummer envelope and Gaussian component if PG model). Columns 7 and 8 report the FWHM and the flux of the additional Gaussian component for PG models, respectively while Col. 9 reports the reduced $\chi^2$ value for each model. \\
In models Pl and PG, all parameters were let free to vary (within the ranges described in the text) and all visibilities were used, that is, we traced the continuum emission in all directions around the peak of the millimeter dust continuum emission. 
PGf reports the parameters of the best-fit PG model we obtained when the envelope parameters were fixed to those from the best-fit Plummer-only model (with all visibilities; reported in Pl). 
PGt reports the parameters of the best-fit PG model we obtained when the Gaussian component flux wa tied to the envelope flux (forcing it to be 10\% of the total flux).
In the Pleq and PGeq  models,  all parameters were let free to vary (within the ranges described in the text), but we only used the visibilities from the uv-plane that traced the direction of the equatorial plane (PA $-80^{\circ}\pm30^{\circ}$ for L1448-2A).\\
\tablefoottext{\lul}{This symbol is used when the parameter value is the upper limit that is allowed by the fitting procedure.}
\tablefoottext{${\ddagger}$} This symbol indicates that the parameter value is the lower limit that is allowed by the fitting procedure.
\fp This symbol indicates that the parameter was fixed during the minimization process.
}
\end{table*}


%
%
\begin{table*}
\caption{L1448-NB1: overview of the modeling results}             
\label{table:l1448nb1}      
\centering                          
\begin{tabular}{c c c c c c c c c}   
\hline\hline                 
Model & DoF & $R_{\rm{out}}$ & $R_{\rm{i}}$ & $p+q$ & $F_{\rm{Tot}}$ & $\Theta_{\rm{Gauss}}$ & $F_{\rm{Gauss}}$ & $\chi^{2}_{\rm{red}}$  \\ 
 &  & ($\arcsec$) & ($\arcsec$) &  & (mJy) & ($\arcsec$) & (mJy) &   \\     
\hline                        
\multicolumn{9}{c}{{\bf{231 GHz}}} \\
   Pl & 28 & 25\lul & 0.01$^{\ddagger}$ & $2.24\pm0.2$ & 3373$\pm$200 & - & - & 3.0 \\      
   PG & 26 & 23$\pm$3 & $0.06\pm0.02$ & $2.22\pm0.2$ & 3176$\pm$380 & $0.01^{\ddagger}$ & 38$\pm$11 & 3.2  \\
   PGf & 30 & 25\fp & 0.01\fp & $2.24$\fp & 3373\fp & 0.01$^{\ddagger}$ & 2 & 2.7  \\ 
   PGt & 27 & 22$\pm$7 & $0.15\pm0.1$ & $1.87\pm0.4$ & 2240$^{\ddagger}$  & 0.15$\pm$0.1 & 224$^{\ddagger}$ & 34  \\ 
   Pleq & 28 & 25\lul & $0.04\pm0.01$ & $2.26\pm0.1$ & 3370$\pm$350 & - & - & 2.5 \\      
   PGeq & 26 & 24$\pm$7 & $0.05\pm0.03$ & $2.25\pm0.2$ & 3370$\pm$1000 & 0.01$^{\ddagger}$ & 31$\pm$13 & 2.7  \\
   \hline                        
\multicolumn{9}{c}{{\bf{94 GHz}}} \\
   Pl & 22 & 19$\pm$3 & $0.2\pm0.1$ & $2.70\pm0.3$ & 226$^{\ddagger}$ & - & - & 7 \\      
   PG & 20 & 23$\pm$3 & $0.94\pm0.1$ & $2.73\pm$0.2 & 243$\pm$18 & 0.94$\pm$0.1 & 39$\pm$4 & 5.0  \\
   Pleq & 19 & 25\lul & 0.01$^{\ddagger}$ & $2.20\pm0.2$ & 353$\pm$39 & - & - & 3.2 \\      
   PGeq & 17 & 25\lul & $0.09\pm0.03$ & $2.61\pm0.2$ & 270$\pm$98 & $0.09\pm0.03$ & 1.0$^{\ddagger}$ & 10  \\
\hline       
\end{tabular}
\tablefoot{
Same as Table \ref{table:l14482a} for L1448-NB1.}
\end{table*}
\begin{table*}
\caption{L1448-NB2: overview of the modeling results}             
\label{table:l1448nb2}      
\centering    
\small{
\begin{tabular}{c c c c c c c c c}        
\hline\hline                 
Model & DoF & $R_{\rm{out}}$ & $R_{\rm{i}}$ & $p+q$ & $F_{\rm{Tot}}$ & $\Theta_{\rm{Gauss}}$ & $F_{\rm{Gauss}}$ & $\chi^{2}_{\rm{red}}$  \\ 
 &  & ($\arcsec$) & ($\arcsec$) &  & (mJy) & ($\arcsec$) & (mJy) &   \\     
\hline                       
\multicolumn{9}{c}{{\bf{231 GHz}}} \\                  
   Pl & 28 & $21\pm9$ & 0.36$\pm$0.09 & $2.5\pm0.3$ & $2433\pm600$ & - & - & 2.2 \\      
   PG & 26 & $16\pm4$ & $0.9\pm0.2$ & $2.51\pm0.4$ & $2200\pm206$ & $0.9\pm0.1$ & $169\pm27$ & 2.7  \\
   PGf & 30 & 21\fp  & 0.36\fp & $2.5$\fp & 2433\fp & $0.32\pm1$ & $4.1\pm4$ & 2.0 \\ 
   PGt & 27 & $15^{\ddagger}$  & $1.0\pm0.3$ & $2.18\pm0.2$ & $2241\pm460$ & $1.00\pm0.2$ & 224\fp & 3.2 \\ 
   Pleq & 28 & $20\pm7$  & $0.14\pm0.05$ & $2.4\pm0.2$ & $2506\pm615$ & - & - & 2.8 \\
   PGeq & 26 & $17.8\pm5$  & $0.9\pm0.2$ & $2.46\pm0.6$ & $2348\pm140$ & $0.8\pm0.3$ & $241\pm37$ &  4.3 \\
   \hline                        
\multicolumn{9}{c}{{\bf{94 GHz}}} \\
   Pl & 22 & 15$^{\ddagger}$  & $0.06\pm0.06$ & $2.3\pm0.1$ & $211\pm13$ & - & - & 0.7 \\      
   PG & 20 & $25\pm2$  & $0.1\pm0.1$ & $2.4\pm0.1$ & $280\pm29$ & $0.1\pm0.1$ & $1^{\ddagger}$ & 1.2  \\
   Pleq & 17 & $17\pm2$  & $0.06\pm0.04$ & $2.4\pm0.2$ & 209$^{\ddagger}$ & - & - & 0.55 \\      
   PGeq & 15 & $25$\lul  & $0.19\pm0.1$ & $2.64\pm0.2$ & $210\pm52$ & $0.2\pm0.1$ & $2.9\pm1$ & 0.72  \\
\hline                                   
\end{tabular}
\tablefoot{Same as Table \ref{table:l1448nb1} for L1448-NB2}.
}
\end{table*}
\begin{table*}
\caption{L1448-NB centered at the barycenter of the two millimeter sources: overview of the modeling results}             
\label{table:l1448nbcent}      
\centering        
\small{
\begin{tabular}{c c c c c c c c c}        
\hline\hline                 
Model & DoF & $R_{\rm{out}}$ & $R_{\rm{i}}$ & $p+q$ & $F_{\rm{Tot}}$ & $\Theta_{\rm{Gauss}}$ & $F_{\rm{Gauss}}$ & $\chi^{2}_{\rm{red}}$  \\ 
 &  & ($\arcsec$) & ($\arcsec$) &  & (mJy) & ($\arcsec$) & (mJy) &   \\     
\hline                       
\multicolumn{9}{c}{{\bf{231 GHz}}} \\ 
   Pl & 28 & $21\pm14$ & $0.6\pm0.15$ & $2.8\pm0.3$ & $2256\pm720$ & - & - & 3.6 \\      
   PG & 26 & $15\pm1.9$ & $0.06\pm0.02$ & $2.0\pm0.1$ & $2397\pm170$ & $0.01^{\ddagger}$ & 1$^{\ddagger}$ & 6.9  \\
   PGf & 31 & $21$\fp & 0.6\fp & $2.8$\fp & $2256$\fp & $0.01^{\ddagger}$ & $10.9\pm5$ & 3.0 \\ 
   PGt & 27 & $15^{\ddagger}$ & $0.9\pm0.2$ & $1.94\pm0.3$ & $2150^{\ddagger}$ & $0.8\pm0.2$ & 215\fp & 11.6 \\ 
   Pleq & 28 & $21\pm24$ & $0.5\pm0.1$ & $2.8\lul$ & $2234\pm1070$ & - & - & 7.4 \\
   PGeq & 26 & $24\pm10$ & $0.6\pm0.1$ & $2.8\lul$ & $2351\pm142$ & $0.5\pm0.2$ & $1^{\ddagger}$ & 7.3 \\
   \hline                        
\multicolumn{9}{c}{{\bf{94 GHz}}} \\
   Pl & 22 & $22\pm5$ & $0.13\pm0.1$ & $2.4\pm0.1$ & $237\pm36$ & - & - & 1.5 \\      
   PG & 20 & $19\pm3$ & $1.0\pm0.2$ & $2.27\pm0.2$ & $227\pm13$ & $0.95\pm0.1$ & $34\pm4$ & 1.0 \\
   Pleq & 17 & $22\pm7$ & $0.1\pm0.1$ & $2.5\pm0.3$ & $217\pm46$ & - & - & 0.7 \\      
   PGeq & 15 & $18\pm2$ & $1.3\pm0.4$ & $2.1\pm0.2$ & $219\pm20$ & $1.1\pm0.3$ & $50\pm5$ & 0.16  \\
\hline                                   
\end{tabular}
\tablefoot{Same as Table \ref{table:l1448nb1}. The modeling visibilities are centered around the barycenter of L1448-NB1 and NB2.
}}
\end{table*}


\begin{table*}
\caption{L1448-C: overview of the modeling results}             
\label{table:l1448c}      
\centering                          
\small{
\begin{tabular}{c c c c c c c c c}        
\hline\hline                 
Model & DoF & $R_{\rm{out}}$ & $R_{\rm{i}}$ & $p+q$ & $F_{\rm{Tot}}$ & $\Theta_{\rm{Gauss}}$ & $F_{\rm{Gauss}}$ & $\chi^{2}_{\rm{red}}$  \\ 
 &  & ($\arcsec$) & ($\arcsec$) &  & (mJy) & ($\arcsec$) & (mJy) &   \\     
\hline                       
\multicolumn{9}{c}{{\bf{231 GHz}}} \\ 
   Pl & 27 & $12.3\pm3$ & $0.01^{\ddagger}$ & $2.54\pm0.2$ & $660^{\ddagger}$ & - & - & 6.8 \\      
   PG & 25 & $14.5\pm4$ & $0.2\pm0.1$ & $1.72\pm0.2$ & $990\lul$ & $0.16\pm0.05$ & $130\pm5$ & 0.7  \\
   PGf & 29 & 12.3\fp & $0.01$\fp &$2.54$\fp & 660\fp & 0.01$^{\ddagger}$ & $11\pm8$ & 6 \\ 
   PGt & 26 & $12\pm7$ & $0.01^{\ddagger}$ & $2.0\pm0.2$ & $871\pm63$ & $0.01^{\ddagger}$ & 87\fp & 1.6 \\ 
   Pleq & 15 & 17\lul & $0.01^{\ddagger}$ & $2.62\pm0.2$ & $660^{\ddagger}$ & - & - & 4.5 \\
   PGeq & 13 & $11\pm10$ & $0.2\pm0.1$ & $1.9\pm0.3$ & $781\pm70$ & $0.2\pm0.1$ & $123\pm6$ & 0.32 \\
   \hline                        
\multicolumn{9}{c}{{\bf{94 GHz}}} \\
   Pl & 22 & $17$\lul  & $0.01^{\ddagger}$ & $2.74\pm0.2$ & 53$^{\ddagger}$ & - & - & 9 \\      
   PG & 20 & $14\pm1$  & $0.05\pm0.03$ & $1.4^{\ddagger}$ & $79\pm2$ & $0.07\pm0.04$ & $18\pm1$ & 0.1  \\
\hline                                   
\end{tabular}
}
\end{table*}


\begin{table*}
\caption{NGC1333 IRAS2A: overview of the modeling results}             
\label{table:iras2a}      
\centering                          
\small{
\begin{tabular}{c c c c c c c c c}        
\hline\hline                 
Model & DoF & $R_{\rm{out}}$ & $R_{\rm{i}}$ & $p+q$ & $F_{\rm{Tot}}$ & $\Theta_{\rm{Gauss}}$ & $F_{\rm{Gauss}}$ & $\chi^{2}_{\rm{red}}$  \\ 
 &  & ($\arcsec$) & ($\arcsec$) &  & (mJy) & ($\arcsec$) & (mJy) &   \\     
\hline                       
\multicolumn{9}{c}{{\bf{231 GHz}}} \\ 
   Pl & 37 & $5.7\pm1$ & $0.01^{\ddagger}$ & $2.5\pm0.2$ & $600^{\ddagger}$ & - & - & 1.3 \\      
   PG & 35 & $7.7\pm1$ & $0.05\pm0.02$ & $2.5\pm0.2$ & $652\pm40$ & $0.01^{\ddagger}$ & $52\pm5$ & 1.3  \\
   PGf & 39 & 5.7\fp & 0.01\fp& $2.5\fp$ & 600\fp & 0.01$^{\ddagger}$ & $16\pm3$ & 1.2 \\ 
   PGt & 36 & $5.3\pm2$ & $ 0.02\pm0.01$ & $2.3\pm0.2$ & $613\pm42$ & $0.02\pm0.1$ & 61\fp & 0.75 \\ 
   Pleq & 17 & $8.2\pm2$ & $0.01^{\ddagger}$ & $2.6\pm0.2$ & $600^{\ddagger}$ & - & - & 0.61 \\
   PGeq & 15 & $6.3\pm2$ & $0.05\pm0.01$ & $2.6\pm0.2$ & $600^{\ddagger}$ & $0.01^{\ddagger}$ & $9\pm10$ & 1.4 \\
   \hline                        
\multicolumn{9}{c}{{\bf{94 GHz}}} \\
   Pl & 21 & $10\lul$ & $0.01^{\ddagger}$ & $2.6\pm0.2$ & $65\pm2$ & - & - & 0.95 \\      
   PG & 19 & $10\lul$ & $0.06\pm0.02$ & $2.4\pm0.2$ & $65\pm2$ & 0.01$^{\ddagger}$ & $9\pm1$ & 0.47  \\
\hline                                   
\end{tabular}
}
\end{table*}


\begin{table*}
\caption{SVS13B: overview of the modeling results}             
\label{table:svs13b}      
\centering                          
\small{
\begin{tabular}{c c c c c c c c c}        
\hline\hline                 
Model & DoF & $R_{\rm{out}}$ & $R_{\rm{i}}$ & $p+q$ & $F_{\rm{Tot}}$ & $\Theta_{\rm{Gauss}}$ & $F_{\rm{Gauss}}$ & $\chi^{2}_{\rm{red}}$  \\ 
 &  & ($\arcsec$) & ($\arcsec$) &  & (mJy) & ($\arcsec$) & (mJy) &   \\     
\hline                       
\multicolumn{9}{c}{{\bf{231 GHz}}} \\ 
   Pl & 26 & $14$\lul & $0.06\pm0.02$ & $2.9$\lul & $446\pm15$ & - & - & 2.5 \\      
   PG & 24 & $9.3\pm3$ & $0.2\pm0.1$ & $2.5\pm0.3$ & $636\pm88$ & $0.19\pm0.1$ & $80\pm7$ & 4.6  \\
   PGf & 29 & 14\fp & 0.06\fp & $2.9$\fp & 446\fp & $0.5\pm1$ & $1^{\ddagger}$ & 2.4 \\ 
   PGt & 25 & $11.3\pm3$ & $0.3\pm0.1$ & $2.1\pm0.2$ & $1070$\lul & $0.30\pm0.1$ & 107\fp & 2.9 \\ 
   Pleq & 11 & $6\pm9$ & $0.04\pm0.015$ & $2.9$\lul & $319\pm240$ & - & - & 1.7 \\
   PGeq & 9 & $8.3\pm3$ & $0.3\pm0.1$ & $2.7\pm0.3$ & $387\pm60$ & $0.25\pm0.1$ & $107\pm15$ & 1.8 \\
   \hline                        
\multicolumn{9}{c}{{\bf{94 GHz}}} \\
   Pl & 19 & $6.7\pm2$ & $0.01^{\ddagger}$ & $2.58\pm0.2$ & $52\pm4$ & - & - & 0.45 \\      
   PG & 17 & $10.3\pm3$ & $0.1\pm0.05$ & $2.4\pm0.2$ & $68\pm4$ & $0.09\pm0.1$ & $10\pm1$ & 0.6  \\
\hline                                   
\end{tabular}
\tablefoot{See Table \ref{table:l14482a} for a description of the models and columns.}
}
\end{table*}


\begin{table*}
\caption{IRAS4A1: overview of the modeling results}             
\label{table:iras4a}      
\centering                          
\small{
\begin{tabular}{c c c c c c c c c}        
\hline\hline                 
Model & DoF & $R_{\rm{out}}$ & $R_{\rm{i}}$ & $p+q$ & $F_{\rm{Tot}}$ & $\Theta_{\rm{Gauss}}$ & $F_{\rm{Gauss}}$ & $\chi^{2}_{\rm{red}}$  \\ 
 &  & ($\arcsec$) & ($\arcsec$) &  & (mJy) & ($\arcsec$) & (mJy) &   \\     
\hline                       
\multicolumn{9}{c}{{\bf{231 GHz}}} \\ 
   Pl & 26 & $3.7\pm0.5$ & $0.1\pm0.05$ & $2.48\pm0.1$ & $3489\pm106$ & - & - & 1.6 \\      
   PG & 24 & $3.8\pm0.6$ & $0.4\pm0.1$ & $2.69\pm0.2$ & $3503\pm60$ & $0.32\pm0.1$ & $348\pm25$ & 1.7  \\
   PGf & 28 & $3.7$\fp & 0.1\fp & 2.48\fp & 3489\fp & $0.01^{\ddagger}$ & $1^{\ddagger}$ & 0.9 \\
   PGt & 25 & $3.7\pm0.3$ & $0.4\pm0.05$ & $2.8\pm0.2$ & $3516\pm600$ & $0.4\pm0.1$ & $351$\fp & 2.0 \\
   Pleq & 11 & $5.9\pm1$ & $0.1\pm0.05$ & $2.7\pm0.2$ & 3900\lul & - & - & 2.5 \\
   PGeq & 9 & $5.7\pm1$ & $0.5\pm0.1$ & $2.8\pm0.2$ & $3900\pm3000$ & $0.46\pm0.1$ & $600\pm45$ & 3.5 \\
   \hline                        
\multicolumn{9}{c}{{\bf{94 GHz}}} \\
   Pl & 21 & $5.4\pm0.8$ & $0.07\pm0.02$ & $2.8\pm0.2$ & $364\pm12$ & - & - & 2.2 \\      
   PG & 19 & $4.5\pm0.3$ & $0.5\pm0.1$ & $2.3\pm0.1$ & $363\pm30$ & $0.5\pm0.1$ & $130\pm9$ & 4.3 \\
\hline                                   
\end{tabular}
\tablefoot{See Table \ref{table:l14482a} for a description of the models and columns.}
}
\end{table*}


\begin{table*}
\caption{IRAS4B: overview of the modeling results}             
\label{table:iras4b}      
\centering                          
\small{
\begin{tabular}{c c c c c c c c c}        
\hline\hline                 
Model & DoF & $R_{\rm{out}}$ & $R_{\rm{i}}$ & $p+q$ & $F_{\rm{Tot}}$ & $\Theta_{\rm{Gauss}}$ & $F_{\rm{Gauss}}$ & $\chi^{2}_{\rm{red}}$  \\ 
 &  & ($\arcsec$) & ($\arcsec$) &  & (mJy) & ($\arcsec$) & (mJy) &   \\     
\hline                       
\multicolumn{9}{c}{{\bf{231 GHz}}} \\ 
   Pl & 28 & $4\pm2$ & $0.145\pm0.1$ & $2.9$\lul & $1448\pm40$ & - & - & 3.3 \\      
   PG & 26 & $3.8\pm1$ & $1.0\pm0.2$ & $2.9$\lul & $1486\pm268$ & $0.53\pm0.1$ & $645\pm35$ & 2.5  \\
   PGf & 30 & $4$\fp & $0.145$\fp & 2.9\fp & 1448\fp & $0.66\pm0.2$ & $216\pm25$ & 1.19 \\
   PGt & 28 & $3.7\pm1$ & $0.29\pm0.1$ & $2.9$\lul & $1736\pm162$ & $0.29\pm0.08$ & $173$ & 7.5  \\
   Pleq & 15 & $3.0^{\ddagger}$ & $0.12\pm0.1$ & $2.9$\lul & $1408\pm46$ & - & - & 2.65 \\
   PGeq & 13 & $3.7\pm3$ & $0.46\pm0.04$ & $2.8\pm0.4$ & $1351\pm68$ & $0.45\pm0.1$ & $562\pm33$ & 2.1 \\
   \hline                        
\multicolumn{9}{c}{{\bf{94 GHz}}} \\
   Pl & 20 & $3.9\pm0.4$ & $0.14\pm0.05$ & $2.9$\lul & $141\pm6$ & - & - & 0.8 \\      
   PG & 19 & $7.8\pm0.5$ & $0.3\pm0.1$ & $2.9$\lul & $155\pm11$ & $0.29\pm0.15$ & $24\pm7$ & 1.8 \\
\hline                                   
\end{tabular}
\tablefoot{See Table \ref{table:l14482a} for a description of the models and columns.}
}
\end{table*}


\begin{table*}
\caption{IRAM04191: overview of the modeling results}             
\label{table:iram04191}      
\centering                          
\small{
\begin{tabular}{c c c c c c c c c}        
\hline\hline                 
Model & DoF & $R_{\rm{out}}$ & $R_{\rm{i}}$ & $p+q$ & $F_{\rm{Tot}}$ & $\Theta_{\rm{Gauss}}$ & $F_{\rm{Gauss}}$ & $\chi^{2}_{\rm{red}}$  \\ 
 &  & ($\arcsec$) & ($\arcsec$) &  & (mJy) & ($\arcsec$) & (mJy) &   \\     
\hline                       
\multicolumn{9}{c}{{\bf{231 GHz}}} \\ 
Pl & 26 & $27\pm2$ & 0.01$^{\ddagger}$ & $1.6\pm0.2$ & $410^{\ddagger}$ & - & - & 2.37 \\      
PG & 24 & $28\pm2$ & $0.2\pm0.3$ & $1.4^{\ddagger}$ & $543\pm10$ & $0.17\pm0.1$ & $3.6\pm1$ & 0.84\\
& &  &  &  &  &  &  \\
\hline                        
\multicolumn{9}{c}{{\bf{94 GHz}}} \\
   Pl & 22 & $29\pm2$ & $0.01^{\ddagger}$ & $1.4^{\ddagger}$ & $44\pm9$ & - & - & 0.86 \\      
   PG & 20 & $29\pm4$ & $1.28\pm1$ & $1.54\pm0.2$ & $33\pm9$ & $0.15\pm0.4$ & $0.2\pm0.1$ & 1.0 \\
\hline                                   
\end{tabular}
\tablefoot{See Table \ref{table:l14482a} for a description of the models and columns.}
}
\end{table*}


\begin{table*}
\caption{L1521F: overview of the modeling results}             
\label{table:l1521f}      
\centering                          
\small{
\begin{tabular}{c c c c c c c c c}        
\hline\hline                 
Model & DoF & $R_{\rm{out}}$ & $R_{\rm{i}}$ & $p+q$ & $F_{\rm{Tot}}$ & $\Theta_{\rm{Gauss}}$ & $F_{\rm{Gauss}}$ & $\chi^{2}_{\rm{red}}$  \\ 
 &  & ($\arcsec$) & ($\arcsec$) &  & (mJy) & ($\arcsec$) & (mJy) &   \\     
\hline                       
\multicolumn{9}{c}{{\bf{231 GHz}}} \\
Pl & 27 & $32\pm2$ & $3.0$\lul & $1.72\pm0.2$ & $1100$\lul & - & - & 2.4  \\      
PG & 25 & $32\pm3$ & $2.6\pm0.8$ & $1.63\pm0.4$ & $1100$\lul & $0.13\pm0.1$ & $1.3\pm0.4$ & 0.79 \\
& &  &  &  &  &  &  \\
\hline                        
\multicolumn{9}{c}{{\bf{94 GHz}}} \\
   Pl & 10 & $37 \pm 8$ & $3.0$\lul & $2.0 \pm 0.2$ & $53^{\ddagger}$ & - & - & 1.35  \\      
   PG & 8 & $39 \pm 10$ & $1.81\pm0.8$ & $1.65 \pm 0.5$ & $53^{\ddagger}$ & $0.16\pm0.2$ & $0.5^{\ddagger}$ & 5 \\
\hline                                   
\end{tabular}
\tablefoot{See Table \ref{table:l14482a} for a description of the models and columns.}
}
\end{table*}


\begin{table*}
\caption{L1527: overview of the modeling results}             
\label{table:l1527}      
\centering                          
\small{
\begin{tabular}{c c c c c c c c c}        
\hline\hline                 
Model & DoF & $R_{\rm{out}}$ & $R_{\rm{i}}$ & $p+q$ & $F_{\rm{Tot}}$ & $\Theta_{\rm{Gauss}}$ & $F_{\rm{Gauss}}$ & $\chi^{2}_{\rm{red}}$  \\ 
 &  & ($\arcsec$) & ($\arcsec$) &  & (mJy) & ($\arcsec$) & (mJy) &   \\     
\hline                       
\multicolumn{9}{c}{{\bf{231 GHz}}} \\
   Pl & 27 & $35$\lul & $0.01^{\ddagger}$ & $2.57\pm0.2$ & $1080^{\ddagger}$ & - & - & 19 \\      
   PG & 25 & $28\pm4$ & $0.4\pm0.2$ & $1.68\pm0.4$ & $1275\pm320$ & $0.4\pm0.1$ & $215\pm14$ & 2.9  \\
   Pleq & 25 & $35$\lul & $0.01^{\ddagger}$ & $2.64\pm0.2$ & $1080^{\ddagger}$ & - & - & 9.1 \\
   PGeq & 23 & $28\pm5$ & $0.4\pm0.2$ & $1.70\pm0.3$ & $1176\pm200$ & $0.4\pm0.1$ & $226\pm13$ & 2.9  \\
   \hline                        
\multicolumn{9}{c}{{\bf{94 GHz}}} \\
   Pl & 22 & $35$\lul & $0.01^{\ddagger}$ & $2.6\pm0.2$ & $85^{\ddagger}$ & - & - & 4.0 \\      
   PG & 20 & $35$\lul & $0.38\pm0.2$ & $1.78\pm0.3$ & $85^{\ddagger}$ & $0.3\pm0.1$ & $23\pm1$ & 0.6 \\
\hline                                   
\end{tabular}
\tablefoot{See Table \ref{table:l14482a} for a description of the models and columns.}
}
\end{table*}


\begin{table*}
\caption{SerpM-S68N: overview of the modeling results}             
\label{table:serpms68n}      
\centering                          
\small{
\begin{tabular}{c c c c c c c c c}        
\hline\hline                 
Model & DoF & $R_{\rm{out}}$ & $R_{\rm{i}}$ & $p+q$ & $F_{\rm{Tot}}$ & $\Theta_{\rm{Gauss}}$ & $F_{\rm{Gauss}}$ & $\chi^{2}_{\rm{red}}$  \\ 
 &  & ($\arcsec$) & ($\arcsec$) &  & (mJy) & ($\arcsec$) & (mJy) &   \\     
\hline                       
\multicolumn{9}{c}{{\bf{231 GHz}}} \\
   Pl & 26 & $15\pm2$ & $0.01^{\ddagger}$ & $2.28\pm0.1$ & $800^{\ddagger}$ & - & - & 1.67 \\      
   PG & 24 & $13.9\pm3$ & $0.1\pm0.1$ & $2.11\pm0.2$ & $800^{\ddagger}$ & $0.11\pm0.1$ & $28\pm11$ & 2.1  \\
   Pleq & 21 & $17.9\pm1$ & $0.01^{\ddagger}$ & $2.27\pm0.1$ & $940$\lul & - & - & 0.5 \\
   PGeq & 19 & $16.6\pm5$ & $0.1\pm0.1$ & $2.11\pm0.2$ & $940$\lul & $0.15\pm0.07$ & $32\pm7$ & 0.7 \\
   \hline                        
\multicolumn{9}{c}{{\bf{94 GHz}}} \\
   Pl & 22 & $22$\lul & $0.03\pm0.03$ & $2.56\pm0.2$ & $35^{\ddagger}$ & - & - & 1.2 \\      
   PG & 20 & $15\pm2$ & $0.5\pm0.1$ & $2.08\pm0.2$ & $35^{\ddagger}$ & $0.52\pm0.1$ & $6.5\pm1$ & 3.4 \\
\hline                                   
\end{tabular}
\tablefoot{See Table \ref{table:l14482a} for a description of the models and columns.}
}
\end{table*}


\begin{table*}
\caption{SerpM-SMM4: overview of the modeling results}             
\label{table:serpmsmm4}      
\centering                          
\small{
\begin{tabular}{c c c c c c c c c}        
\hline\hline                 
Model & DoF & $R_{\rm{out}}$ & $R_{\rm{i}}$ & $p+q$ & $F_{\rm{Tot}}$ & $\Theta_{\rm{Gauss}}$ & $F_{\rm{Gauss}}$ & $\chi^{2}_{\rm{red}}$  \\ 
 &  & ($\arcsec$) & ($\arcsec$) &  & (mJy) & ($\arcsec$) & (mJy) &   \\     
\hline                       
\multicolumn{9}{c}{{\bf{231 GHz}}} \\
   Pl & 24 & $25$\lul & $0.06\pm0.05$ & $2.8$\lul & $1700^{\ddagger}$ & - & - & 2.7  \\      
   PG & 23 & $22\pm5$ & $0.7\pm0.1$ & $1.8\pm0.5$ & $1900\pm1000$ & $0.70\pm0.2$ & $595\pm35$ & 0.14 \\
   \hline                        
\multicolumn{9}{c}{{\bf{94 GHz}}} \\
   Pl & 22 & $15^{\ddagger}$ & $0.01^{\ddagger}$ & $2.8$\lul & $108\pm6$ & - & - & 3.9 \\      
   PG & 21 & $25$\lul & $0.65\pm0.1$ & $1.6^{\ddagger}$ & $103^{\ddagger}$ & $0.62\pm0.1$ & $61\pm3$ & 0.34 \\
\hline                                   
\end{tabular}
\tablefoot{See Table \ref{table:l14482a} for a description of the models and columns.}
}
\end{table*}


\begin{table*}
\caption{SerpS-MM18: overview of the modeling results}             
\label{table:ssmm18}      
\centering                          
\small{
\begin{tabular}{c c c c c c c c c}        
\hline\hline                 
Model & DoF & $R_{\rm{out}}$ & $R_{\rm{i}}$ & $p+q$ & $F_{\rm{Tot}}$ & $\Theta_{\rm{Gauss}}$ & $F_{\rm{Gauss}}$ & $\chi^{2}_{\rm{red}}$  \\ 
 &  & ($\arcsec$) & ($\arcsec$) &  & (mJy) & ($\arcsec$) & (mJy) &   \\     
\hline                       
\multicolumn{9}{c}{{\bf{231 GHz}}} \\
   Pl & 27 & $16\pm2$ & $0.01^{\ddagger}$ & $2.24\pm0.2$ & $2208\pm190$ & - & - & 1.8  \\      
   PG & 25 & $15.5\pm6$ & $0.13\pm0.07$ & $2.17\pm0.2$ & $2327\pm55$ & $0.128\pm0.08$ & $76\pm4$ & 0.68 \\
   \hline                        
\multicolumn{9}{c}{{\bf{94 GHz}}} \\
   Pl & 22 & $20$\lul & $0.015\pm0.05$ & $2.40\pm0.2$ & $114^{\ddagger}$ & - & - & 0.87 \\      
   PG & 20 & $11\pm2$ & $0.04\pm0.02$ & $2.24\pm0.2$ & $114^{\ddagger}$ & $0.01^{\ddagger}$ & $1^{\ddagger}$ & 2.78 \\
\hline                                   
\end{tabular}
\tablefoot{See Table \ref{table:l14482a} for a description of the models and columns.}
}
\end{table*}


\begin{table*}
\caption{SerpS-MM22: overview of the modeling results}             
\label{table:ssmm22}      
\centering                          
\small{
\begin{tabular}{c c c c c c c c c}        
\hline\hline                 
Model & DoF & $R_{\rm{out}}$ & $R_{\rm{i}}$ & $p+q$ & $F_{\rm{Tot}}$ & $\Theta_{\rm{Gauss}}$ & $F_{\rm{Gauss}}$ & $\chi^{2}_{\rm{red}}$  \\ 
 &  & ($\arcsec$) & ($\arcsec$) &  & (mJy) & ($\arcsec$) & (mJy) &   \\     
\hline                       
\multicolumn{9}{c}{{\bf{231 GHz}}} \\
   Pl & 27 & $11\pm2$ & $0.01^{\ddagger}$  & $2.56\pm0.2$ & $135$\lul & - & - & 1.8  \\      
   PG & 25 & $10\pm4$ & $0.26\pm0.08$ & $2.00\pm0.3$ & $148\pm9$ & $0.25\pm0.08$ & $31\pm4$ & 1.1 \\
   \hline                        
\multicolumn{9}{c}{{\bf{94 GHz}}} \\
   Pl & 22 & $25$\lul & $0.01^{\ddagger}$ & $2.71\pm0.3$ & $10^{\ddagger}$ & - & - & 1.16 \\      
   PG & 20 & $19\pm3$ & $0.31\pm0.07$ & $1.98\pm0.3$ & $10^{\ddagger}$ & $0.31\pm0.07$ & $3.2\pm0.5$ & 0.59 \\
\hline                                   
\end{tabular}
\tablefoot{See Table \ref{table:l14482a} for a description of the models and columns.}
}
\end{table*}


\begin{table*}
\caption{L1157: overview of the modeling results}             
\label{table:l1157}      
\centering                          
\small{
\begin{tabular}{c c c c c c c c c}        
\hline\hline                 
Model & DoF & $R_{\rm{out}}$ & $R_{\rm{i}}$ & $p+q$ & $F_{\rm{Tot}}$ & $\Theta_{\rm{Gauss}}$ & $F_{\rm{Gauss}}$ & $\chi^{2}_{\rm{red}}$  \\ 
 &  & ($\arcsec$) & ($\arcsec$) &  & (mJy) & ($\arcsec$) & (mJy) &   \\     
\hline                       
\multicolumn{9}{c}{{\bf{231 GHz}}} \\
   Pl & 24 & $12\pm10$ & $0.01^{\ddagger}$ & $2.68\pm0.2$ & $520\pm170$ & - & - & 1.29 \\      
   PG & 22 & $6.8\pm2$ & $0.05\pm0.04$ & $2.50\pm0.2$ & $494\pm35$ & $0.02\pm0.2$ & $56\pm6$ & 0.76  \\
   PGf & 26 & $12$\fp & $0.01$\fp & 2.68\fp & 520\fp & $0.01$\lul & $1.8\pm2$ & 1.2 \\
   PGt & 23 & $6.76\pm1$ & $0.01^{\ddagger}$ & $2.45\pm0.2$ & $551\pm17$ & $0.01^{\ddagger}$ & $55.1$ & 0.95  \\
   Pleq & 13 & $10.3\pm7$ & $0.01^{\ddagger}$ & $2.63\pm0.2$ & $573\pm150$ & - & - & 0.31 \\
   PGeq & 11 & $8.1\pm2$ & $0.03\pm0.02$ & $2.43\pm0.3$ & $659\pm29$ & $0.01$\lul & $57.7\pm4$ & 0.19 \\
   \hline                        
\multicolumn{9}{c}{{\bf{94 GHz}}} \\
   Pl & 15 & $16\pm5$ & $0.01^{\ddagger}$ & $2.65\pm0.2$ & $61$\lul & - & - & 1.4 \\      
   PG & 13 & $8.2\pm2$ & $0.05\pm0.04$ & $2.24\pm0.3$ & $61$\lul & $0.01$\lul & $9\pm1$ & 2.5 \\
\hline                                   
\end{tabular}
\tablefoot{See Table \ref{table:l14482a} for a description of the models and columns.}
}
\end{table*}


\begin{table*}
\caption{GF9-2: overview of the modeling results}             
\label{table:gf92}      
\centering                          
\small{
\begin{tabular}{c c c c c c c c c}        
\hline\hline                 
Model & DoF & $R_{\rm{out}}$ & $R_{\rm{i}}$ & $p+q$ & $F_{\rm{Tot}}$ & $\Theta_{\rm{Gauss}}$ & $F_{\rm{Gauss}}$ & $\chi^{2}_{\rm{red}}$  \\ 
 &  & ($\arcsec$) & ($\arcsec$) &  & (mJy) & ($\arcsec$) & (mJy) &   \\     
\hline                       
\multicolumn{9}{c}{{\bf{231 GHz}}} \\
   Pl & 27 & $40$\lul & $0.01^{\ddagger}$ & $2.36\pm0.2$ & $179$\lul & - & - & 5.0 \\      
   PG & 25 & $37\pm3$ & $0.18\pm0.07$ & $1.67\pm0.3$ & $407$\lul & $0.18\pm0.06$ & $11.8\pm1$ & 0.55  \\
   \hline                        
\multicolumn{9}{c}{{\bf{94 GHz}}} \\
   Pl & 17 & $40$\lul & $0.01^{\ddagger}$ & $2.20\pm0.2$ & $27^{\ddagger}$ & - & - & 1.8 \\      
   PG & 15 & $34\pm5$ & $0.15\pm0.07$ & $1.71\pm0.3$ & $39\pm20$ & $0.01^{\ddagger}$ & $1.3\pm0.5$ & 0.45 \\
\hline                                   
\end{tabular}
\tablefoot{See Table \ref{table:l14482a} for a description of the models and columns.}
}
\end{table*}

\end{appendix}

\end{document}